\documentclass{egpubl}

 \JournalSubmission    
 \electronicVersion 

\ifpdf \usepackage[pdftex]{graphicx} \pdfcompresslevel=9
\else \usepackage[dvips]{graphicx} \fi

\PrintedOrElectronic

\usepackage{t1enc,dfadobe}

\usepackage{egweblnk}
\usepackage{cite}

\usepackage{algorithm}
\usepackage{algorithmic}
\usepackage{array}
\usepackage{amsfonts,amsmath}
\usepackage{tikz}
\usepackage{overpic}
\usepackage{wrapfig}
\usepackage{bm}
\usepackage{pgfplots}
\usepackage{pgf,tikz}

\newcolumntype{C}[1]{>{\centering\arraybackslash}p{#1}}

\newcommand{\X}{\bm{\Psi}}

\newcommand{\A}{\mathbf{A}}
\newcommand{\B}{\mathbf{B}}
\newcommand{\Z}{\mathbf{Z}}
\newcommand{\Id}{\mathbf{I}}
\newcommand{\dx}{\,\mathrm{d}x}
\newcommand{\Q}{\mathbf{Q}}
\newcommand{\M}{\mathcal{X}}
\newcommand{\N}{\mathcal{Y}}

\title[Localized Manifold Harmonics for Spectral Shape Analysis]%
      {Localized Manifold Harmonics for Spectral Shape Analysis}

\author[S. Melzi, E. Rodol\`{a}, U. Castellani, M. Bronstein]
{\parbox{\textwidth}{\centering S. Melzi$^{1}$, E. Rodol\`{a}$^{2,3}$, U. Castellani$^{1}$, M. M. Bronstein$^{2,4,5,6}$}
        \\
{\parbox{\textwidth}
      {\centering 
         $^1$University of Verona, Italy \\
         $^2$USI Lugano, Switzerland\\ 
         $^3$Sapienza University of Rome, Italy \\
         $^4$Tel Aviv University, Israel \\
         $^5$Intel Perceptual Computing, Israel\\
         $^6$IAS, TU Munich, Germany
       }
}}

\newcommand{\rev}[1]{{\color{black}{#1}}}

\begin{document}

 \teaser{
  \includegraphics[width=1.0\linewidth]{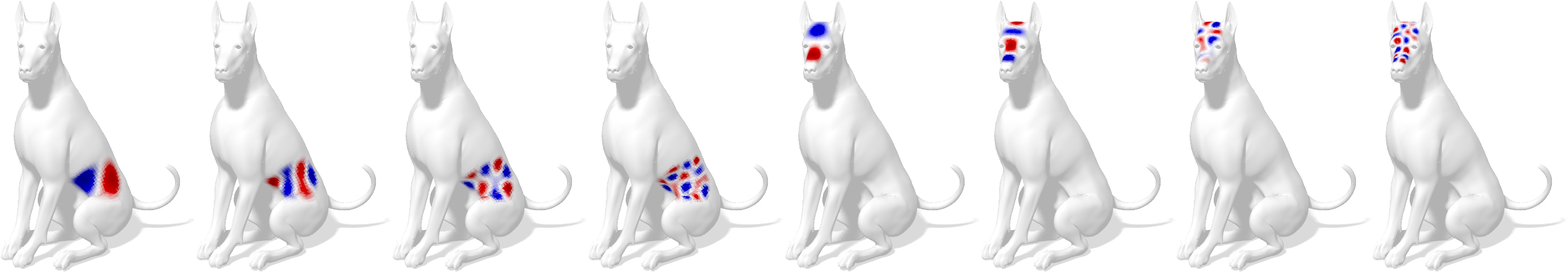}
  \vspace{-2ex}
  \centering
   \caption{A few Localized Manifold Harmonics (LMH) on two different regions of the dog shape. By changing the region location on the surface, our model provides an ordered set of {\em localized} harmonic functions (i.e., defined on the entire surface, but strongly concentrated on the selected region). In this figure the localized harmonics are clearly visible across different frequencies. The LMH constitute a valid alternative to the classical manifold harmonics and can be used in conjunction with those, or as a drop-in replacement in typical spectral shape analysis tasks.}
 \label{fig:teaser}
 }

\maketitle

\begin{abstract}
The use of Laplacian eigenfunctions is ubiquitous in a wide range of computer graphics and geometry processing applications. In particular, Laplacian eigenbases allow generalizing the classical Fourier analysis to manifolds. 
A key drawback of such bases is their inherently global nature, as the Laplacian eigenfunctions carry geometric and topological structure of the entire manifold. 
%
%
In this paper, we introduce a new framework for local spectral shape analysis. We show how to efficiently construct localized orthogonal bases by solving an optimization problem that in turn can be posed as the eigendecomposition of a new operator obtained by a modification of the standard Laplacian. We study the theoretical and computational aspects of the proposed framework and showcase our new construction on the classical problems of shape approximation and correspondence. We obtain significant improvement compared to classical Laplacian eigenbases as well as other alternatives for constructing localized bases.
\end{abstract}


\section{Introduction}\label{sec:intro}
Spectral methods are ubiquitously used in 3D shape analysis and geometry processing communities for a wide range of applications ranging from constructing local shape descriptors \cite{HKS1}, shape retrieval \cite{jain2007spectral,bronstein2011shape} and correspondence between deformable shapes \cite{ovsjanikov2012functional,quasiHarmonic} to mesh filtering \cite{Levy08,Taubin}, \rev{remeshing \cite{dong2006spectral}} and compression \cite{Karni:2000:SCM:344779.344924}. 
The centerpiece of such methods is the construction of an orthogonal basis for the space of functions defined on a manifold, allowing to generalize classical Fourier analysis to non-Euclidean domains. 
Typically, such bases are constructed by the diagonalization of the Laplace-Beltrami operator \cite{Levy06}.  
The choice of Laplacian eigenbasis is convenient for several reasons. First, it is intrinsic and thus invariant to manifold parametrization and its isometric deformations \cite{Levy:2010:SMP:1837101.1837109}. Second, it allows to be agnostic to a specific shape representation, as the Laplace-Beltrami operator can be discretized on meshes, point clouds, volumes, etc. Third, Laplacian eigenbasis turns to be optimal for approximating functions with bounded variation \cite{AflBreKim15} and in many applications only the first few eigenfunctions are sufficient to achieve a good approximation. Finally, in the discrete setting, the computation of the Laplacian eigenbasis has relatively low complexity due to the sparse structure of the Laplacian matrix.

One of the key disadvantages of the Laplacian eigenbases is their {\em global} support. Thus, representing local structures requires using (potentially, infinitely) many basis functions. 
In many applications, one wishes to have a local basis that allows to limit the analysis to specific parts of the shape. 
The recently proposed {\em compressed manifold harmonics} \cite{ozolicnvs2013compressed,Neumann,KovGlaBro16,bronstein2016consistent} attempt to construct local orthogonal bases that approximately diagonalize the Laplace-Beltrami operator. The main disadvantage of this framework is the inability to explicitly control the localization of the basis functions. 
%
%
Moreover, the basis is computed by solving an optimization problem on the Stiefel manifold of orthogonal matrices which does not guarantee a global solution. 

\noindent\textbf{Contribution. }
In this paper, we consider a new type of intrinsic operators whose spectral decomposition provides a local basis.
Similarly to related constructions like \cite{elliptic}, the new basis is smooth, local, and orthogonal; it is localized at specified regions of the shape, explicitly controllable; and it is efficiently computed by solving a standard eigendecomposition, thus coming with global optimality guarantees. The key novelty of our approach comes from its capability to integrate the \emph{global} information obtained by the Laplacian eigenfunctions with \emph{local} details given by our new basis. To this end, the localized basis is constructed in an incremental way, such that the new functions are orthogonal to some given set of functions (e.g., standard Laplacian eigenfunctions). Due to the aforementioned properties, we name our new basis {\em Localized Manifold Harmonics (LMH)}.

The rest of the paper is structured as follows. 
In Section~\ref{sec:related} we review related works.
Section~\ref{sec:bg} discusses the classical notions of spectral analysis on manifolds and the construction of manifold harmonic bases. 
In Section~\ref{sec:PM}, we introduce our construction of localized manifold harmonics and study their properties in Section~\ref{sec:prop}. 
In Section~\ref{sec:impl} we discuss the implementation details, and in Section~\ref{sec:apps} we show experimental results. 
We exemplify our construction in several applications including shape approximation and matching, where it shows superior performance compared to standard Laplacian eigenbases, as well as other alternatives for constructing localized bases. 
Finally, Section~\ref{sec:concl} concludes the paper discussing the advantages and limitations of our framework and possible future research directions. 
Proofs of theorems appear as Appendices in the supplementary materials. 


\section{Related work}
\label{sec:related}
The Laplace-Beltrami operator is so ubiquitous in computer graphics and geometry processing that it has earned the title of the ``Swiss army knife'' of geometry processing \cite{SolCraVou14}. 
The seminal work of Taubin realized the similarity between the Laplacian spectral decomposition and classical Fourier analysis \cite{Taubin}. 
Karni and Gotsman used Laplacians for spectral mesh compression \cite{Karni:2000:SCM:344779.344924}. 
An influential paper of Levy \cite{Levy06} showed how a wide range of applications can be addressed in the Laplacian spectral domain, igniting the interest in spectral approaches in the computer graphics field.


Many popular shape descriptors such as heat- \cite{HKS1,HKS2} and wave- \cite{WKS} kernel signatures, global point signatures \cite{rustamov2007laplace}, and  shape DNAs \cite{reuter} were constructed in the spectral domain. 
Coifman et al. introduced the notion of diffusion distances \cite{coifman2006diffusion} on non-Euclidean domains, also constructed considering the spectral decomposition of heat operators. 
More recently, anisotropic versions of Laplacians and their heat kernels were considered \cite{andreux2014anisotropic,boscaini2016learning}. \rev{Hildebrandt et al. considered Hessians of surface energies as an alternative family of operators \cite{hildebrandt2012modal} incorporating extrinsic curvature information.}

Ovsjanikov et al. \cite{ovsjanikov2012functional} introduced the functional maps framework to find correspondences between functions rather than points on the shapes, and used Laplacian eigenfunctions as natural choice for the basis to represent such maps (later, this choice was theoretically justified by Aflalo et al. \cite{AflBreKim15}, who showed that Laplacian eigenbases are optimal for representing classes of functions with bounded variation).  
Kovnatsky et al. proposed the construction of compatible quasi-harmonic bases on collections of shapes using simultaneous diagonalization of Laplacians \cite{quasiHarmonic}. The approach was later extended by Litany et al. \cite{fspm} to shapes having missing parts, giving rise to ad-hoc localized harmonic bases for correspondence problems.



A key drawback of Laplacian eigenbases is their global structure, which has adverse effects in numerous applications. In spectral shape deformation, it is hard to concentrate the analysis on local parts of the shape. 
In shape correspondence, the dependence on the Laplacian eigenfunctions on the global structure of the shape makes it hard to cope with topological noise and missing parts. Recently, to introduce local analysis on non-Euclidean manifolds, the Windowed Fourier Transform (WFT) has been proposed for graphs \cite{Shumann} and shapes \cite{boscaini2015learning, MRCB16}. Although these methods improved the encoding of local parts, the authors still proposed Laplacian eigenfunctions as a basis to compute the spectral components, and therefore it was still hard to perform well in the challenging scenarios mentioned above. 

As a possible remedy, Ozoli{\c{n}}{\v{s}} et al. \cite{ozolicnvs2013compressed} introduced {\em compressed modes}, a construction of local orthogonal bases that approximately diagonalize the Laplacian. The key idea of this method is the addition of a sparsity-promoting $L_1$-norm to the Dirichlet energy (the combined effect of smoothness and sparsity results in localization of the basis functions). \rev{Rustamov \cite{rustamov2011multiscale} previously used a similar regularization to construct local biharmonic kernels for function interpolation.}
Neumann et al. \cite{Neumann} applied the approach of \cite{ozolicnvs2013compressed} to problems in computer graphics. 
%
%
Kovnatsky et al. \cite{KovGlaBro16} showed an efficient way of computing compressed manifold modes, while Bronstein et al. \cite{bronstein2016consistent} proposed a more theoretically sound approach for the computation of $L_{1}$-norm on manifolds.


\rev{Closely related to our method is the recent approach of Choukroun et al. \cite{elliptic}, who considered the spectral decomposition of an elliptic operator realized as a diagonal update to the standard Laplacian. Differently from \cite{elliptic}, our solutions are simultaneously {\em localized} and {\em orthogonal} to the globally-supported Laplacian eigenbasis, leading to important practical consequences in several applications.}

The key idea of this paper is the construction of localized bases by spectral decomposition of a modified Laplacian operator, crafted especially to provide eigenfunctions with local support. Our new operator inherits the important properties of the original Laplacian such as isometry invariance. In particular, it has a clear Fourier-like meaning that makes its use well interpretable. Differently from \cite{Neumann,KovGlaBro16,bronstein2016consistent,rustamov2011multiscale} which impose locality through an $L_{1}$ constraint, we allow an explicit indication of the local support of each function. This improves the versatility in controlling the local analysis, especially for semantically-guided interventions.     
Another important difference from other methods is that our new basis is computed by solving a standard eigendecomposition problem avoiding the need for more complex optimization methods.

\section{Background}\label{sec:bg}

\noindent\textbf{Manifolds. }
We model shapes as smooth two-dimensional manifolds $\M$ (possibly with a boundary $\partial \M$) embedded into $\mathbb{R} ^{3}$. 
%
Locally around point $x$, the manifold is homeomorphic to the {\em tangent space} (or {\em plane}) $T_x \M$. 
The disjoint union of all the tangent spaces is the {\em tangent bundle} $T\M$. 
We further equip the manifold with a {\em Riemannian metric}, defined as an inner product $\langle \cdot, \cdot \rangle_{T_x \M} : T_x \M \times T_x \M \to \mathbb{R}$ on the tangent space depending smoothly on $x$.  
Properties expressed solely in terms of the metric are called {\em intrinsic}. In particular, {\em isometric} (metric-preserving) deformations of the embedded manifold preserve all intrinsic structures. 

Let $f:\M \rightarrow \mathbb{R}$ and $F:\M \rightarrow T\M$ denote real {\em scalar} and {\em tangent vector fields} on the manifold, respectively. We can define inner products
\begin{eqnarray}
\label{eq:inner1}
\langle f, g \rangle_{L^2(\M)} &=& \int_{\M} f(x)g(x) \mathrm{d}x ; \\
\langle F, G \rangle_{L^2(T\M)} &=& \int_{\M} \langle F(x), G(x) \rangle_{T_x \M} \mathrm{d}x;   
\label{eq:inner2}
\end{eqnarray}
and denote them by $L^2(\M) = \{ f: \M \rightarrow \mathbb{R} \,\,\, \mathrm{s.t.} \,\,\,  \langle f, f \rangle_{L^2(\M)} < \infty \}$ and $L^2(T\M) = \{ F: \M \rightarrow T\M \,\,\, \mathrm{s.t.} \,\,\,  \langle F, F \rangle_{L^2(T\M)} < \infty \}$ the respective Hilbert spaces of square-integrable functions (here, $\mathrm{d}x$ is the area element induced by the metric).


\noindent\textbf{Laplace-Beltrami operator. }
In classical calculus, the notion of derivative describes how the value of a function $f$ changes with an infinitesimal change of its argument $x$. 
Due to the lack of vector space structure on the manifold (meaning that we cannot add two points, $x+dx$), we need to define the {\em differential} of $f$ as an operator $\mathrm{d}f : T\M \rightarrow \mathbb{R}$ acting on tangent vector fields. 
At each point $x$, the differential is a linear functional $\mathrm{d}f(x) = \langle \nabla f(x), \, \cdot \, \rangle_{T_x \M}$ acting on tangent vectors $F(x) \in T_x \M$, which model a small displacement around $x$. The change of the function value as the result of this displacement is given by applying the differential to the tangent vector, $\mathrm{d}f(x) F(x) = \langle \nabla_{\M} f(x), F(x) \rangle_{T_x \M}$, and can be thought of as an extension of the notion of the classical directional derivative. 
The operator $ \nabla_{\M} f : L^2(\M) \rightarrow L^2(T\M)$ in the above definition is called the {\em intrinsic gradient}, and is similar to the classical notion of the gradient defining the direction of the steepest change of the function at a point. 

\rev{The {\em intrinsic divergence} $ \mathrm{div}_{\M} : L^2(T\M) \rightarrow L^2(\M)$ is defined as an operator adjoint to the intrinsic gradient, $\langle F, \nabla_{\M} f \rangle_{L^2(T\M)} = \langle -\mathrm{div}_{\M}, f \rangle_{L^2(\M)}$, where $f \in L^2(\M)$ and $F \in L^2(T\M)$ are some scalar and vector fields, respectively. 
The positive semi-definite {\em Laplace-Beltrami operator} (or {\em manifold Laplacian}) is defined as $\Delta_{\M} f = -\mathrm{div}_{\M} (\nabla_{\M} f)$, generalizing the corresponding notion from Euclidean spaces to manifolds. The Laplacian is self-adjoint,}
%
%
%
\begin{eqnarray}
\langle \nabla_{\M} f, \nabla_{\M} g \rangle_{L^2(T\M)} =  \langle \Delta_{\M} f, g \rangle_{L^2(\M)} = \langle f, \Delta_{\M} g \rangle_{L^2(\M)}.
\label{eq:adjoint1}
\end{eqnarray} 
Geometrically, the Laplace-Beltrami operator can be interpreted as the \rev{(normalized)} difference between the average of a function on an infinitesimal sphere around a point and the value of the function at the point itself.


\noindent\textbf{Spectral analysis on manifolds. }
Due to self-adjointness, on a compact manifold $\M$ with boundary $\partial \M$, the Laplace-Beltrami operator admits an orthonormal eigendecomposition \cite{chavel1984eigenvalues}
%
%
 \begin{eqnarray}
\Delta_{\M} \phi_i(x) = \lambda_i \phi_i(x)  & \,\,\,\,\,& x \in \mathrm{int}(\M) \\
\langle \nabla_{\M} \phi_i(x) , \hat{n}(x) \rangle = 0 &\,\,\,\,\,& x \in \partial\M, 
\label{eq:bc}
\end{eqnarray}
with Neumann boundary conditions~(\ref{eq:bc}), where 
$\hat{n}$ is the normal vector to the boundary.
Here, $0=\lambda_1 \leq \lambda_2 \leq \hdots$ is a countable set of non-negative real eigenvalues and $\phi_1, \phi_2, \hdots$ are the corresponding orthonormal eigenfunctions satisfying $\langle \phi_i, \phi_j \rangle_{L^2(\M)} = \delta_{ij}$. 
%
%

The Laplacian eigenfunctions form an orthonormal basis for $L^2(\M)$ referred to as {\em manifold harmonics (MH)}. 
A function $f\in L^2(\M)$ can therefore be expressed as the {\em Fourier series} 
\begin{equation}\label{eq:fourier}
f(x) = \sum_{i\geq 1} \underbrace{\langle \phi_i,f \rangle_{L^2(\M)}}_{\hat{f}_i} \phi_i(x), 
\end{equation}
where $\hat{f}_i $ are the Fourier coefficients (or the forward Fourier transform), and the 
the synthesis $\sum_{i\ge 1} \hat{f}_i \phi_i(x)$ is the inverse Fourier transform. 
The eigenvalues $\lambda_i$ can be interpreted as frequencies in the classical harmonic analysis; 
thus, truncating the series \eqref{eq:fourier} to the first $k$ terms will result in a band-limited (with bandwidth $\lambda_k$) representation of $f$.
%

Given a scalar field $f\in L^2(\M)$, the {\em Dirichlet energy} 
\begin{equation}
\label{eq:dirichlet}
\mathcal{E}_S(f):= \langle \nabla_\M f, \nabla_\M f \rangle_{L^2(T\M)} = \langle f, \Delta f\rangle_{L^2(\M)} 
\end{equation}
measures how `smooth' the field is.  It is possible to show that the Laplacian eigenbasis is the solution to the optimization problem
\begin{eqnarray}
\label{eq:courant}
\min_{\psi_1, \hdots, \psi_k}  \,\, \sum_{i=1}^k \mathcal{E}_S(\psi_i)  & \mathrm{s.t.} & \langle \psi_i, \psi_j \rangle_{L^2(\M)} = \delta_{ij}
\end{eqnarray}
%
and thus can be considered as the smoothest possible orthonormal eigenbasis. Furthermore, the eigenvalues can be obtained as the values of the Dirichlet energy,  $\mathcal{E}_S(\phi_i) = \lambda_i$. 

\section{Localized manifold harmonics}\label{sec:PM}
A notable drawback of classical spectral analysis on manifolds lies in its inherently ``global'' nature. In fact, despite the Laplacian itself being a local (differential) operator, its eigenfunctions and eigenvalues carry geometric and topological information about the entire manifold \cite{chavel1984eigenvalues,reuter,rustamov2007laplace}. 
As a practical consequence, operations that should be local by design are often affected by such global effects (see examples in Section~\ref{sec:apps}).

\begin{figure}[b]
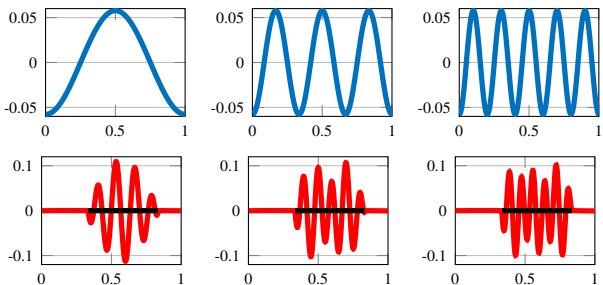

  \centering
  \setlength{\tabcolsep}{0pt}
  \begin{tabular}{ccc}
  \input{./MH1.tikz}&
  \input{./MH2.tikz}&
  \input{./MH3.tikz}
  \\
  \input{./newLMH1.tikz}&
  \input{./newLMH2.tikz}&
  \input{./newLMH3.tikz}
  \end{tabular}
\caption{\label{fig:euc}Classical (top row) and localized (bottom row) harmonics in 1D under Neumann boundary conditions. \rev{Note that the localized harmonics are orthogonal to those in the first row}. The selected region $R\subset[0,1]$ is marked as a black segment.}
\end{figure}

In this paper, we introduce a new framework for spectral shape analysis that is designed to be at the same time {\em local} and {\em compatible} with the existing spectral constructions. In practice, our approach boils down to the computation of the eigenfunctions of a new operator, which is realized as a simple update to the classical manifold Laplacian -- thus fully retaining the computational efficiency and theoretical guarantees of the resulting optimization process.

\noindent\textbf{Definition. }
Let us be given a manifold $\M$, a region $R\subseteq\M$ thereof, a set of orthonormal functions $\phi_1, \hdots, \phi_{k'}$ (e.g. the first $k'$ Laplacian eigenfunctions), and an integer $k$. 
We seek a new set $\psi_1, \hdots, \psi_k$ of functions that are {\em smooth}, {\em orthonormal}, and {\em localized} on $R$, as the solution to the following optimization problem:
\rev{
%
\begin{align}\label{eq:problem}
\min_{\psi_1, \hdots, \psi_k} & \sum_{j=1}^k \mathcal{E}_S(\psi_j) + \mu_R \mathcal{E}_R(\psi_j)\\
\mathrm{s.t.}~~&\langle \psi_i, \psi_j\rangle_{L^2(\M)} = \delta_{ij} \quad i,j = 1,\hdots, k\\
&\langle \psi_i,\phi_j\rangle_{L^2(\M)} = 0\quad \,\,\, i = 1,\hdots, k; \, j = 1\hdots, k'
\label{eq:orthophi}
\end{align}
where the constraints \eqref{eq:orthophi} demand the basis functions to be {\em orthogonal} to the subspace $\mathrm{span}\{\phi_1, \hdots, \phi_{k'} \}$. As we will see in what follows, it allows constructing an incremental set of functions that are orthogonal to a given set of standard Laplacian eigenfunctions.

%
%
}
The first term $\mathcal{E}_S$ is the Dirichlet functional \eqref{eq:dirichlet} promoting the {\em smoothness} of the new basis. 
The term 
\begin{equation}\label{eq:loc}
\mathcal{E}_R (f) := \int_\M (f(x) (1-u(x)))^2 \dx\,,
\end{equation}
is a quadratic penalty promoting the localization of the basis functions on the given region $R\subseteq\M$. 
Here $u:\M\to [0,1]$ is a membership function such that $u(x)=1$ for $x\in R$ and $u(x)=0$ otherwise. Note that we let function $u$ assume a continuum of values in $[0,1]$, implementing the notion of ``soft'' membership (the choice between binary and soft $u$ is application-dependent). 

We refer to the solutions of problem \eqref{eq:problem} as {\em localized manifold harmonics (LMH)}. Figure~\ref{fig:euc} provides an illustration of LMH in the $[0,1]$ interval, while Figures~\ref{fig:teaser} and \ref{fig:harmonics} depict a few examples of such bases on 2D manifolds.

\def\imagetop#1{\vtop{\null\hbox{#1}}}
\begin{figure}[t]
  \centering
  \setlength{\tabcolsep}{0pt}
  \begin{tabular}{cc}
  \imagetop{\begin{overpic}
  [trim=0cm 0cm 0cm 0cm,clip,width=0.85\linewidth]{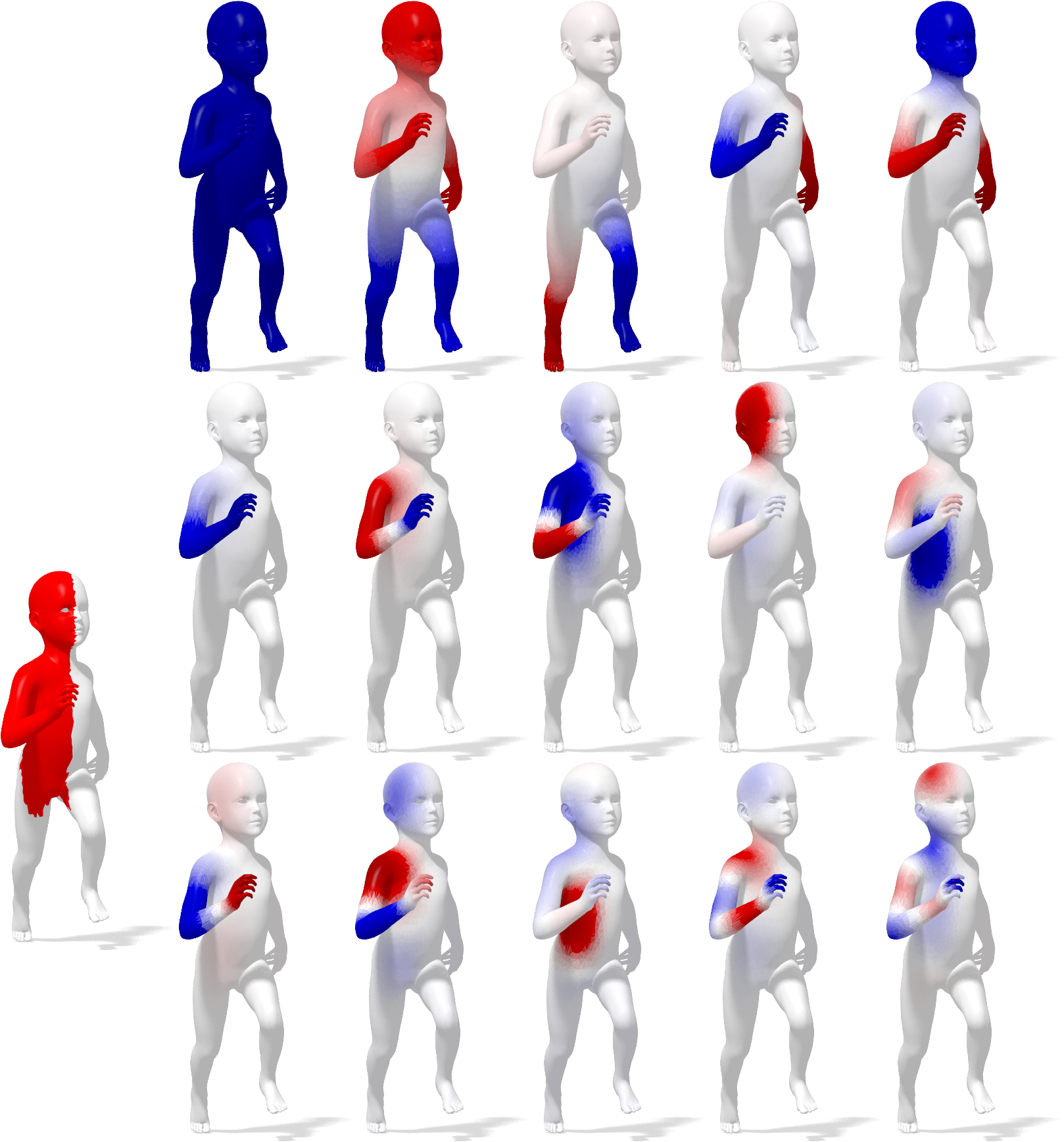}
  \put(4,15){\footnotesize $R$}
  \end{overpic}}
  &
  \hspace{-0.4cm}
  \imagetop{\begin{tabular}{c}
  \\\\
%
%
\definecolor{mycolor1}{rgb}{0.00000,0.44700,0.74100}%
\definecolor{mycolor2}{rgb}{0.85000,0.32500,0.09800}%
\definecolor{mycolor3}{rgb}{0.92900,0.69400,0.12500}%
\begin{tikzpicture}

\begin{axis}[%
width=0.1\linewidth,
height=0.15\linewidth,
scale only axis,
xmin=1,
xmax=10,
ymin=0,
ymax=0.14,
ymajorgrids=true,
ytick={0,0.1},
ylabel={\footnotesize $\lambda$},
ylabel style={at={(0.93,0.5)}},
axis background/.style={fill=white},
every x tick label/.append style={font=\color{black}, font=\footnotesize},
every y tick label/.append style={font=\color{black}, font=\footnotesize},
axis x line*=bottom,
axis y line*=left
]
\addplot [color=mycolor1,solid,line width=2.0pt,forget plot]
  table[row sep=crcr]{%
1	0.0000000000000000\\
2	0.0014817417531242\\
3	0.0023059780780538\\
4	0.0025644891512067\\
5	0.0031640938654766\\
6	0.0058180306660553\\
7	0.0117599701704012\\
8	0.0137408255691844\\
9	0.0171962096571384\\
10	0.0204880801755599\\
};
\addplot [color=mycolor1,solid,mark=*,mark options={solid,fill=mycolor1,draw=mycolor1},forget plot]
  table[row sep=crcr]{%
1	0.0000000000000000\\
2	0.0014817417531242\\
3	0.0023059780780538\\
4	0.0025644891512067\\
5	0.0031640938654766\\
6	0.0058180306660553\\
7	0.0117599701704012\\
8	0.0137408255691844\\
9	0.0171962096571384\\
10	0.0204880801755599\\
};
\addplot [color=mycolor2,dotted,line width=1.0pt,forget plot]
  table[row sep=crcr]{%
1	0.0024909040285600\\
2	0.0158990384734583\\
3	0.0311500774568827\\
4	0.0365598074167109\\
5	0.0647700643239714\\
6	0.0769859452748116\\
7	0.0905811562091755\\
8	0.0974802838096865\\
9	0.1071286875113193\\
10	0.1185731802242381\\
};
\addplot [color=mycolor2,solid,mark options={solid,fill=mycolor2,draw=mycolor2},forget plot]
  table[row sep=crcr]{%
1	0.0024909040285600\\
2	0.0158990384734583\\
3	0.0311500774568827\\
4	0.0365598074167109\\
5	0.0647700643239714\\
6	0.0769859452748116\\
7	0.0905811562091755\\
8	0.0974802838096865\\
9	0.1071286875113193\\
10	0.1185731802242381\\
};
\addplot [color=mycolor3,dotted,line width=1.0pt,forget plot]
  table[row sep=crcr]{%
1	0.0614286515220834\\
2	0.0888279998884785\\
3	0.0907325902166534\\
4	0.0962062112060567\\
5	0.1064987012578537\\
6	0.1185194790195390\\
7	0.1240092539790511\\
8	0.1353854892357272\\
9	0.1472968475943827\\
10	0.1631158574872740\\
};
\addplot [color=mycolor3,solid,mark options={solid,fill=mycolor3,draw=mycolor3},forget plot]
  table[row sep=crcr]{%
1	0.0614286515220834\\
2	0.0888279998884785\\
3	0.0907325902166534\\
4	0.0962062112060567\\
5	0.1064987012578537\\
6	0.1185194790195390\\
7	0.1240092539790511\\
8	0.1353854892357272\\
9	0.1472968475943827\\
10	0.1631158574872740\\
};
\end{axis}
\end{tikzpicture}%
\vspace{0.7cm}\\
%
%
\definecolor{mycolor1}{rgb}{0.00000,0.44700,0.74100}%
\definecolor{mycolor2}{rgb}{0.85000,0.32500,0.09800}%
\definecolor{mycolor3}{rgb}{0.92900,0.69400,0.12500}%
\begin{tikzpicture}

\begin{axis}[%
width=0.1\linewidth,
height=0.15\linewidth,
scale only axis,
xmin=1,
xmax=10,
ymin=0,
ymax=0.14,
ymajorgrids=true,
ytick={0,0.1},
ylabel={\footnotesize $\lambda$},
ylabel style={at={(0.93,0.5)}},
axis background/.style={fill=white},
every x tick label/.append style={font=\color{black}, font=\footnotesize},
every y tick label/.append style={font=\color{black}, font=\footnotesize},
axis x line*=bottom,
axis y line*=left
]
\addplot [color=mycolor1,dotted,line width=1.0pt,forget plot]
  table[row sep=crcr]{%
1	0.0000000000000000\\
2	0.0014817417531242\\
3	0.0023059780780538\\
4	0.0025644891512067\\
5	0.0031640938654766\\
6	0.0058180306660553\\
7	0.0117599701704012\\
8	0.0137408255691844\\
9	0.0171962096571384\\
10	0.0204880801755599\\
};
\addplot [color=mycolor1,solid,mark options={solid,fill=mycolor1,draw=mycolor1},forget plot]
  table[row sep=crcr]{%
1	0.0000000000000000\\
2	0.0014817417531242\\
3	0.0023059780780538\\
4	0.0025644891512067\\
5	0.0031640938654766\\
6	0.0058180306660553\\
7	0.0117599701704012\\
8	0.0137408255691844\\
9	0.0171962096571384\\
10	0.0204880801755599\\
};
\addplot [color=mycolor2,solid,line width=2.0pt,forget plot]
  table[row sep=crcr]{%
1	0.0024909040285600\\
2	0.0158990384734583\\
3	0.0311500774568827\\
4	0.0365598074167109\\
5	0.0647700643239714\\
6	0.0769859452748116\\
7	0.0905811562091755\\
8	0.0974802838096865\\
9	0.1071286875113193\\
10	0.1185731802242381\\
};
\addplot [color=mycolor2,solid,mark=*,mark options={solid,fill=mycolor2,draw=mycolor2},forget plot]
  table[row sep=crcr]{%
1	0.0024909040285600\\
2	0.0158990384734583\\
3	0.0311500774568827\\
4	0.0365598074167109\\
5	0.0647700643239714\\
6	0.0769859452748116\\
7	0.0905811562091755\\
8	0.0974802838096865\\
9	0.1071286875113193\\
10	0.1185731802242381\\
};
\addplot [color=mycolor3,dotted,line width=1.0pt,forget plot]
  table[row sep=crcr]{%
1	0.0614286515220834\\
2	0.0888279998884785\\
3	0.0907325902166534\\
4	0.0962062112060567\\
5	0.1064987012578537\\
6	0.1185194790195390\\
7	0.1240092539790511\\
8	0.1353854892357272\\
9	0.1472968475943827\\
10	0.1631158574872740\\
};
\addplot [color=mycolor3,solid,mark options={solid,fill=mycolor3,draw=mycolor3},forget plot]
  table[row sep=crcr]{%
1	0.0614286515220834\\
2	0.0888279998884785\\
3	0.0907325902166534\\
4	0.0962062112060567\\
5	0.1064987012578537\\
6	0.1185194790195390\\
7	0.1240092539790511\\
8	0.1353854892357272\\
9	0.1472968475943827\\
10	0.1631158574872740\\
};
\end{axis}
\end{tikzpicture}%
\vspace{0.8cm}\\
%
%
\definecolor{mycolor1}{rgb}{0.00000,0.44700,0.74100}%
\definecolor{mycolor2}{rgb}{0.85000,0.32500,0.09800}%
\definecolor{mycolor3}{rgb}{0.92900,0.69400,0.12500}%
\begin{tikzpicture}

\begin{axis}[%
width=0.1\linewidth,
height=0.15\linewidth,
scale only axis,
xmin=1,
xmax=10,
ymin=0,
ymax=0.14,
ymajorgrids=true,
ytick={0,0.1},
ylabel={\footnotesize $\lambda$},
ylabel style={at={(0.93,0.5)}},
axis background/.style={fill=white},
every x tick label/.append style={font=\color{black}, font=\footnotesize},
every y tick label/.append style={font=\color{black}, font=\footnotesize},
axis x line*=bottom,
axis y line*=left
]
\addplot [color=mycolor1,dotted,line width=1.0pt,forget plot]
  table[row sep=crcr]{%
1	0.0000000000000000\\
2	0.0014817417531242\\
3	0.0023059780780538\\
4	0.0025644891512067\\
5	0.0031640938654766\\
6	0.0058180306660553\\
7	0.0117599701704012\\
8	0.0137408255691844\\
9	0.0171962096571384\\
10	0.0204880801755599\\
};
\addplot [color=mycolor1,solid,mark options={solid,fill=mycolor1,draw=mycolor1},forget plot]
  table[row sep=crcr]{%
1	0.0000000000000000\\
2	0.0014817417531242\\
3	0.0023059780780538\\
4	0.0025644891512067\\
5	0.0031640938654766\\
6	0.0058180306660553\\
7	0.0117599701704012\\
8	0.0137408255691844\\
9	0.0171962096571384\\
10	0.0204880801755599\\
};
\addplot [color=mycolor2,dotted,line width=1.0pt,forget plot]
  table[row sep=crcr]{%
1	0.0024909040285600\\
2	0.0158990384734583\\
3	0.0311500774568827\\
4	0.0365598074167109\\
5	0.0647700643239714\\
6	0.0769859452748116\\
7	0.0905811562091755\\
8	0.0974802838096865\\
9	0.1071286875113193\\
10	0.1185731802242381\\
};
\addplot [color=mycolor2,solid,mark options={solid,fill=mycolor2,draw=mycolor2},forget plot]
  table[row sep=crcr]{%
1	0.0024909040285600\\
2	0.0158990384734583\\
3	0.0311500774568827\\
4	0.0365598074167109\\
5	0.0647700643239714\\
6	0.0769859452748116\\
7	0.0905811562091755\\
8	0.0974802838096865\\
9	0.1071286875113193\\
10	0.1185731802242381\\
};
\addplot [color=mycolor3,solid,line width=2.0pt,forget plot]
  table[row sep=crcr]{%
1	0.0614286515220834\\
2	0.0888279998884785\\
3	0.0907325902166534\\
4	0.0962062112060567\\
5	0.1064987012578537\\
6	0.1185194790195390\\
7	0.1240092539790511\\
8	0.1353854892357272\\
9	0.1472968475943827\\
10	0.1631158574872740\\
};
\addplot [color=mycolor3,solid,mark=*,mark options={solid,fill=mycolor3,draw=mycolor3},forget plot]
  table[row sep=crcr]{%
1	0.0614286515220834\\
2	0.0888279998884785\\
3	0.0907325902166534\\
4	0.0962062112060567\\
5	0.1064987012578537\\
6	0.1185194790195390\\
7	0.1240092539790511\\
8	0.1353854892357272\\
9	0.1472968475943827\\
10	0.1631158574872740\\
};
\end{axis}
\end{tikzpicture}%
  \end{tabular}}
  \vspace{-0.1cm}
  \end{tabular}
  \caption{\label{fig:harmonics}Localized manifold harmonics for the red region shown on the left. We show the first $k'=5$ standard Laplacian eigenfunctions (top row), the first $k=5$ LMH with the orthogonality term disabled ($\mu_\perp=0$, middle row), and the first $k=5$ LMH obtained by optimizing the full energy \eqref{eq:ener} (bottom row). Note that the latter harmonics are orthogonal to the first $k'$ Laplacian eigenfunctions. We also show the generalized eigenvalues associated with each of the three cases.}
\end{figure}

\rev{
\noindent\textbf{Relaxed problem. }
For practical reasons, in what follows we will consider a relaxed variant of \eqref{eq:problem}, in which we replace the hard constraints \eqref{eq:orthophi} by a large penalty:
\begin{align}\label{eq:prel}
\min_{\psi_1, \hdots, \psi_k} \sum_{j=1}^k \mathcal{E}(\psi_j)\quad
\mathrm{s.t.}~~\langle \psi_i, \psi_j\rangle_{L^2(\M)} = \delta_{ij}\,,
\end{align}
where
\begin{align}\label{eq:ener}
\mathcal{E}(\psi_j) &= \mathcal{E}_S(\psi_j) + \mu_R \mathcal{E}_R(\psi_j) + \mu_\perp \mathcal{E}_\perp(\psi_j)\,,\\
\mathcal{E}_\perp(f) &:= \sum_{i=1}^{k'} |\langle \phi_i , f \rangle_{L^2(\M)}|^2\,.\label{eq:ortho}
\end{align}
Note that problems \eqref{eq:prel} and \eqref{eq:problem} are equivalent as $\mu_\perp\to\infty$. An empirical evaluation of the equivalence of the two formulations will be provided in Section \ref{sec:impl}.
}

\noindent\textbf{Discretization. }
In the discrete setting, the manifold $\M$ is sampled at $n$ points $x_1, \hdots, x_n$ and is approximated by a triangular mesh $(V,E,F)$ constructed upon these points, where $V=\{1, \hdots, n\}$, $E=E_\mathrm{i} \cup E_\mathrm{b}$ and $F$ are the vertices, edges, and faces of the mesh, respectively ($E_\mathrm{i}$ and $E_\mathrm{b}$ denote the interior and boundary edges, respectively). 
The discretization of the Laplace-Beltrami operator $\Delta_\M$ takes the form of an $n \times n$ sparse matrix $\mathbf{L}=-\A^{-1}\mathbf{W}$ according to the standard lumped linear FEM \cite{macneal1949solution}. The {\em mass matrix} $\A$ is a diagonal matrix of area elements $a_i=\frac{1}{3} \sum_{jk:ijk\in F} A_{ijk}$, where $A_{ijk}$ denotes the area of triangle $ijk$. The {\em stiffness matrix} $\mathbf{W}$ contains the {\em cotangent weights} 
\begin{eqnarray}
\label{eq:cotan}
w_{ij} & = & \left\{ 
		\begin{array}{lc}
			(\cot \alpha _{ij} + \cot \beta _{ij})/2 &   ij \in E_\mathrm{i};  \\		
			(\cot \alpha _{ij})/2 &   ij \in E_\mathrm{b};  \\		
			-\sum_{k\neq i} w_{ik}    & i = j; \\
			0 & \mathrm{else}; 
		\end{array}
\right.
\end{eqnarray}
where $\alpha_{ij},\beta_{ij}$ denote the angles $\angle ikj, \angle jhi$ of the triangles sharing the edge $ij$.
Scalar fields $f \in L^2(\M)$ are represented as $n$-dimensional vectors $\mathbf{f} = (f(x_1), \hdots, f(x_n))^\top$; the inner products $\langle f,g\rangle_{L^2(\M)}$ are discretized by area-weighted dot products $\mathbf{f}^\top \A \mathbf{g}$. 

We now turn to the discretization of problem \eqref{eq:prel}. Let $\X\in\mathbb{R}^{n\times k}$ be a matrix containing our discretized basis functions $\psi_1, \hdots, \psi_k$ as its columns, and same way, let $\bm{\Phi}\in\mathbb{R}^{n\times k'}$ be a matrix of the first $k'$ Laplacian eigenfunctions $\phi_1, \hdots, \phi_{k'}$. 
The total energy is discretized as $\sum_{j=1}^k\mathcal{E}(\psi_j) =\mathcal{E}(\X)$, comprising purely quadratic terms
\begin{align}
\mathcal{E}_S(\X) &= \mathrm{tr} (\X^\top \mathbf{W} \X)\\
\mathcal{E}_R(\X) &= \mathrm{tr} (\X^\top \A\, \mathrm{diag}(\mathbf{v}) \X)\label{eq:localR}\\
\mathcal{E}_\perp(\X) &= \mathrm{tr} (\X^\top \A \underbrace{\bm{\Phi}\bm{\Phi}^\top\A}_{\mathbf{P}_{k'}} \X)
\end{align}
%
\rev{where $\mathbf{v}$ denotes the discrete version of $v(x) \equiv (1-u(x))^2$.} 
%
%
In other words, the locality penalty \eqref{eq:loc} is implemented as a diagonal update to the standard Laplacian; while the term promoting  orthogonality to $\bm{\Phi}$ \eqref{eq:ortho} is realized as a rank-$k'$ projector $\mathbf{P}_{k'}$.

Due to the linearity of the trace, the discrete version of problem \eqref{eq:prel} can be expressed as 
\begin{align}\label{eq:eval}
\min_{\X\in\mathbb{R}^{n\times k}} ~~\mathrm{tr}(\X^\top \Q_{v,k'} \X)\quad
\mathrm{s.t.} ~~\X^\top \A \X = \Id\,,
\end{align}
where the matrix 
\begin{equation}\label{eq:Q}
\Q_{v,k'} = \mathbf{W} + \mu_R\A\mathrm{diag}(\mathbf{v}) + \mu_\perp\A\mathbf{P}_{k'}
\end{equation} 
is symmetric and positive semi-definite (we make the dependency on $v,k'$ explicit as a subscript). Problem \eqref{eq:eval} is equivalent to the {\em generalized eigenvalue problem} 
\begin{equation}\label{eq:geigen}
\Q_{v,k'} \X = \A \X \bm{\Lambda}, 
\end{equation}
(see Theorem~1.2 of \cite{sameh1982trace}). We stress that the new operator $\Q_{v,k'}$ is intrinsic, and so are its eigenfunctions.  

As shown later in Section~\ref{sec:impl}, a {\em global} optimum of this problem can be found by classical Arnoldi-like methods. 
\rev{Note that global solutions to the original constrained problem \eqref{eq:problem} can also be easily computed (see Appendix A), however throughout this paper we favor the relaxed formulation for computational efficiency reasons. We refer to Section \ref{sec:impl} for comparisons.}

\begin{figure}[t]
  \centering  
  \begin{overpic}
  [trim=0cm 0cm 0cm 0cm,clip,width=1\linewidth]{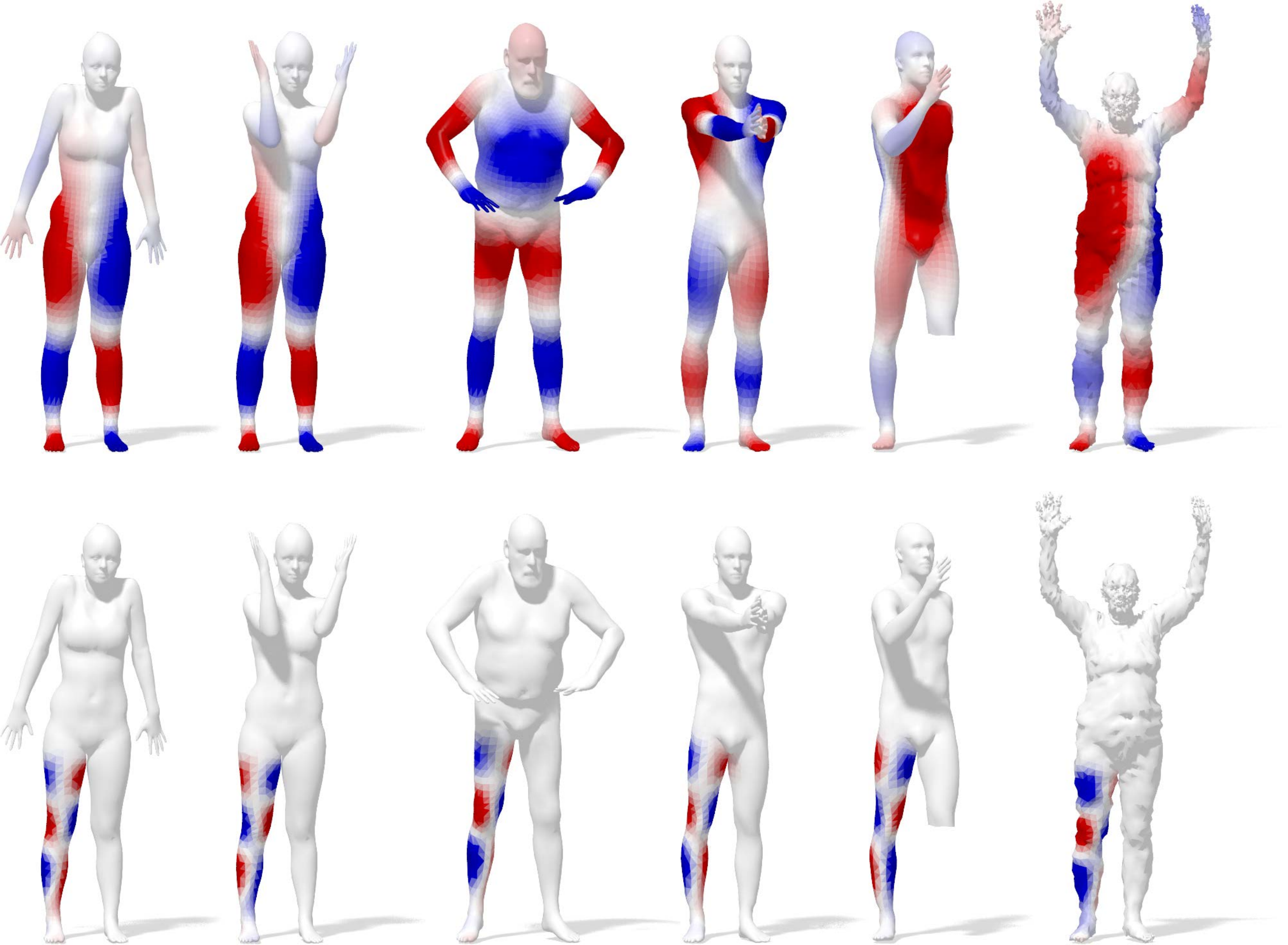}
  \put(47,70){\footnotesize MH}
  \put(45,32){\footnotesize LMH}
  \put(19,-4){\footnotesize Pose}
  \put(35,-4){\footnotesize Subject}
  \put(49,-4){\footnotesize Topology}
  \put(65,-4){\footnotesize Partiality}
  \put(80,-4){\footnotesize Gaussian}
  
  \end{overpic}
\vspace{-0.1cm}  
  
  \caption{\label{fig:robustness} Laplacian eigenfunction $\phi_{12}$ (top) and localized manifold harmonic  $\psi_{12}$  (bottom) under different shape transformations. From left to right: near-isometry (different pose), non-isometric deformation (different subject), topological noise (glued hands), missing part, and geometric noise. LMH is more stable under such deformations compared to the standard MH.}
\end{figure}

\section{Properties of LMH}\label{sec:prop}
In this section, we discuss the main theoretical properties and computational aspects of our framework.

\noindent\textbf{Basis functions. }
As mentioned before, our localized basis functions are orthonormal eigenfunctions of a matrix \eqref{eq:geigen} obtained by modification of the Laplacian. Indeed, by setting $\mu_R=\mu_\perp=0$, the solution of \eqref{eq:eval} is attained by the first $k$ standard Laplacian eigenfunctions. Similarly, by setting $\mu_R=0$ (no locality) and for $\mu_\perp\to\infty$, problem \eqref{eq:eval} can be equivalently rewritten as
\begin{align}
\min_{\X\in\mathbb{R}^{n\times k}} ~~\mathrm{tr}(\X^\top \mathbf{W} \X)\quad
\mathrm{s.t.} ~~(\X \, \bm{\Phi})^\top \A (\X \, \bm{\Phi})= \Id\,,
\end{align}
whose minimizers are the standard Laplacian eigenfunctions $\bm{\psi}_1 = \bm{\phi}_{k'+1}, \hdots, \bm{\psi}_k = \bm{\phi}_{k'+k}$.  

For $\mu_R>0$ and $\mu_\perp>0$, we obtain a new set of $k$ functions $\bm{\psi}_1, \hdots, \bm{\psi}_k$ localized to a given region $R\subseteq\M$. These functions effectively {\em extend} the Laplacian eigenbasis, in the sense that the new set 
$\bm{\phi}_1, \hdots, \bm{\phi}_{k'}, \bm{\psi}_1, \hdots, \bm{\psi}_k$ 
forms an {\em orthonormal basis} for a $k+k'$-dimensional subspace of $L^2(\M)$. Importantly, the new basis is still isometry-invariant, and is designed to effectively represent functions with support restricted to the given region (see Figure~\ref{fig:robustness}). Compared to only using the `global' Laplacian eigenbasis, the new representation provides a more parsimonious model: fewer localized harmonics are needed to capture the high-frequency content within $R$, than the number of harmonics that would be needed in the global basis. The localized nature of this construction allows to mitigate considerably the non-local effects associated with the adoption of the global basis (influence of topological noise, etc).

Finally, disabling the orthogonal penalty ($\mu_R>0$, $\mu_\perp=0$) would lead to the set of (now possibly linearly dependent) $k+k'$ functions spanning a $k''\le k+k'$-dimensional subspace of $L^2(\M)$. This \rev{may} result in a redundant representation of functions supported on $R\subseteq \M$\rev{, e.g., whenever a standard Laplacian eigenfunction has also support in $R$. We refer to the experimental section for a deeper analysis of the effect of orthogonality on the representation quality}.

\begin{figure}[t]
  \centering
  \setlength{\tabcolsep}{0pt}
  \begin{tabular}{cc}
    \begin{minipage}{0.85\columnwidth}
%
%
\definecolor{cool_blue}{rgb}{0.00000,0.44700,0.74100}%
\definecolor{cool_yellow}{rgb}{0.92900,0.69400,0.12500}%
\definecolor{cool_red}{rgb}{0.85000,0.32500,0.09800}%
\begin{tikzpicture}

\begin{axis}[%
width=0.55\linewidth,
height=0.6\linewidth,
scale only axis,
every outer x axis line/.append style={black},
xmin=0,
xmax=70,
xlabel={\footnotesize Index},
every x tick label/.append style={font=\color{black}, font=\footnotesize},
every y tick label/.append style={font=\color{black}, font=\footnotesize},
every outer y axis line/.append style={black},
ymin=0,
ymax=5000,
ymajorgrids,
ytick={0,1000,2000,3000,4000,5000},
yticklabels = {0,$10$,$20$,$30$,$40$,$50$},
ylabel={\footnotesize Eigenvalue},
xlabel style={at={(0.5,0.05)}},
ylabel style={at={(0.14,0.5)}},
axis background/.style={fill=white},
axis x line*=bottom,
axis y line*=left,
legend style={at={(0.03,0.97)},anchor=north west,legend cell align=left,align=left,draw=black}
]
\addplot [color=cool_blue,solid,line width=2.0pt]
  table[row sep=crcr]{%
21	384.870955135668\\
22	410.407841409432\\
23	597.843209996136\\
24	773.264586299968\\
25	816.16598371819\\
26	833.981001127976\\
27	1049.5463566277\\
28	1072.34502429198\\
29	1102.29336945762\\
30	1104.99142238103\\
31	1211.45134169557\\
32	1293.39253972954\\
33	1379.63776398074\\
34	1405.70054509302\\
35	1580.55191622776\\
36	1610.19875796444\\
37	1646.15443390012\\
38	1681.08101428457\\
39	1746.5268710412\\
40	1947.55490563138\\
41	1967.93092448097\\
42	2063.3013171879\\
43	2135.24671322839\\
44	2173.09695658171\\
45	2287.64609999776\\
46	2387.04609659037\\
47	2425.28483542359\\
48	2453.89888075288\\
49	2516.14098462511\\
50	2565.30183936024\\
51	2591.7875002931\\
52	2722.66743223643\\
53	2763.90240719154\\
54	2920.54580997275\\
55	2956.03739043009\\
56	3001.75722586785\\
57	3062.87671489566\\
58	3130.17734318894\\
59	3190.38971939529\\
60	3297.66476080372\\
61	3312.98336659193\\
62	3465.54443283416\\
63	3552.38193559459\\
64	3610.89210425787\\
65	3625.64105809816\\
66	3673.60424598008\\
67	3737.44304614021\\
68	3848.16794233689\\
69	3960.30497380435\\
70	4014.99102208149\\
};
\addlegendentry{\footnotesize LMH on $R_1$};

\addplot [color=cool_yellow,solid,line width=2.0pt]
  table[row sep=crcr]{%
21	346.963827335793\\
22	398.008773145393\\
23	423.476723458444\\
24	445.995808959804\\
25	462.718472596112\\
26	550.407455359736\\
27	587.545764235\\
28	647.157925120532\\
29	678.160453923147\\
30	697.135738093104\\
31	708.022373838705\\
32	765.492450939106\\
33	777.824104096067\\
34	842.695457688795\\
35	926.914681669686\\
36	951.810769140759\\
37	982.74804409789\\
38	1015.87516500578\\
39	1029.80283181505\\
40	1047.2796608916\\
41	1054.27320769571\\
42	1087.04918922312\\
43	1108.48546645706\\
44	1162.24390730698\\
45	1182.08237399702\\
46	1197.49944061599\\
47	1212.67324373783\\
48	1230.6442677454\\
49	1280.74646473428\\
50	1288.67333173057\\
51	1320.86430752014\\
52	1327.89042026072\\
53	1357.68740410154\\
54	1458.23008039856\\
55	1479.57909804475\\
56	1509.05186560871\\
57	1561.92513508695\\
58	1568.90669475618\\
59	1611.84378254494\\
60	1612.8867375263\\
61	1666.69310179206\\
62	1668.6864932404\\
63	1685.33792419083\\
64	1707.9634356058\\
65	1720.58259076326\\
66	1760.20345070587\\
67	1808.38205627669\\
68	1827.45828827572\\
69	1900.24365440995\\
70	1936.02361321172\\
};
\addlegendentry{\footnotesize LMH on $R_2$};

\addplot [color=cool_blue,dashed,line width=2.0pt]
  table[row sep=crcr]{%
1	6.78497141554536e-14\\
2	54.0865528134872\\
3	92.9558007977542\\
4	195.192433193682\\
5	287.666087542766\\
6	360.576786048715\\
7	401.487910967433\\
8	414.968039819515\\
9	459.245014863843\\
10	618.790592572252\\
11	679.872127753204\\
12	740.433598305773\\
13	848.493804242392\\
14	921.625436321699\\
15	1029.04596371713\\
16	1067.88676785573\\
17	1102.40150408717\\
18	1122.896587685\\
19	1254.34314410595\\
20	1290.61063903334\\
21	1324.55753360005\\
22	1491.51616807861\\
23	1588.31355621855\\
24	1605.43807640024\\
25	1625.48168364583\\
26	1733.70434398026\\
27	1827.39600832951\\
28	1947.62396009737\\
29	2019.85768620765\\
30	2086.8593462428\\
31	2125.48455700673\\
32	2182.61080594649\\
33	2252.60801286236\\
34	2315.89774100389\\
35	2410.8631170748\\
36	2463.57664305918\\
37	2577.20634241684\\
38	2612.19084278953\\
39	2705.53766015801\\
40	2746.34415263412\\
41	2777.09080786112\\
42	2863.93577960792\\
43	2889.39428361085\\
44	2967.00754291663\\
45	3093.71810781536\\
46	3164.27753985459\\
47	3294.64787063277\\
48	3321.37899783042\\
49	3363.22959876488\\
50	3496.1100756255\\
51	3524.68573338351\\
52	3616.43625786321\\
53	3702.20660313604\\
54	3728.60968228716\\
55	3873.70585009788\\
56	3933.74235681527\\
57	3963.4812749091\\
58	3978.66455761494\\
59	4088.86695610532\\
60	4122.17024118791\\
61	4197.67275009567\\
62	4240.71735623705\\
63	4333.78706182758\\
64	4417.86228831064\\
65	4446.77394973982\\
66	4515.23420191116\\
67	4567.07450492281\\
68	4689.93379830251\\
69	4721.55900214178\\
70	4822.52937492351\\
};
\addlegendentry{\footnotesize $\lambda(\Delta_{R_1\subset\M})$};

\addplot [color=cool_yellow,dashed,line width=2.0pt]
  table[row sep=crcr]{%
1	1.57383634390943e-15\\
2	26.7941663325686\\
3	28.9916405739076\\
4	49.5587999570412\\
5	119.009521843582\\
6	130.926734641149\\
7	134.234298465684\\
8	205.415499505859\\
9	260.604141006607\\
10	269.594048889446\\
11	338.054526386373\\
12	357.370812467338\\
13	367.516302486553\\
14	407.946580948955\\
15	419.414363021096\\
16	451.368373871872\\
17	495.159194157778\\
18	523.414780412348\\
19	549.898341771898\\
20	583.204558118906\\
21	623.296158259439\\
22	708.2347699161\\
23	726.132820124736\\
24	758.871682442712\\
25	822.155216950983\\
26	862.561714193898\\
27	878.199356719437\\
28	906.32254196524\\
29	925.132330025846\\
30	980.22495132336\\
31	1023.0130864511\\
32	1038.68356101197\\
33	1053.73000991427\\
34	1092.68051888382\\
35	1102.83695824987\\
36	1110.70280056654\\
37	1144.52151695048\\
38	1173.70974075408\\
39	1183.75047232404\\
40	1198.80795666856\\
41	1217.13292666119\\
42	1243.6563818148\\
43	1284.08558425673\\
44	1325.23421041599\\
45	1353.10982082127\\
46	1361.01784735434\\
47	1378.72546786694\\
48	1401.43952757537\\
49	1429.23472789598\\
50	1445.08890559724\\
51	1484.13391625936\\
52	1514.28664809078\\
53	1555.69304381089\\
54	1572.24182377408\\
55	1648.30454597113\\
56	1686.43340452878\\
57	1722.39021169658\\
58	1753.88214488057\\
59	1770.05560110802\\
60	1835.17058551712\\
61	1890.11084905733\\
62	1913.22646357307\\
63	1920.25019979483\\
64	1951.1528911814\\
65	1986.34729119768\\
66	2008.41991882771\\
67	2012.33266090182\\
68	2029.67426667999\\
69	2067.36023184867\\
70	2106.37798021326\\
};
\addlegendentry{\footnotesize $\lambda(\Delta_{R_2\subset\M})$};

\addplot [color=cool_red,solid,line width=2.0pt]
  table[row sep=crcr]{%
1	3.1694848026502e-14\\
2	4.87507178819213\\
3	6.3261742599453\\
4	9.30256739865693\\
5	13.9450194048203\\
6	19.9100327424224\\
7	39.9862829059263\\
8	40.3050827658758\\
9	57.3094348662391\\
10	61.007710871497\\
11	87.8770388627712\\
12	101.653905228337\\
13	118.736393885299\\
14	131.144749926961\\
15	135.32573262194\\
16	171.979368108208\\
17	184.75571102838\\
18	194.612549557522\\
19	217.481524383442\\
20	235.222225065863\\
};
\addlegendentry{\footnotesize $\lambda(\Delta_\M)$};

\addplot [color=cool_blue,only marks,mark=o,mark options={solid},forget plot]
  table[row sep=crcr]{%
21	384.870955135668\\
};
\addplot [color=cool_yellow,only marks,mark=o,mark options={solid},forget plot]
  table[row sep=crcr]{%
21	346.963827335793\\
};
\addplot [color=cool_red,only marks,mark=o,mark options={solid},forget plot]
  table[row sep=crcr]{%
21	251.431746550473\\
};
\addplot [color=black,dotted,forget plot]
  table[row sep=crcr]{%
21	251.431746550473\\
21	384.870955135668\\
};
\addplot [color=black,dotted,forget plot]
  table[row sep=crcr]{%
21	251.431746550473\\
21	346.963827335793\\
};
\end{axis}
\end{tikzpicture}%
    \end{minipage}
    &\hspace{-1.7cm}
    \begin{minipage}{0.86\columnwidth}
%
%
\definecolor{cool_blue}{rgb}{0.00000,0.44700,0.74100}%
\definecolor{cool_red}{rgb}{0.85000,0.32500,0.09800}%
\definecolor{cool_yellow}{rgb}{0.92900,0.69400,0.12500}%
\begin{tikzpicture}

\begin{axis}[%
width=0.3\linewidth,
height=0.25\linewidth,
scale only axis,
every outer x axis line/.append style={black},
every x tick label/.append style={font=\color{black}},
xmin=17,
xmax=23,
xtick={17,18,19,20,21,22,23},
xticklabels = {,,,$k'$},
every x tick label/.append style={font=\color{black}, font=\footnotesize},
every y tick label/.append style={font=\color{black}, font=\footnotesize},
every outer y axis line/.append style={black},
ymin=160,
ymax=450,
ymajorgrids,
ytick={200,300,400},
yticklabels = {2,3,4},
xlabel style={at={(0.5,0.05)}},
ylabel style={at={(0.1,0.5)}},
axis background/.style={fill=white},
axis x line*=bottom,
axis y line*=left,
]
\addplot [color=cool_blue,solid,line width=2.0pt]
  table[row sep=crcr]{%
21	384.870955135668\\
22	410.407841409432\\
23	597.843209996136\\
24	773.264586299968\\
25	816.16598371819\\
26	833.981001127976\\
27	1049.5463566277\\
28	1072.34502429198\\
29	1102.29336945762\\
30	1104.99142238103\\
31	1211.45134169557\\
32	1293.39253972954\\
33	1379.63776398074\\
34	1405.70054509302\\
35	1580.55191622776\\
36	1610.19875796444\\
37	1646.15443390012\\
38	1681.08101428457\\
39	1746.5268710412\\
40	1947.55490563138\\
41	1967.93092448097\\
42	2063.3013171879\\
43	2135.24671322839\\
44	2173.09695658171\\
45	2287.64609999776\\
46	2387.04609659037\\
47	2425.28483542359\\
48	2453.89888075288\\
49	2516.14098462511\\
50	2565.30183936024\\
51	2591.7875002931\\
52	2722.66743223643\\
53	2763.90240719154\\
54	2920.54580997275\\
55	2956.03739043009\\
56	3001.75722586785\\
57	3062.87671489566\\
58	3130.17734318894\\
59	3190.38971939529\\
60	3297.66476080372\\
61	3312.98336659193\\
62	3465.54443283416\\
63	3552.38193559459\\
64	3610.89210425787\\
65	3625.64105809816\\
66	3673.60424598008\\
67	3737.44304614021\\
68	3848.16794233689\\
69	3960.30497380435\\
70	4014.99102208149\\
};

\addplot [color=cool_yellow,solid,line width=2.0pt]
  table[row sep=crcr]{%
21	346.963827335793\\
22	398.008773145393\\
23	423.476723458444\\
24	445.995808959804\\
25	462.718472596112\\
26	550.407455359736\\
27	587.545764235\\
28	647.157925120532\\
29	678.160453923147\\
30	697.135738093104\\
31	708.022373838705\\
32	765.492450939106\\
33	777.824104096067\\
34	842.695457688795\\
35	926.914681669686\\
36	951.810769140759\\
37	982.74804409789\\
38	1015.87516500578\\
39	1029.80283181505\\
40	1047.2796608916\\
41	1054.27320769571\\
42	1087.04918922312\\
43	1108.48546645706\\
44	1162.24390730698\\
45	1182.08237399702\\
46	1197.49944061599\\
47	1212.67324373783\\
48	1230.6442677454\\
49	1280.74646473428\\
50	1288.67333173057\\
51	1320.86430752014\\
52	1327.89042026072\\
53	1357.68740410154\\
54	1458.23008039856\\
55	1479.57909804475\\
56	1509.05186560871\\
57	1561.92513508695\\
58	1568.90669475618\\
59	1611.84378254494\\
60	1612.8867375263\\
61	1666.69310179206\\
62	1668.6864932404\\
63	1685.33792419083\\
64	1707.9634356058\\
65	1720.58259076326\\
66	1760.20345070587\\
67	1808.38205627669\\
68	1827.45828827572\\
69	1900.24365440995\\
70	1936.02361321172\\
};

\addplot [color=cool_blue,dashed,line width=2.0pt]
  table[row sep=crcr]{%
1	6.78497141554536e-14\\
2	54.0865528134872\\
3	92.9558007977542\\
4	195.192433193682\\
5	287.666087542766\\
6	360.576786048715\\
7	401.487910967433\\
8	414.968039819515\\
9	459.245014863843\\
10	618.790592572252\\
11	679.872127753204\\
12	740.433598305773\\
13	848.493804242392\\
14	921.625436321699\\
15	1029.04596371713\\
16	1067.88676785573\\
17	1102.40150408717\\
18	1122.896587685\\
19	1254.34314410595\\
20	1290.61063903334\\
21	1324.55753360005\\
22	1491.51616807861\\
23	1588.31355621855\\
24	1605.43807640024\\
25	1625.48168364583\\
26	1733.70434398026\\
27	1827.39600832951\\
28	1947.62396009737\\
29	2019.85768620765\\
30	2086.8593462428\\
31	2125.48455700673\\
32	2182.61080594649\\
33	2252.60801286236\\
34	2315.89774100389\\
35	2410.8631170748\\
36	2463.57664305918\\
37	2577.20634241684\\
38	2612.19084278953\\
39	2705.53766015801\\
40	2746.34415263412\\
41	2777.09080786112\\
42	2863.93577960792\\
43	2889.39428361085\\
44	2967.00754291663\\
45	3093.71810781536\\
46	3164.27753985459\\
47	3294.64787063277\\
48	3321.37899783042\\
49	3363.22959876488\\
50	3496.1100756255\\
51	3524.68573338351\\
52	3616.43625786321\\
53	3702.20660313604\\
54	3728.60968228716\\
55	3873.70585009788\\
56	3933.74235681527\\
57	3963.4812749091\\
58	3978.66455761494\\
59	4088.86695610532\\
60	4122.17024118791\\
61	4197.67275009567\\
62	4240.71735623705\\
63	4333.78706182758\\
64	4417.86228831064\\
65	4446.77394973982\\
66	4515.23420191116\\
67	4567.07450492281\\
68	4689.93379830251\\
69	4721.55900214178\\
70	4822.52937492351\\
};

\addplot [color=cool_yellow,dashed,line width=2.0pt]
  table[row sep=crcr]{%
1	1.57383634390943e-15\\
2	26.7941663325686\\
3	28.9916405739076\\
4	49.5587999570412\\
5	119.009521843582\\
6	130.926734641149\\
7	134.234298465684\\
8	205.415499505859\\
9	260.604141006607\\
10	269.594048889446\\
11	338.054526386373\\
12	357.370812467338\\
13	367.516302486553\\
14	407.946580948955\\
15	419.414363021096\\
16	451.368373871872\\
17	495.159194157778\\
18	523.414780412348\\
19	549.898341771898\\
20	583.204558118906\\
21	623.296158259439\\
22	708.2347699161\\
23	726.132820124736\\
24	758.871682442712\\
25	822.155216950983\\
26	862.561714193898\\
27	878.199356719437\\
28	906.32254196524\\
29	925.132330025846\\
30	980.22495132336\\
31	1023.0130864511\\
32	1038.68356101197\\
33	1053.73000991427\\
34	1092.68051888382\\
35	1102.83695824987\\
36	1110.70280056654\\
37	1144.52151695048\\
38	1173.70974075408\\
39	1183.75047232404\\
40	1198.80795666856\\
41	1217.13292666119\\
42	1243.6563818148\\
43	1284.08558425673\\
44	1325.23421041599\\
45	1353.10982082127\\
46	1361.01784735434\\
47	1378.72546786694\\
48	1401.43952757537\\
49	1429.23472789598\\
50	1445.08890559724\\
51	1484.13391625936\\
52	1514.28664809078\\
53	1555.69304381089\\
54	1572.24182377408\\
55	1648.30454597113\\
56	1686.43340452878\\
57	1722.39021169658\\
58	1753.88214488057\\
59	1770.05560110802\\
60	1835.17058551712\\
61	1890.11084905733\\
62	1913.22646357307\\
63	1920.25019979483\\
64	1951.1528911814\\
65	1986.34729119768\\
66	2008.41991882771\\
67	2012.33266090182\\
68	2029.67426667999\\
69	2067.36023184867\\
70	2106.37798021326\\
};

\addplot [color=cool_red,solid,line width=2.0pt]
  table[row sep=crcr]{%
1	3.1694848026502e-14\\
2	4.87507178819213\\
3	6.3261742599453\\
4	9.30256739865693\\
5	13.9450194048203\\
6	19.9100327424224\\
7	39.9862829059263\\
8	40.3050827658758\\
9	57.3094348662391\\
10	61.007710871497\\
11	87.8770388627712\\
12	101.653905228337\\
13	118.736393885299\\
14	131.144749926961\\
15	135.32573262194\\
16	171.979368108208\\
17	184.75571102838\\
18	194.612549557522\\
19	217.481524383442\\
20	235.222225065863\\
};

\addplot [color=cool_blue,only marks,mark=o,mark options={solid},forget plot]
  table[row sep=crcr]{%
21	384.870955135668\\
};
\addplot [color=cool_yellow,only marks,mark=o,mark options={solid},forget plot]
  table[row sep=crcr]{%
21	346.963827335793\\
};
\addplot [color=cool_red,only marks,mark=o,mark options={solid},forget plot]
  table[row sep=crcr]{%
20	235.222225065863\\
21	235.222225065863\\
};
\addplot [color=black,dotted,line width=1.5pt,forget plot]
  table[row sep=crcr]{%
21	235.222225065863\\
21	346.963827335793\\
};
\end{axis}
\end{tikzpicture}%
      \\
      \begin{overpic}
      [trim=0cm 0cm 0cm 0cm,clip,width=0.3\linewidth]{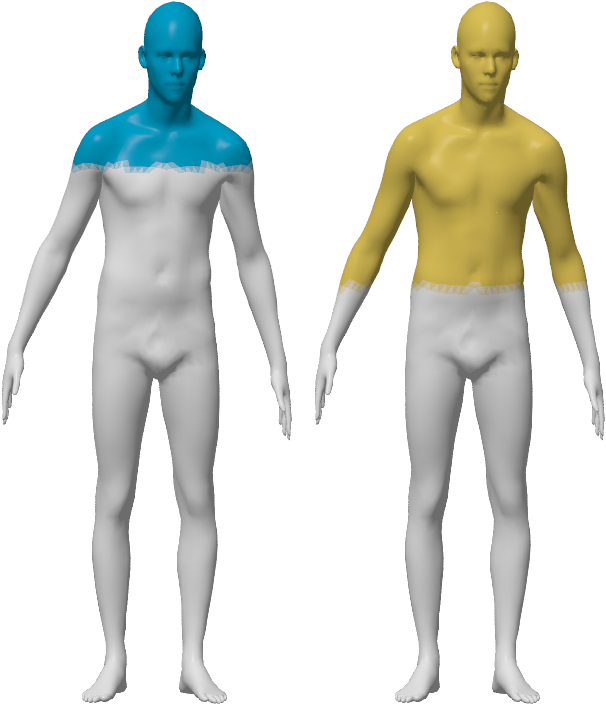}
      \put(37,80){\footnotesize $R_1$}
	  \put(82,80){\footnotesize $R_2$}
      \end{overpic}
    \end{minipage}
	\vspace{-0.15cm}
  \end{tabular}
\caption{\label{fig:gap}{\em Left}: LMH spectra grow linearly with rate inversely proportional to the area of the region, and are bounded from above by the standard Laplacian eigenvalues of submanifold $R\subseteq\M$. {\em Top-right}: Enlargement of the left plot around the index $k'=20$. We illustrate the spectral gap (dotted) between $\lambda_{k'}(\mathbf{W})$ and $\lambda_1(\mathbf{Q}_{v,k'})$; note that the gap is different among the two regions.}
\end{figure}

\noindent\textbf{Spectrum. }
By the interpretation of \eqref{eq:eval} as a generalized eigenvalue problem \eqref{eq:geigen}, we obtain a natural notion of {\em spectrum} associated with the LMH, namely given by $\lambda_j(\Q) = \bm{\psi}_j^\top \Q \bm{\psi}_j, j=1,\dots,k$. Indeed, since at the optimum $\mathcal{E}_R(\bm{\psi}_j) \approx \mathcal{E}_\perp(\bm{\psi}_j)\approx 0$ for all $j$, we have $\bm{\psi}_j^\top \Q \bm{\psi}_j \approx\bm{\psi}_j^\top \mathbf{W} \bm{\psi}_j$, i.e., the Dirichlet energy of $\bm{\psi}_j$ as in the classical setting~\eqref{eq:dirichlet}. This provides us with a natural ordering of the basis functions; it remains to see in what measure do the localized harmonics bring additional (higher frequency) information to the global basis formed by the first $k'$ Laplacian eigenfunctions $\bm{\phi}_1, \hdots, \bm{\phi}_{k'}$. A first answer is provided by the following 
\newtheorem{theorem}{Theorem}
\begin{theorem}\textbf{(spectral gap). }
Let $(\phi_i,\lambda_i(\mathbf{W}))_{i=1}^n$ be the eigenpairs of the standard Laplacian, and let $\Q_{v,k'}$ be defined as in \eqref{eq:Q}. Then, for \rev{large enough $\mu_\perp$,} any non-negative $v$ and any choice of $k' \leq n-1$, we have
%
$\lambda_{k'}(\mathbf{W}) \le \lambda_1 (\Q_{v,k'})$, 
%
with equality holding iff $\bm{\phi}_{k'+1}(x)=0$ whenever $v(x)\neq 0$.
\label{thm:gap}
\end{theorem}
{\em Proof}. See supplementary material.

This theorem ensures the existence of a non-negative gap between the two spectra, i.e., the new basis functions do not introduce any redundancy, and the gap is the smallest possible by the global optimality of \eqref{eq:eval}. In other words, the localized basis ``picks up'' where the global basis ``left off'' (see Figure~\ref{fig:gap} for examples). 
Note that the last condition on $v$ in Theorem~\ref{thm:gap} is almost never realized in practice: equality is obtained only when the Laplacian eigenfunction $\bm{\phi}_{k'+1}$ is localized to the same region indicated by $v$.

Interestingly, for a special class of functions $v$ the spectrum $\lambda_1(\Q_{v,k'})\le \lambda_2(\Q_{v,k'})\le\dots$ follows a well-defined behavior, as remarked below.

\noindent\textbf{Observation. }
Let $v$ be a {\em binary} indicator function supported on some (possibly disconnected) region $R\subseteq\M$. Then,
%
$\lambda_{i}(\Q_{v,k'}) - \lambda_{1}(\Q_{v,k'}) \propto i/\sqrt{\mathrm{Area}(R)}$
%
as $i\to\infty$.

The observation above can be thought of as a generalization of Weyl's asymptotic law \cite{chavel1984eigenvalues} to sub-regions of $\M$ (see Figure~\ref{fig:gap}).
 
\noindent\textbf{Comparison to standard Laplacian on parts. }
%
Perhaps the most direct way to achieve locality is to consider the given region $R\subseteq\M$ as a separate manifold with boundary $\partial R$ and Laplacian $\Delta_R$, and then compute the  eigen-decomposition $\mathbf{W}^R \X^R = \A^R \X^R \bm{\Lambda}^R$ of $\Delta_R$ (note that $\mathbf{W}^R,\A^R$ can be obtained as submatrices of $\mathbf{W},\A$ followed by normalization and by fixing the weights along $\partial R$). The eigenfunctions $\psi^R_i$ can then be extended to the entire $\M$ by means of zero-padding, 
$$
\tilde{\psi}^R_i(x) = \left \{
\begin{array}{cc}
\psi^R_i(x) & x\in R  \\
0 & \mathrm{else} 
\end{array}
\right.
$$


A first difference between this and our approach lies in the fact that $\langle \tilde{\psi}_i^R, \phi_j\rangle_{L^2(\M)} \neq \delta_{ij}$ in general, i.e., the extended partial eigenfunctions do not ``complete'' the global basis and there is no separation of spectra (guaranteed in our case by Theorem~\ref{thm:gap}), leading in turn to a redundant representation.

\begin{figure}[t]
  \centering
  \begin{overpic}
  [trim=0cm 0cm 0cm 0cm,clip,width=1.0\linewidth]{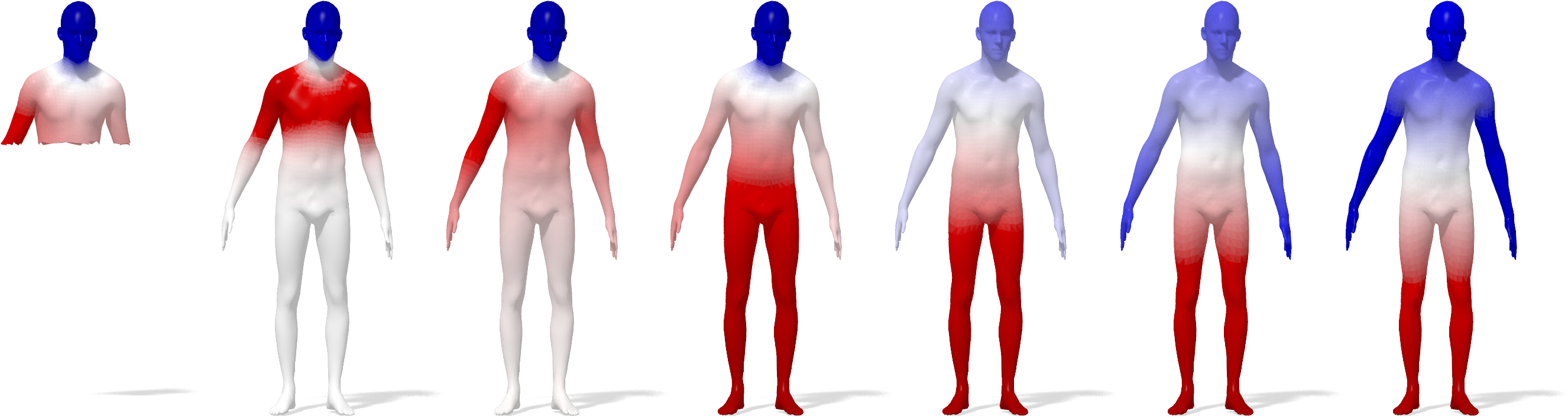}
  \put(13,-4){\footnotesize $\mu_R=350$}
  \put(32.5,-4){\footnotesize $50$}
  \put(47,-4){\footnotesize $25$}
  \put(61,-4){\footnotesize $15$}
  \put(76.5,-4){\footnotesize $5$}
  \put(90.5,-4){\footnotesize $0$}
  \put(8,24){\footnotesize $\phi_i^R$}
  \put(23.5,24){\footnotesize $\psi_{j_1}$}
  \put(37.5,24){\footnotesize $\psi_{j_2}$}
  \put(52,24){\footnotesize $\psi_{j_3}$}
  \put(66,24){\footnotesize $\psi_{j_4}$}
  \put(80.5,24){\footnotesize $\psi_{j_5}$}
  \put(95,24){\footnotesize $\phi_k$}
  \end{overpic}
  \vspace{-0.1cm}
  \caption{\label{fig:completion}
  Our model allows to smoothly transition from a localized solution equivalent to a standard Laplacian eigenfunction $\phi_i^R$ on a {\em partial} shape \rev{with Neumann boundary conditions} (first column), to a globally supported solution equivalent to a standard Laplacian eigenfunction $\phi_k$ on the {\em full} shape (last column). Each $\psi$ is obtained by solving a different problem with a different $\mu_R$ and $\mu_\perp=0$. Note that the resulting ``interpolating'' harmonics $\psi_{j_{1,\dots,5}}$ do not necessarily correspond to the same eigenvalue.}
\end{figure}

Secondly, our approach is more general in that we allow ``soft'' regions represented by allowing $v$ to obtain values in the interval $[0,1]$, which is obviously not achievable by extracting sub-regions.
This latter property is especially important in applications where a sharp (binary) selection would lead to undesirable boundary effects around the region of interest (see Section~\ref{sec:apps} for examples).

Finally, we stress that the standard Laplacian $\Delta_{R\subseteq\M}$ may have an eigenspace in common with our operator with $\mu_\perp=0$, a binary $v$ on $R$, and large enough $\mu_R$. 
%
%
In turn, the full Laplacian $\Delta_\M$ is always obtained for $\mu_R=0$. A remarkable manifestation of this fact is given by the ``interpolation effect'' shown in Figure~\ref{fig:completion}.
Note that the observation above is not true in general, since we do not impose any boundary conditions w.r.t. $R$ in our problem (indeed, we allow $R$ to be soft), while all eigenfunctions of $\Delta_R$ always satisfy specific boundary conditions such as \eqref{eq:bc}.
Despite the loose connection, this observation allows us to complement the lower bound of Theorem~\ref{thm:gap} by the following
\begin{theorem}\textbf{(upper bound). }
Let $\mathbf{W}^R$ be the stiffness matrix associated with the submanifold $R\subseteq\M$ and define $v$ as the binary indicator function of $\M\setminus R$. Then, 
%
$\lambda_i(\mathbf{Q}_{v,k'}) \le \lambda_{i+k'}(\mathbf{W}^R)$ 
%
for any $k \leq n$.
\label{thm:upper}
\end{theorem}
{\em Proof}. See supplementary material.

See Figure~\ref{fig:gap} for an example.

\noindent\textbf{Comparison to compressed manifold modes. }
Ozoli{\c{n}}{\v{s}} et al. \cite{ozolicnvs2013compressed} proposed computing {\em compressed manifold modes (CMM)} as solutions to 
\begin{align}\label{eq:cmm}
\min_{\X\in\mathbb{R}^{n\times k}} ~~\mathrm{tr}(\X^\top \mathbf{W} \X)+\mu\|\X\|_1\quad
\mathrm{s.t.} ~~\X^\top \A \X = \Id\,.
\end{align}
%
Problem \eqref{eq:cmm} makes use of a sparsity-inducing $L_1$ prior which, together with the smoothness promoted by the Dirichlet term, leads to the resulting functions having compact support controlled by parameter $\mu$. It is important to note that this model does not allow to explicitly control the modes location. As shown in \cite{Neumann}, these functions tend to concentrate around areas like shape protrusions and ridges.
\rev{While different in its nature, the CMM model \eqref{eq:cmm} admits a computational procedure which shares some similarities with ours.} Assume that the solution of \eqref{eq:cmm} for a given $\mu$ is a set of $k$ functions supported on regions $R_1,\dots,R_k$ \rev{(obtained {\em a posteriori})}, represented by the soft indicators $1-v_1,\dots,1-v_k$, and let $\mu_\perp=0$. Then, the application of our framework using the matrices $\mathbf{Q}_{v_i,0}$ \rev{corresponds to one iteration of} the iterative reweighting scheme proposed for the efficient computation of CMMs in \cite{bronstein2016consistent}.

\rev{
\noindent\textbf{Comparison to elliptic operator. }
Concurrently with our work, Choukroun et al. \cite{elliptic} considered a family of elliptic operators of the form $\mathbf{H} = \mathbf{W} + \mathbf{V}$, where the {\em potential} $\mathbf{V}$ is a diagonal operator akin to our localization term \eqref{eq:localR}. The same approach was recently followed in \cite{liu17} to obtain localized basis functions around points of interest on the surface.

Differently from these approaches, we seek for localized basis functions that {\em simultaneously} lie in a subspace orthogonal to $\mathrm{span}\{\phi_1,\dots,\phi_{k'}\}$, where $\phi_i$ are the first $k'$ standard Laplacian eigenfunctions. In other words, we seek to ``augment'' the global basis by introducing a local refinement, while the aforementioned works attempt to construct a complete basis in agreement with the input potential. 
This is a crucial difference that has noticeable effects in practice, as we will demonstrate in Section~\ref{sec:apps}.
}

\section{Implementation}\label{sec:impl}
\noindent\textbf{Optimization. }
As shown in Section~\ref{sec:PM}, computing our localized basis functions boils down to solving a generalized eigenvalue problem
%
$\Q\X = \A\X\bm{\Lambda}$, 
%
with $\Q = \mathbf{W}+\mu_R\, \A\mathrm{diag}(\mathbf{v})+\mu_\perp \A\bm{\Phi} \bm{\Phi}^\top \A$. 
We note that computing $\Q$ explicitly involves the construction of a dense $n\times n$ matrix $\A\bm{\Phi} \bm{\Phi}^\top \A$, which may become prohibitive for large meshes. 
However, \rev{we avoid this computation} altogether by noticing that the $\mu_\perp$-term has very low rank $k'\ll n$. This condition allows the application of exact update formulas throughout the optimization of problem \eqref{eq:eval}, which can be solved efficiently and globally as detailed in Appendix B.
%
%
\begin{table}[t!] 
\resizebox{1\columnwidth}{!}{
\begin{tabular}{|C{0.90cm}|C{0.90cm}|C{0.90cm}|C{0.90cm}|C{0.90cm}|C{0.90cm}|C{0.90cm}|}\hline 
 & \multicolumn{2}{c|}{ \footnotesize{ $\mathbf{k = 100}$}} & \multicolumn{2}{c|}{ \footnotesize{ $\mathbf{k = 200}$}} & \multicolumn{2}{c|}{ \footnotesize{ $\mathbf{k = 300}$}} \\\hline 
 & \footnotesize{ \textbf{hard}} & \footnotesize{ \textbf{soft}} & \footnotesize{ \textbf{hard}} & \footnotesize{ \textbf{soft}} &\footnotesize{ \textbf{hard}} & \footnotesize{ \textbf{soft}} \\\hline 
\textbf{\footnotesize{$\mathbf{ \sim 120}$K}} &  \footnotesize{-} &  \footnotesize{112.3s} &  \footnotesize{-} & \footnotesize{200.5s} &  \footnotesize{-} &  \footnotesize{304.9s}
 \\\hline
\footnotesize{\textbf{$\mathbf{\sim 12}$K}} &  \footnotesize{40.8s} &  \footnotesize{7.6s} &  \footnotesize{66.7s} &  \footnotesize{16.4s} &  \footnotesize{79.3s} &  \footnotesize{25.0s} 
\\\hline
\footnotesize{\textbf{$\mathbf{\sim 1.2K}$}} &  \footnotesize{1.1s} & \footnotesize{0.7s} &  \footnotesize{2.2s} & \footnotesize{1.0s} &  \footnotesize{4.6s} &  \footnotesize{1.4s}
 \\\hline \end{tabular}
} 
\caption{\label{fig:timesKIDS} \rev{Runtime comparison for global optimization of our problem under hard \eqref{eq:problem} and soft constraints \eqref{eq:prel} across different mesh resolutions (number of faces) and basis size $k$. Tests denoted by `-' could not run due to memory limitations (see Appendix A).}}
\end{table}

\begin{figure}[tbh]
  \centering
  \setlength{\tabcolsep}{0pt}
  \begin{tabular}{cc}
  \imagetop{
	  \begin{overpic}
	  [trim=0cm 0cm 0cm 0cm,clip,width=0.56\linewidth]{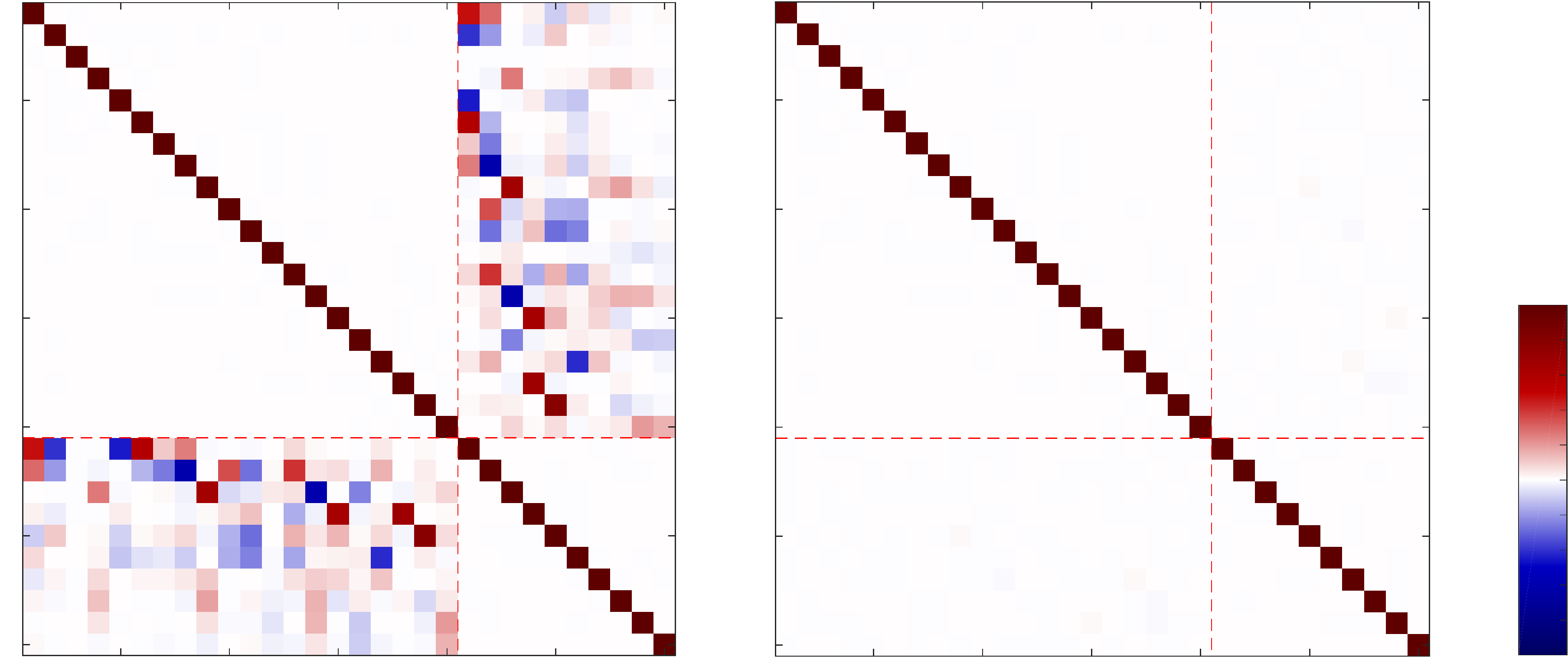}
	  \put(1,-8.5){\footnotesize $1$}
	  \put(26,-8.5){\footnotesize $k'$}
	  \put(43,-8.5){\footnotesize $k$}
	  \put(13,-8.5){\footnotesize $\bm{\Phi}$}
	  \put(34.5,-8.5){\footnotesize $\X$}
	  \put(101,20){\footnotesize $+1$}
	  \put(101,10){\footnotesize $0$}
	  \put(101,1){\footnotesize $-1$}
	  \put(1,-2){\line(1,0){28}}
	  \put(31,-2){\line(1,0){12}}
	  \end{overpic}
  }
  &
  \hspace{0.3cm}
  \imagetop{
%
%
\begin{tikzpicture}

\begin{axis}[%
width=0.25\linewidth,
height=0.24\linewidth,
scale only axis,
separate axis lines,
every outer x axis line/.append style={black},
every x tick label/.append style={font=\color{black}, font=\footnotesize},
every y tick label/.append style={font=\color{black}, font=\footnotesize},
xmode=log,
xmin=100,
xmax=10000000,
xminorticks=true,
xlabel={\footnotesize $\mu_\perp$},
every outer y axis line/.append style={black},
ymin=0,
ymax=120,
ylabel={\footnotesize Error},
ylabel style={at={(0.25,0.5)}},
xlabel style={at={(0.5,0.1)}},
axis background/.style={fill=white}
]
\addplot [color=blue,solid,line width=2.0pt,forget plot]
  table[row sep=crcr]{%
100	106.925365270206\\
500	103.881756028422\\
1000	98.633735074658\\
5000	60.6079800893538\\
10000	38.5917130822397\\
50000	16.0143073629384\\
100000	9.12461962078076\\
200000	4.87316247123897\\
300000	3.31858544731935\\
400000	2.51464303766818\\
500000	2.02438243261125\\
1000000	1.02480747333889\\
5000000	0.206931856398059\\
10000000	0.10360414886626\\
};
\end{axis}
\end{tikzpicture}%
  }
  \vspace{-0.25cm}
  \end{tabular}
  \caption{\label{fig:ortho}\rev{{\em Left}:} Matrix of inner products $(\langle b_i,b_j \rangle_{L^2(\M)})$, where $b_i,b_j\in\{\phi_1,\dots,\phi_{k'},\psi_1,\dots,\psi_k\}$. Here $\psi_i$ are the optimal localized basis functions computed with $\mu_\perp=10^{-1}$ (first column) and $\mu_\perp=10^5$ (second column). \rev{{\em Right}: We plot the discrepancy between solutions to the exact \eqref{eq:problem} and relaxed \eqref{eq:prel} problems as a function of $\mu_\perp$, measured as the $L_2$ distance between the resulting spectra.}}
\end{figure}

\rev{\noindent\textbf{Timing. }
In Table~\ref{fig:timesKIDS} we report the runtime (in seconds) required by our method as executed on an Intel 3.6 GHz Core i7 cpu with 16GB ram. We compare the execution time for the exact problem with hard orthogonality constraints \eqref{eq:problem} and the relaxed problem \eqref{eq:prel}. Note that while the latter relaxation is significantly more efficient, the two formulations yielded numerically close solutions in all our experiments (see Figure~\ref{fig:ortho}).

%
%
%
%
}

\noindent\textbf{Choice of parameters. }
%
Parameters $\mu_R$ and $k'$ control the locality and the number of global harmonics to use, respectively, and are application-dependent (see Section~\ref{sec:apps}). 
Parameter $\mu_\perp$ enforces orthogonality w.r.t. the standard Laplacian eigenfunctions, and should be chosen large enough so that the orthogonality constraints are satisfied. \rev{In our experiments we used $\mu_R\approx10^2$ and $\mu_\perp \approx 10^5$; see Figure~\ref{fig:ortho} for a quantitative evaluation on the choice of $\mu_\perp$.}

\begin{figure}[t]
  \centering
  \begin{overpic}
  [trim=0cm 0cm 0cm 0cm,clip,width=1.0\linewidth]{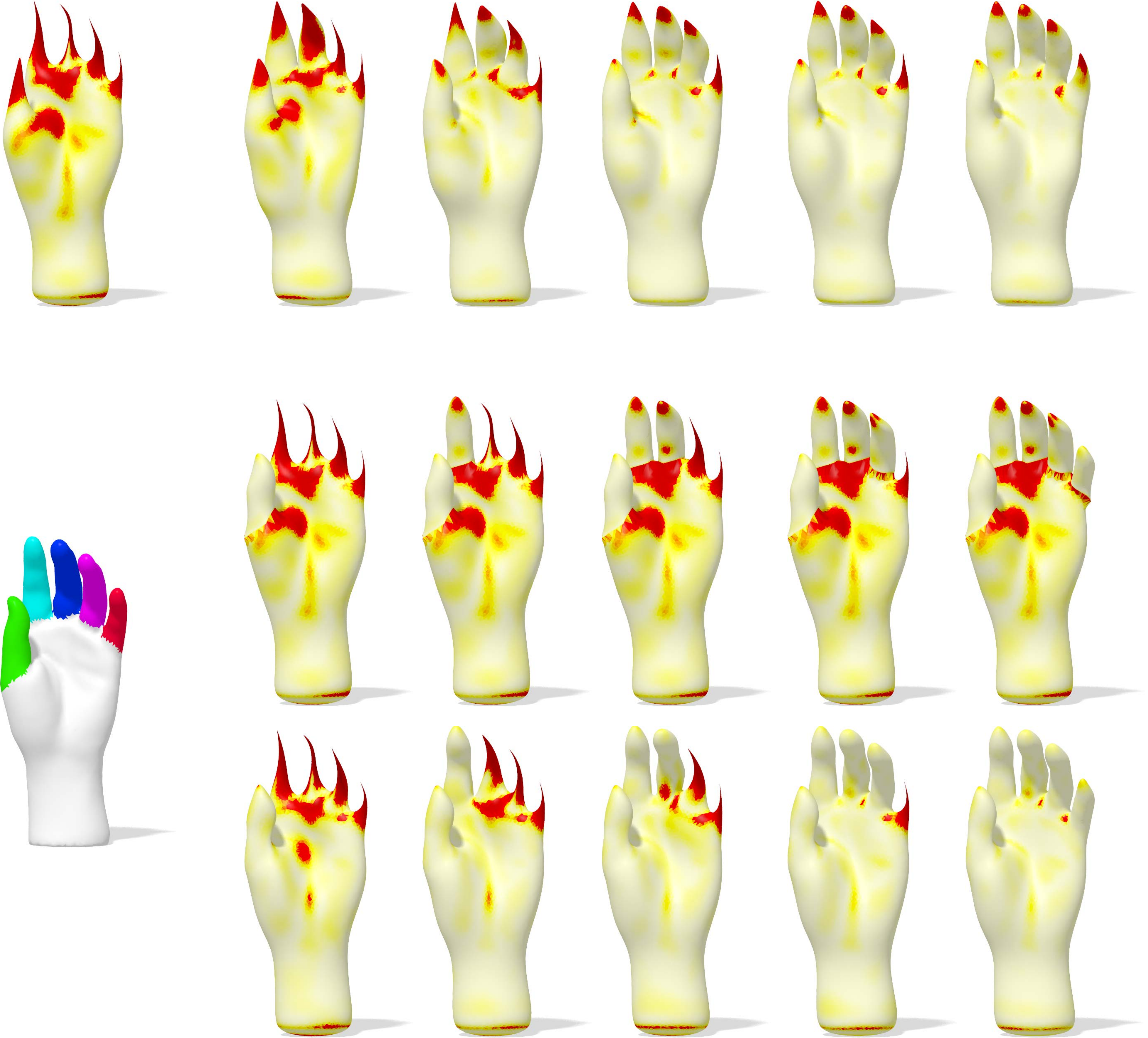}
  \put(1,60){\footnotesize $k' = 50$}
  \put(23,60){\footnotesize $k'+10$}
  \put(38,60){\footnotesize $k'+20$}
  \put(54,60){\footnotesize $k'+30$}
  \put(70,60){\footnotesize $k'+40$}
  \put(86,60){\footnotesize $k'+50$}
  

  

  \put(16,87){\footnotesize MH}
  \put(15.6,53.5){\footnotesize PMH}
  \put(15.6,24.5){\footnotesize LMH}

  \put(16,75){\footnotesize $\cdots$}

  \put(0,13){\footnotesize regions $R_i$}
  \end{overpic}
  \vspace{-0.4cm}
  \caption{\label{fig:hand_reconstruction}{\em First row}: Surface reconstruction via \eqref{eq:xyz} using the first $k'$ to $k'+50$ Laplacian eigenfunctions. {\em Second row}: Each finger $R_i\subset\M$ is treated as a separate sub-manifold, and the eigenfunctions of the ``partial'' Laplacians $\Delta_{R_i}$ are used to update the initial reconstruction by adding 10 harmonics per finger. 
  {\em Third row}: Reconstruction via \eqref{eq:locxyz} with 10 localized harmonics per finger. Note the significantly higher accuracy of LMH despite using the same number of harmonics as MH and PMH. The heatmap encodes reconstruction error, growing from white to dark red.
}
\end{figure}

\section{Applications}\label{sec:apps}
Localized manifold harmonics are a general tool that can be employed as a drop-in replacement for, or in conjunction with the classical manifold harmonics ubiquitous in spectral shape analysis. In this Section we showcase their application in two broad tasks in graphics: spectral shape processing and shape correspondence.

\noindent\textbf{Spectral shape processing. }
%
%
%
In this context, the surface $\M$ is represented as \rev{a vector-valued function $\mathbf{x}:\M\to\mathbb{R}^3$}, encoding the spatial coordinates of its embedding in $\mathbb{R}^{3}$; transformations to the surface geometry are then phrased as filtering operations applied to the coordinate functions. Vallet and L\'evy~\cite{Levy08} proposed to perform such filtering in the Fourier domain, where the coordinates $\mathbf{x}$ are expressed as linear combinations of Laplacian eigenfunctions,\rev{
\begin{equation}\label{eq:xyz}
\mathbf{x}=\sum_{i\ge 1}\langle\phi_i,\mathbf{x}\rangle_{L^2(\M)} \phi_i\,,
\end{equation}
}
\rev{where, with some abuse of notation, $\langle\phi_i,\mathbf{x}\rangle_{L^2(\M)} = (\langle\phi_i,x_1\rangle_{L^2(\M)}, \hdots, \langle\phi_i,x_3 \rangle_{L^2(\M)})$.}
By truncating the summation to the first $k'$ terms, one obtains a band-limited representation of the surface. The representation is coarse for small $k'$, while finer details are captured for large values of $k'$; see Figure~\ref{fig:hand_reconstruction} (top row) for an example.
The expression in \eqref{eq:xyz} provides an effective way for representing and manipulating simple shapes with smoothly varying coordinate functions, which can be compactly represented in the first few harmonics. Conversely, this representation is much less efficient for surfaces having details at smaller scales.

\begin{figure}[t]
  \centering
  \begin{overpic}
  [trim=0cm 0cm 0cm 0cm,clip,width=1.0\linewidth]{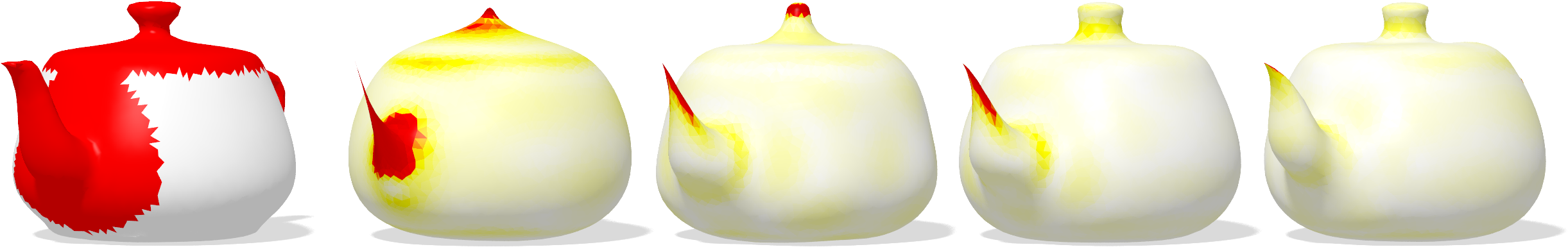}
  \end{overpic}
  \begin{overpic}
  [trim=0cm 0cm 0cm 0cm,clip,width=1.0\linewidth]{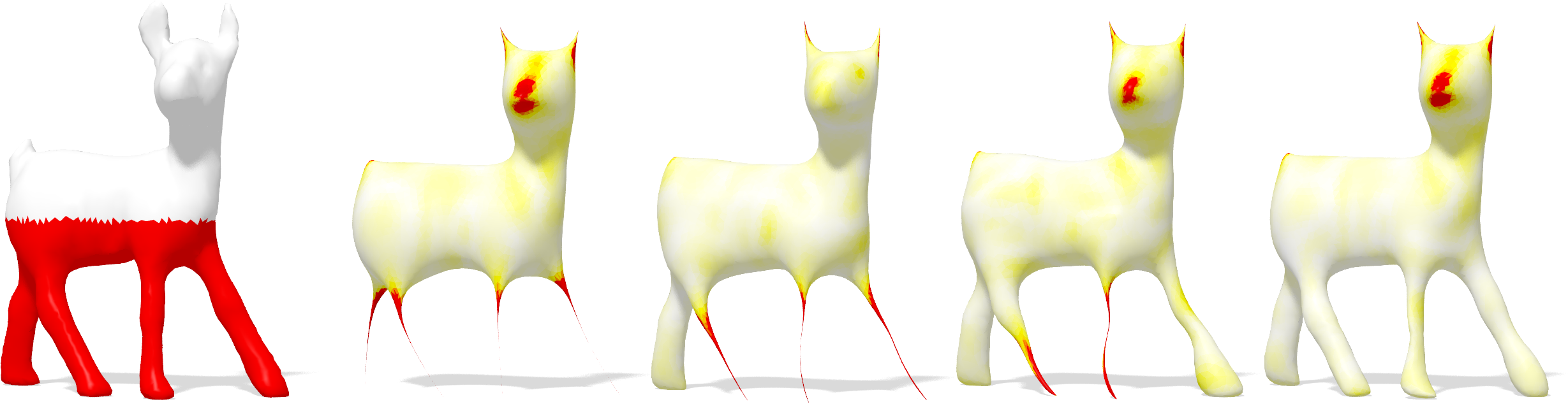}

  \put(81, -3.5){\footnotesize $30+40$ LMH}
  \put(61, -3.5){\footnotesize $30+40$ LMH}
  \put(64.5, -6.5){\footnotesize $\mu_{\perp}=0$ }
    
  \put(45, -3.5){\footnotesize $70$ MH}
  \put(25.5, -3.5){\footnotesize $30$ MH}

  \put(4.5,-3.5){\footnotesize region}
  \end{overpic}
  \caption{\label{fig:stag_no_ortho}\rev{Comparison between manifold harmonics (MH) and localized manifold harmonics (LMH) without and with the orthogonality term \eqref{eq:ortho} using $k'=30$ and $k=40$.}}
\end{figure}

\begin{figure*}[t]
  \centering
  \setlength{\tabcolsep}{0pt}
  \begin{tabular}{ccc}
 \vspace{0.3cm}  
    \begin{minipage}{0.333\linewidth}
%
%
\definecolor{mycolor1}{rgb}{0.29412,0.54471,0.74941}%
\definecolor{mycolor2}{rgb}{0.37176,0.71765,0.36118}%
\definecolor{mycolor3}{rgb}{0.90471,0.19176,0.19882}%
\begin{tikzpicture}

\begin{axis}[%
width=0.7\linewidth,
height=0.7\linewidth,
at={(0.758in,0.481in)},
scale only axis,
xmin=1,
xmax=71,
ymode=log,
ymin=0,
ymax=500,
yminorticks=true,
axis background/.style={fill=white},
legend style={legend cell align=left,align=left,draw=white!15!black},
xmajorgrids,
ymajorgrids,
xtick={1, 15, 29, 43, 57, 71},
xticklabels = {\footnotesize{$70$}, \footnotesize{$84$}, \footnotesize{$98$},\footnotesize{$112$},\footnotesize{$126$},\footnotesize{$140$},},
ytick={25, 50, 75, 100, 500},
yticklabels = {\footnotesize{$\frac{1}{4}\xi$}, \footnotesize{$\frac{1}{2}\xi$}, \footnotesize{$\frac{3}{4}\xi$}, \footnotesize{$\xi$}, \footnotesize{$5\xi$}},
xlabel style={yshift=0.7em},
ylabel style={yshift=-1.5em},
xlabel = {\footnotesize{basis dimension}},
ylabel = {\footnotesize{reconstruction error}},
title style={yshift= - 0.7em},
title = {\footnotesize{Ant}},
]
\addplot [color=mycolor1,solid,line width=2.0pt]
  table[row sep=crcr]{%
1	98.5332806357132\\
2	97.9974260046485\\
3	97.6619129225106\\
4	97.3269733377794\\
5	97.1073498229459\\
6	96.7214943119101\\
7	96.281132683041\\
8	95.7184246038803\\
9	95.3638112336872\\
10	95.0897566893083\\
11	94.6660894690421\\
12	94.3338934684026\\
13	94.1495237453253\\
14	93.9403384168289\\
15	93.6946654572285\\
16	93.5183263072941\\
17	93.2545643030918\\
18	93.1728860365618\\
19	92.9460986469975\\
20	92.7863278916224\\
21	92.6498533728508\\
22	92.4523743166266\\
23	92.2769865686264\\
24	92.0794850351116\\
25	91.9888546792081\\
26	91.7471591086445\\
27	91.467243432981\\
28	91.3601025840214\\
29	90.7453216144187\\
30	90.0324409550854\\
31	89.6865599393978\\
32	89.1545643414546\\
33	88.4055889902372\\
34	87.6388468877547\\
35	87.0381887702279\\
36	86.1686095105644\\
37	85.1137364105777\\
38	84.3011390359166\\
39	83.4653168310426\\
40	82.9571916874758\\
41	82.0932994056109\\
42	81.1700872859064\\
43	80.1941580990003\\
44	79.4325873597311\\
45	78.7282786391777\\
46	77.5328075324004\\
47	76.803995772688\\
48	75.3868766690314\\
49	74.5287453296353\\
50	74.0298132468191\\
51	73.332958568378\\
52	72.8268780246456\\
53	72.0755803618086\\
54	71.486558413738\\
55	70.9405670632072\\
56	70.0084450974366\\
57	69.4790233817295\\
58	68.5811238672122\\
59	67.5622298171247\\
60	66.947384197875\\
61	65.9565037728198\\
62	65.0825067420335\\
63	64.5118726928277\\
64	63.9461523072786\\
65	63.2843371455717\\
66	62.3984308997107\\
67	61.5479471928726\\
68	60.9607927930563\\
69	60.2167993984729\\
70	58.8253842717735\\
71	58.1307080074492\\
};
\addlegendentry{{\footnotesize MH}};

\addplot [color=mycolor2,solid,line width=2.0pt]
  table[row sep=crcr]{%
1	102.933438728574\\
2	141.930433235738\\
3	182.224649029315\\
4	221.311074025672\\
5	399.477595819536\\
6	439.517987937619\\
7	475.9198257722\\
8	488.803776200466\\
9	455.773411214606\\
10	422.076353791934\\
11	389.389101117897\\
12	265.360946201534\\
13	230.371493628358\\
14	197.586953604456\\
15	185.770324169393\\
16	181.999875677397\\
17	177.598278469138\\
18	173.221977678337\\
19	137.722801053692\\
20	134.824844508016\\
21	133.3241439396\\
22	133.09077030093\\
23	131.825239797684\\
24	130.717443743161\\
25	129.760081947479\\
26	119.342452358662\\
27	118.115882117574\\
28	116.635530291815\\
29	115.699120489136\\
30	115.377451338744\\
31	114.841723787099\\
32	114.317993134963\\
33	109.396377639076\\
34	109.039376141019\\
35	108.743158168462\\
36	107.190465562195\\
37	106.146823122566\\
38	105.391630231867\\
39	104.479040059089\\
40	103.251476288321\\
41	102.387598916547\\
42	101.380530883462\\
43	99.4555662056605\\
44	97.7389359969847\\
45	96.1196696037549\\
46	94.2955168839632\\
47	94.0998024691085\\
48	92.6076370996156\\
49	90.7287727853889\\
50	89.6608147763082\\
51	87.322485535585\\
52	84.9534402712702\\
53	82.3414028178339\\
54	81.9657996283836\\
55	78.8942248113938\\
56	76.4963041004564\\
57	75.8789887919541\\
58	73.9815441279137\\
59	71.6635163443668\\
60	69.8639121866665\\
61	69.1264977357441\\
62	66.8028579835676\\
63	64.677796559854\\
64	64.6246858330368\\
65	63.1750141090962\\
66	61.6891204361439\\
67	60.5393555676283\\
68	59.5390870760243\\
69	58.694867726239\\
70	57.7421682242381\\
71	57.6665497202944\\
};
\addlegendentry{{\footnotesize PMH}};

\addplot [color=mycolor3,solid,line width=2.0pt]
  table[row sep=crcr]{%
1	98.5332806357132\\
2	97.8143657352287\\
3	97.3095079517064\\
4	96.6832587325906\\
5	95.993256240772\\
6	95.377645363586\\
7	94.5608270003763\\
8	93.3922517895281\\
9	90.7437086600847\\
10	88.0682720619667\\
11	85.1812333456616\\
12	83.5549765346782\\
13	80.832952244676\\
14	78.1680965261633\\
15	75.8442206704642\\
16	73.1785373051478\\
17	70.2766992492224\\
18	67.6941452240898\\
19	65.7400943681425\\
20	62.5039554991114\\
21	59.626712160393\\
22	58.0143296101936\\
23	57.0694225629962\\
24	55.870204375797\\
25	54.8833967652196\\
26	52.214798623534\\
27	51.434430807439\\
28	50.5467281092159\\
29	50.4948143180572\\
30	49.4923531652221\\
31	48.6306407620129\\
32	47.7582407950518\\
33	45.3920673988693\\
34	44.7020984057043\\
35	43.9063515985914\\
36	43.8152773765627\\
37	43.1961362990637\\
38	42.6009585999277\\
39	42.0872168861086\\
40	40.4433130092428\\
41	39.971921376636\\
42	39.5387788697126\\
43	39.522982149152\\
44	39.044996971246\\
45	38.5766060424837\\
46	38.1339826242366\\
47	37.0637117596197\\
48	36.5198985450423\\
49	36.22184976974\\
50	36.0234505409003\\
51	35.5725055559907\\
52	34.9712099016084\\
53	34.4364990920864\\
54	33.5974424092925\\
55	33.0074161399281\\
56	32.6784840276882\\
57	32.468244843355\\
58	32.2622950028855\\
59	32.0058539841992\\
60	31.8087086391485\\
61	31.1095065633155\\
62	30.9235333437911\\
63	30.741613081154\\
64	30.7016662263698\\
65	30.5583251170527\\
66	30.3860241894388\\
67	30.239837635261\\
68	29.8921808812104\\
69	29.6935352409975\\
70	29.5827604883366\\
71	29.3961327076163\\
};
\addlegendentry{{\footnotesize LMH}};

\end{axis}
\end{tikzpicture}%
     \vspace{0.3cm}
      
	  \begin{overpic}
      	[trim=0cm 0cm 0cm 0cm,clip,width=0.95\linewidth]{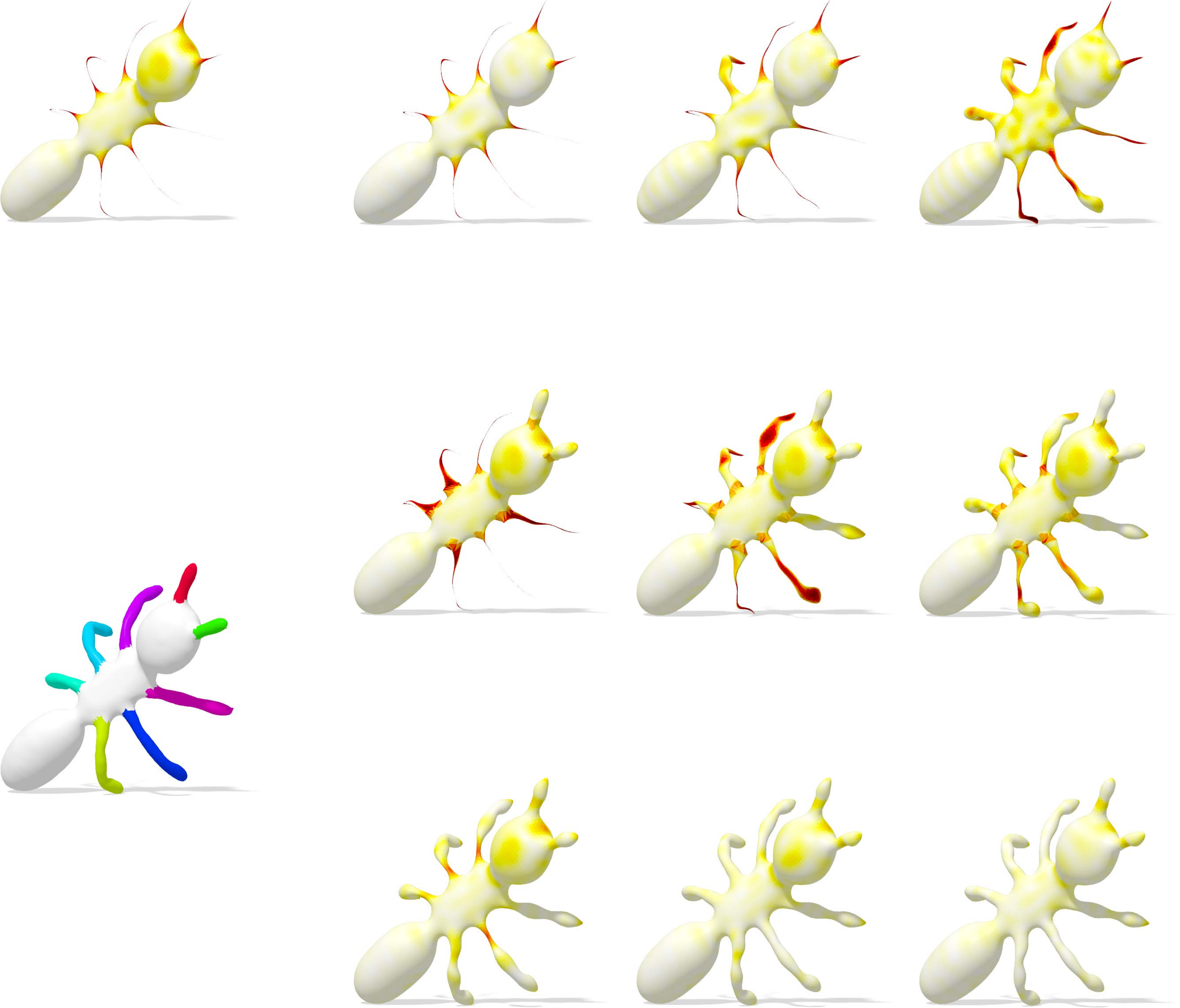}
		\put(21.5,47){$\ldots$}
	    \put(21.5,71){$\ldots$}
	  	\put(21.5,14.5){$\ldots$}

		\put(1,81){\footnotesize  $\xi$ }	  	
	  	\put(1.5,13){\footnotesize regions}
	  	

      	\put(52,61){\footnotesize MH}
	  	\put(48,27){\footnotesize PMH}
	  	\put(48,-6){\footnotesize LMH}
      \end{overpic}      
    \end{minipage}
    & 
      \begin{minipage}{0.333\linewidth}
%
%
\definecolor{mycolor1}{rgb}{0.29412,0.54471,0.74941}%
\definecolor{mycolor2}{rgb}{0.37176,0.71765,0.36118}%
\definecolor{mycolor3}{rgb}{0.90471,0.19176,0.19882}%
\begin{tikzpicture}

\begin{axis}[%
width=0.7\linewidth,
height=0.7\linewidth,
at={(0.758in,0.481in)},
scale only axis,
xmin=1,
xmax=81,
ymode=log,
ymin=0,
ymax=500,
yminorticks=true,
axis background/.style={fill=white},
legend style={legend cell align=left,align=left,draw=white!15!black},
xmajorgrids,
ymajorgrids,
xtick={ 1, 17, 33, 49, 65, 81},
xticklabels = {\footnotesize{$120$}, \footnotesize{$136$}, \footnotesize{$152$},\footnotesize{$168$},\footnotesize{$184$},\footnotesize{$200$},},
ytick={50, 75, 100, 500},
yticklabels = {\footnotesize{$\frac{1}{2}\xi$}, \footnotesize{$\frac{3}{4}\xi$}, \footnotesize{$\xi$}, \footnotesize{$5\xi$}},
xlabel style={yshift=0.7em},
ylabel style={yshift=-1.5em},
xlabel = {\footnotesize{basis dimension}},
ylabel = {\footnotesize{reconstruction error}},
title style={yshift= - 0.7em},
title = {\footnotesize{Chair}},
]
\addplot [color=mycolor1,solid,line width=2.0pt]
  table[row sep=crcr]{%
1	100\\
2	99.2739008173646\\
3	98.4944257290124\\
4	97.8328166550107\\
5	97.4429270534218\\
6	96.8703787857987\\
7	96.2574905636762\\
8	95.7004733899108\\
9	95.2722893427502\\
10	94.6392061034102\\
11	93.9288995354399\\
12	93.3273903540763\\
13	92.5453705624734\\
14	91.7819760262254\\
15	90.8866272100645\\
16	90.2649503708412\\
17	89.6365247380262\\
18	89.0568461946167\\
19	88.6957777389304\\
20	88.289592255342\\
21	87.8651097873932\\
22	87.2876481509955\\
23	86.9145643863373\\
24	86.5475858164563\\
25	86.1711505174031\\
26	85.7701950777449\\
27	85.411990197688\\
28	84.9733144008162\\
29	84.381219406488\\
30	84.0791061285382\\
31	83.7332375281443\\
32	83.4045558271321\\
33	83.167906030966\\
34	82.8916698185336\\
35	82.3742340917204\\
36	81.8484569694545\\
37	81.3731059754101\\
38	80.8903933096519\\
39	80.4411414565462\\
40	79.9537152506012\\
41	79.3005753964776\\
42	79.052507687281\\
43	78.5488885441103\\
44	78.0540696438439\\
45	77.7875614804238\\
46	77.1246864839581\\
47	76.4422142353162\\
48	75.9605770696193\\
49	75.4108860391882\\
50	74.7820834683119\\
51	74.2640620962014\\
52	73.7119161491647\\
53	73.3800737243892\\
54	73.0236135927916\\
55	72.1862333988078\\
56	71.4721289247549\\
57	70.6046775793263\\
58	69.8526194390996\\
59	68.9581477063002\\
60	67.9464709177672\\
61	67.3660045459216\\
62	66.62555925319\\
63	66.101139917634\\
64	65.8704572760832\\
65	65.6220745220616\\
66	65.2882809498021\\
67	64.8312990257935\\
68	64.3836797397958\\
69	63.8463831250164\\
70	63.4726499726925\\
71	62.9157850260713\\
72	62.4083327614251\\
73	61.8440589035866\\
74	61.46023754132\\
75	61.0415008869181\\
76	60.3983298722136\\
77	59.7326952716461\\
78	59.4421743435049\\
79	59.1987923605484\\
80	58.8820729129112\\
81	58.2759128516953\\
};
\addlegendentry{{\footnotesize MH}};

\addplot [color=mycolor2,solid,line width=2.0pt]
  table[row sep=crcr]{%
1	100\\
2	119.428104364288\\
3	138.373061456698\\
4	164.422543594248\\
5	230.633517372064\\
6	296.030589140218\\
7	349.713479969337\\
8	403.813728002638\\
9	456.871228252032\\
10	440.376015866669\\
11	424.389480785858\\
12	401.215303685184\\
13	351.395369671572\\
14	299.791881943818\\
15	252.120386218561\\
16	204.658213138481\\
17	158.337226779469\\
18	158.257117442629\\
19	158.039581354854\\
20	157.784590727935\\
21	149.81786950404\\
22	144.210279357404\\
23	143.142822675592\\
24	141.620949600836\\
25	140.160446822857\\
26	139.354228124096\\
27	138.574633660661\\
28	137.427202986211\\
29	131.688084656959\\
30	126.922883542082\\
31	124.727246767693\\
32	122.553687549319\\
33	120.290939238355\\
34	119.478774575453\\
35	118.597717759069\\
36	118.020337679644\\
37	117.985441860054\\
38	117.772195537275\\
39	117.583369193097\\
40	117.41317381217\\
41	117.271099274742\\
42	116.08558988622\\
43	114.848549315211\\
44	113.765076733324\\
45	113.066546229646\\
46	112.251320674679\\
47	111.273791512155\\
48	110.27575485352\\
49	109.307273918188\\
50	108.159426996285\\
51	107.348699583888\\
52	106.751384042649\\
53	106.257034367019\\
54	105.73033856315\\
55	104.831495933902\\
56	103.997640240611\\
57	103.123979411392\\
58	102.172496955613\\
59	101.226752234204\\
60	100.369007839382\\
61	99.4931988780976\\
62	98.6368198639823\\
63	97.2958777606776\\
64	95.9411800161881\\
65	94.681604517483\\
66	94.2239334848489\\
67	93.6306227609305\\
68	92.9757962960366\\
69	91.9544642646556\\
70	90.4661383570537\\
71	89.1402444926796\\
72	87.619549162575\\
73	85.728560170041\\
74	85.3808990997138\\
75	85.0015461573726\\
76	84.3831551478852\\
77	83.0238494940397\\
78	81.5342142765639\\
79	80.092246077769\\
80	78.4981332158398\\
81	76.7427645137171\\
};
\addlegendentry{{\footnotesize PMH}};

\addplot [color=mycolor3,solid,line width=2.0pt]
  table[row sep=crcr]{%
1	100\\
2	98.9338100997128\\
3	97.8802817804082\\
4	96.9899580588158\\
5	96.1068674667742\\
6	95.3472147283569\\
7	94.3188142870944\\
8	93.3860097562135\\
9	92.3299476486738\\
10	90.9418856569812\\
11	89.5200752473302\\
12	88.1099732175439\\
13	86.3093572907455\\
14	84.5552284601438\\
15	82.6243707460302\\
16	80.9316523754032\\
17	78.9713387217043\\
18	78.7153645532245\\
19	78.412459105174\\
20	77.7787833687313\\
21	76.4957984049487\\
22	75.0555726387225\\
23	73.7538745303493\\
24	72.56628094557\\
25	71.546126586932\\
26	71.3598046426223\\
27	71.068226994272\\
28	70.5846727883727\\
29	69.95841819493\\
30	69.3546590454181\\
31	68.7089110301523\\
32	68.1471637015094\\
33	67.8067489694972\\
34	67.5120823329853\\
35	67.3884173723067\\
36	67.1741362836686\\
37	66.53769691189\\
38	66.1064428596795\\
39	65.7223050627459\\
40	65.2551229859743\\
41	64.7126686879156\\
42	64.4048340540323\\
43	64.1738904280595\\
44	63.8677810507703\\
45	63.3378206371528\\
46	63.08377345461\\
47	62.6039374078352\\
48	62.1359380709104\\
49	61.7665538125891\\
50	61.4201532086503\\
51	61.1179967182984\\
52	60.6874066968894\\
53	60.1599453489224\\
54	59.7350763099589\\
55	59.1372439374886\\
56	58.5600739586982\\
57	58.1473836274538\\
58	57.8031065021439\\
59	57.4643515560164\\
60	56.9317320391823\\
61	56.4545811706019\\
62	56.166484537257\\
63	55.9482692523501\\
64	55.7734240884798\\
65	55.6067835569466\\
66	55.4085831074501\\
67	55.1614846305956\\
68	54.7832730702783\\
69	54.4003785058341\\
70	54.227043376117\\
71	54.0851473989601\\
72	53.979279177116\\
73	53.8722350431634\\
74	53.749210148895\\
75	53.5302932890231\\
76	53.3238619752997\\
77	53.1972914650587\\
78	53.0041774300366\\
79	52.8321679742589\\
80	52.647963549311\\
81	52.5361824892682\\
};
\addlegendentry{{\footnotesize LMH}};

\end{axis}
\end{tikzpicture}%
      \vspace{0.3cm}

	  \begin{overpic}
      	[trim=0cm 0cm 0cm 0cm,clip,width=0.95\linewidth]{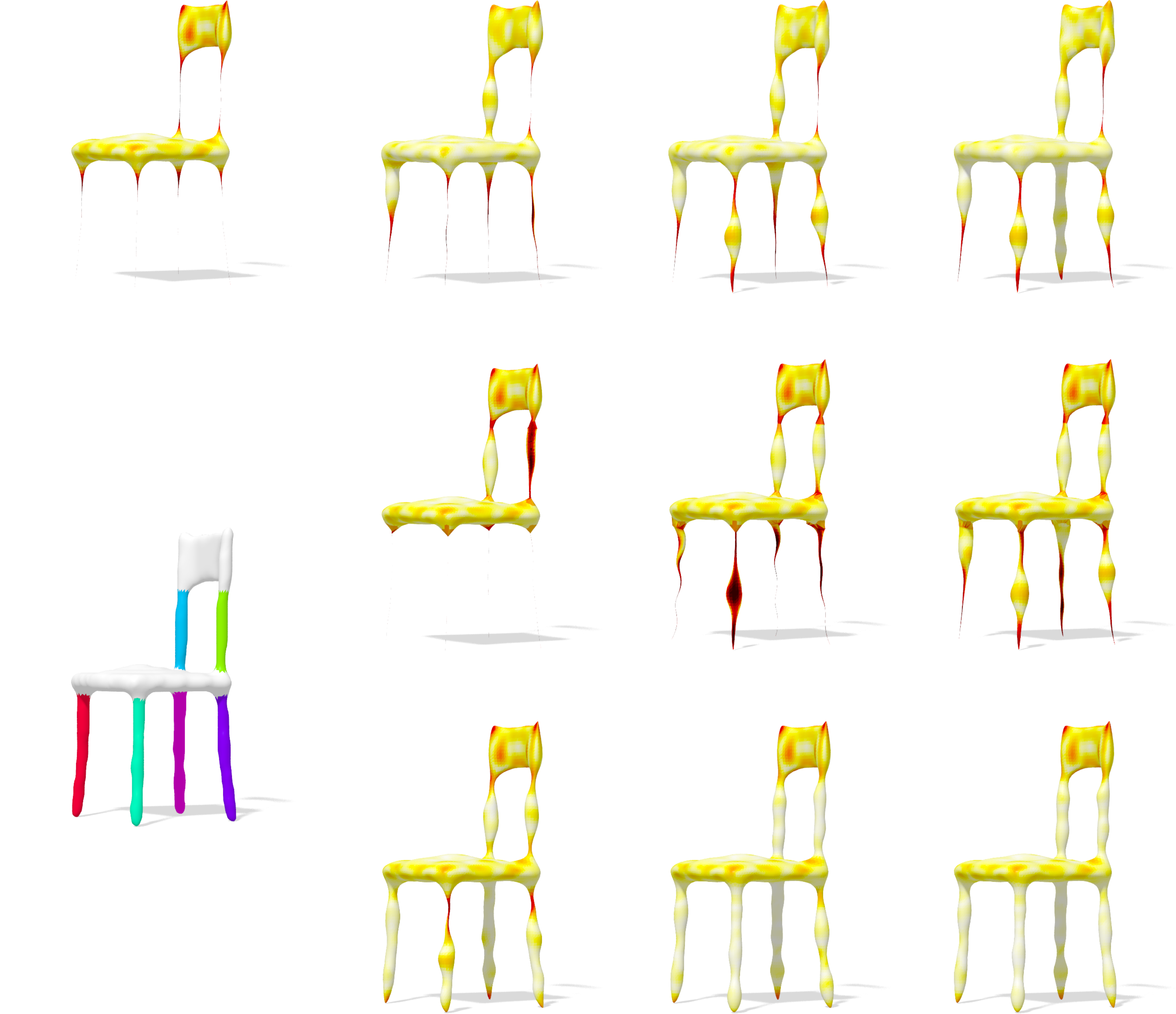}
      	
      	\put(24.5,76.5){$\ldots$}
	  	\put(24.5,45){$\ldots$}
	  	\put(24.5,14.5){$\ldots$}

		\put(5,80.5){\footnotesize  $\xi$ }	  	
	  	\put(6,11){\footnotesize regions}
	  	

      	\put(51,57){\footnotesize MH}
	  	\put(48,25.5){\footnotesize PMH}
	  	\put(48,-6){\footnotesize LMH}
      \end{overpic}      
    \end{minipage}
    & 
      \begin{minipage}{0.333\linewidth}
%
%
\definecolor{mycolor1}{rgb}{0.29412,0.54471,0.74941}%
\definecolor{mycolor2}{rgb}{0.37176,0.71765,0.36118}%
\definecolor{mycolor3}{rgb}{0.90471,0.19176,0.19882}%
\begin{tikzpicture}

\begin{axis}[%
width=0.7\linewidth,
height=0.7\linewidth,
at={(0.758in,0.481in)},
scale only axis,
xmin=1,
xmax=26,
ymode=log,
ymin=0,
ymax=300, 
yminorticks=true,
axis background/.style={fill=white},
legend style={legend cell align=left,align=left,draw=white!15!black},
xmajorgrids,
ymajorgrids,
xtick={ 1, 6, 11, 15, 21, 26},
xticklabels = {\footnotesize{$30$}, \footnotesize{$35$}, \footnotesize{$40$},\footnotesize{$45$},\footnotesize{$50$},\footnotesize{$55$},},
ytick={50, 75, 100, 300},
yticklabels = {\footnotesize{$\frac{1}{2}\xi$}, \footnotesize{$\frac{3}{4}\xi$}, \footnotesize{$\xi$}, \footnotesize{$3\xi$}},
xlabel style={yshift=0.7em},
ylabel style={yshift=-1.5em},
xlabel = {\footnotesize{basis dimension}},
ylabel = {\footnotesize{reconstruction error}},
title style={yshift= - 0.7em},
title = {\footnotesize{Hand}},
]
\addplot [color=mycolor1,solid,line width=2.0pt]
  table[row sep=crcr]{%
1	99.9999999999999\\
2	96.6055397571516\\
3	94.5360617737965\\
4	92.3671033233132\\
5	90.9578805221626\\
6	89.7474891695196\\
7	88.8508325333396\\
8	87.5416411328768\\
9	86.6785883820833\\
10	85.6366301685043\\
11	84.4723597638406\\
12	82.4372093374853\\
13	81.2682970370658\\
14	79.1516104292189\\
15	76.8501414711209\\
16	75.549344562104\\
17	73.7338645906561\\
18	71.397318025437\\
19	68.617549898159\\
20	66.6644702941983\\
21	63.3448235519272\\
22	61.1687011288899\\
23	58.4593866100394\\
24	56.6208664943856\\
25	55.2252943721264\\
26	53.5687996991838\\
};
\addlegendentry{{\footnotesize MH}};

\addplot [color=mycolor2,solid,line width=2.0pt]
  table[row sep=crcr]{%
1	100\\
2	160.681534278313\\
3	196.585649026319\\
4	236.332222799905\\
5	264.761238926612\\
6	281.799184130273\\
7	236.013001609423\\
8	203.141242872855\\
9	168.237243979366\\
10	145.451767448492\\
11	131.674231294578\\
12	127.74553552767\\
13	127.355298920128\\
14	126.12839682942\\
15	123.184233044907\\
16	120.879523114285\\
17	116.737162911903\\
18	114.2926770211\\
19	111.801195174014\\
20	109.601298094901\\
21	107.735444722528\\
22	102.843531772646\\
23	99.0262913254356\\
24	95.3256351996608\\
25	92.0698878931173\\
26	89.6753286720098\\
};
\addlegendentry{{\footnotesize PMH}};

\addplot [color=mycolor3,solid,line width=2.0pt]
  table[row sep=crcr]{%
1	99.9999999999999\\
2	96.5644838906092\\
3	93.5359974190441\\
4	89.9870732323538\\
5	88.334149208346\\
6	87.1833321729431\\
7	83.3873250313561\\
8	78.3707244712463\\
9	73.3146858395293\\
10	69.3568413172931\\
11	66.5496795102258\\
12	64.3461299592866\\
13	61.6851338244328\\
14	59.8541085424299\\
15	57.4171079382803\\
16	55.2418077032641\\
17	54.8037451978785\\
18	54.4794250200309\\
19	53.9640907368593\\
20	53.3814552810621\\
21	53.1516959990497\\
22	52.7113644244946\\
23	52.2224346935995\\
24	51.5580869948475\\
25	50.9604425640857\\
26	50.704965275161\\
};
\addlegendentry{{\footnotesize LMH}};

\end{axis}
\end{tikzpicture}%
      \vspace{0.3cm}
      
	  \begin{overpic}
      	[trim=0cm 0cm 0cm 0cm,clip,width=0.95\linewidth]{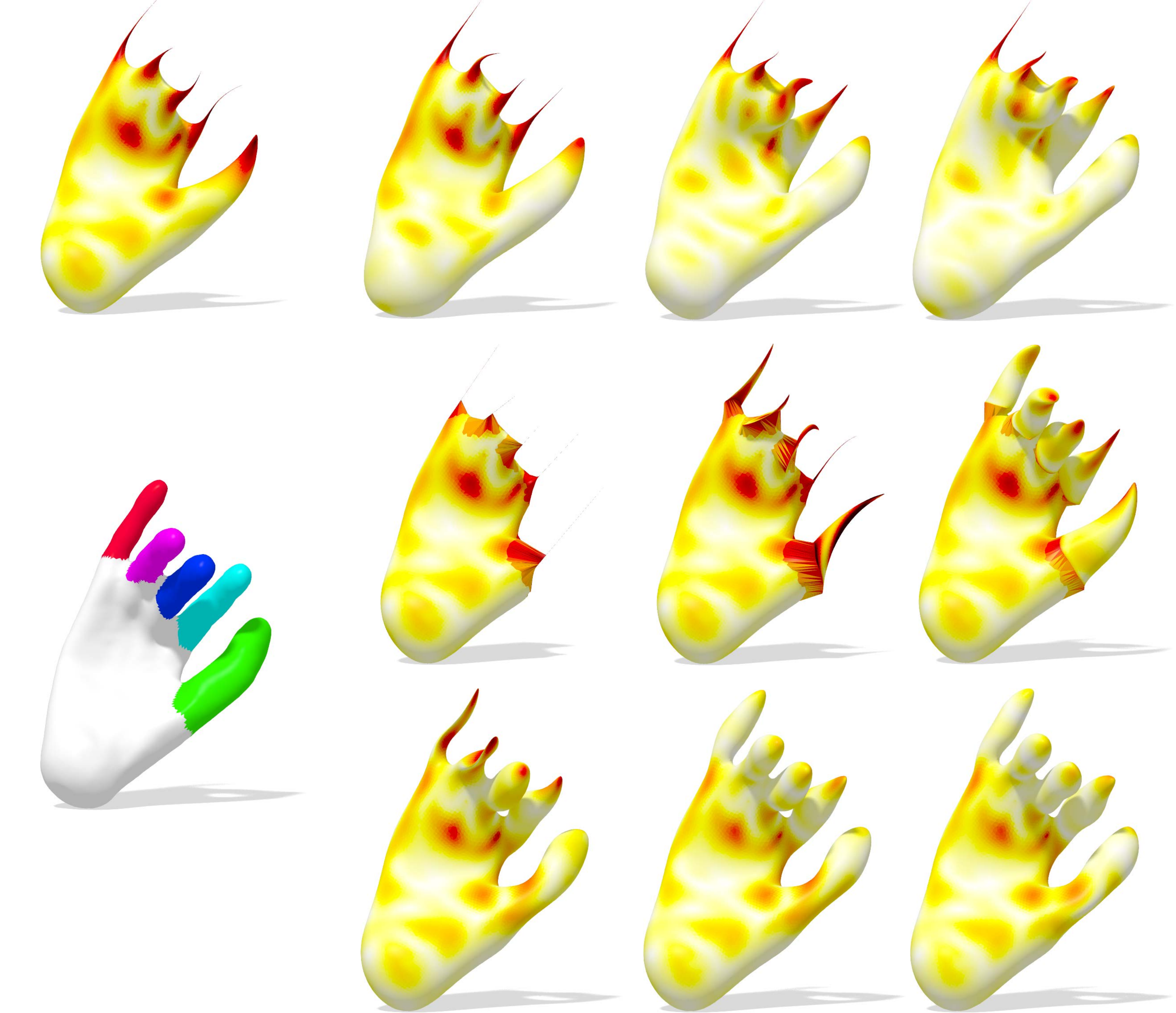}
      	
      	\put(24.5,76.5){$\ldots$}
	  	\put(24.5,45){$\ldots$}
	  	\put(24.5,14.5){$\ldots$}

		\put(3,80){\footnotesize  $\xi$ }	  	
	  	\put(5,12.2){\footnotesize regions}
	  	
	  	
      	\put(53.5,54){\footnotesize MH}
	  	\put(50.5,24.5){\footnotesize PMH}
	  	\put(50.5,-5){\footnotesize LMH}
      \end{overpic}      
    \end{minipage}
	\vspace{-0.1cm}
  \end{tabular}
\caption{\label{fig:rec_err}Comparisons among MH, PMH and LMH in surface representation. For each class we report the average reconstruction error at increasing number of basis functions ($x$-axis). Here, $\xi$ denotes the error obtained by MH with $k'=50$. The heatmaps encode reconstruction error, growing from white to dark red.}
\end{figure*}

Assume a given set of regions, identifying areas of the shape with geometric detail. By computing localized harmonics $\{\psi_j\}_j$ on the given regions, we obtain a representation of the surface geometry:
\rev{
\begin{equation}\label{eq:locxyz}
\mathbf{x}\approx\sum_{i=1}^{k'}\langle\phi_i, \mathbf{x} \rangle_{L^2(\M)} \phi_i + \sum_{j=1}^k\langle\psi_j, \mathbf{x}\rangle_{L^2(\M)} \psi_j\,,
\end{equation}
}
where $\phi_1, \hdots, \phi_{k'}$ are the standard Laplacian eigenfunctions and $\langle \phi_i,\psi_j\rangle_{L^2(\M)} \approx 0$ for all $i=1,\hdots,k'$ and $j = 1,\hdots, k$. 
For a fixed number of terms in the series, the expression \eqref{eq:locxyz} yields a more accurate approximation of the original surface than \eqref{eq:xyz}, since the localized basis functions capture the high-frequency content more quickly (as also manifested in the rapid growth of the spectrum, see Figure~\ref{fig:gap}). We refer to Figure~\ref{fig:hand_reconstruction} for a detailed illustration of this behavior.
\rev{In Figure~\ref{fig:stag_no_ortho}, we demonstrate the effect of the lack of orthogonality ($\mu_\perp=0$) on the reconstruction quality, and} in Figure~\ref{fig:rec_err} provide a quantitative evaluation \rev{on pre-segmented meshes} from the Princeton segmentation benchmark \cite{Chen:2009:ABF}.
%
%
%
%

%

In these tests, we compare with standard manifold harmonics (MH) and ``partial'' manifold harmonics (PMH). The latter approach consists in reconstructing the surface indicated by each region $R_i\subset\M$ separately by using the eigenfunctions of the Laplacian $\Delta_{R_i}$; the reconstructed part is then ``glued'' back to the full shape. We measure the point-wise reconstruction error by the Euclidean distance between each reconstructed vertex and its corresponding point in the original surface.
%
%
%


%
\begin{figure}[tb]
  \centering
  \begin{overpic}
  [trim=0cm 0cm 0cm 0cm,clip,width=1.0\linewidth]{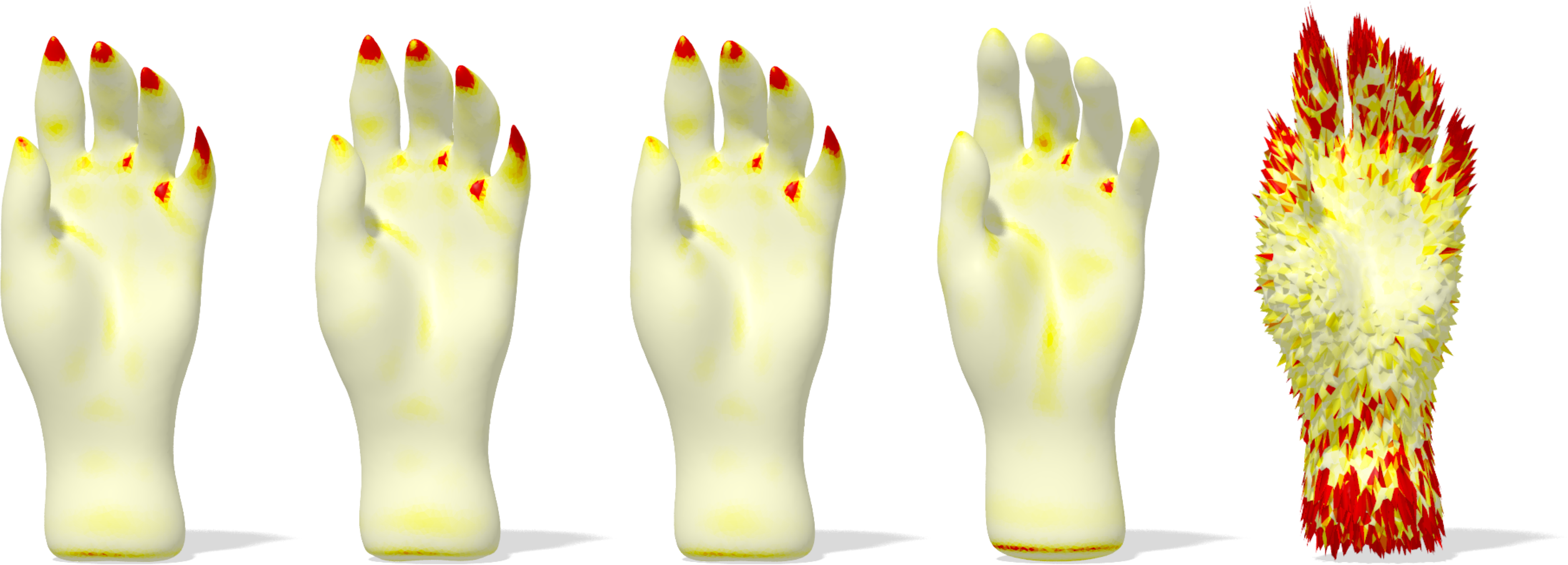}
  \put(84, -3){\footnotesize CMM}
    \put(84, -6.5){\footnotesize $0.2574$}
  \put(64, -3){\footnotesize LMH}
    \put(63.5, -6.5){\footnotesize $\mathbf{0.1123}$}
  \put(39.7, -3){\footnotesize LMH ($v_\mathrm{EO}$)}
      \put(43.2, -6.5){\footnotesize $0.1147$}
  \put(26, -3){\footnotesize EO}
      \put(23.5, -6.5){\footnotesize $0.1205$}
  \put(5, -3){\footnotesize MH}
    \put(3.4, -6.5){\footnotesize $0.1217$}
   \end{overpic}
\caption{\label{fig:all_pipelines} \rev{Comparison between different pipelines using a fixed number of basis functions (equal to 100 for all methods -- LMH uses 50 global and 50 localized harmonics in both experiments). We report the reconstruction error below each method.}}
\end{figure}

\begin{figure}[tb]
  \centering
  \begin{overpic}
  [trim=0cm 0cm 0cm 0cm,clip,width=1.0\linewidth]{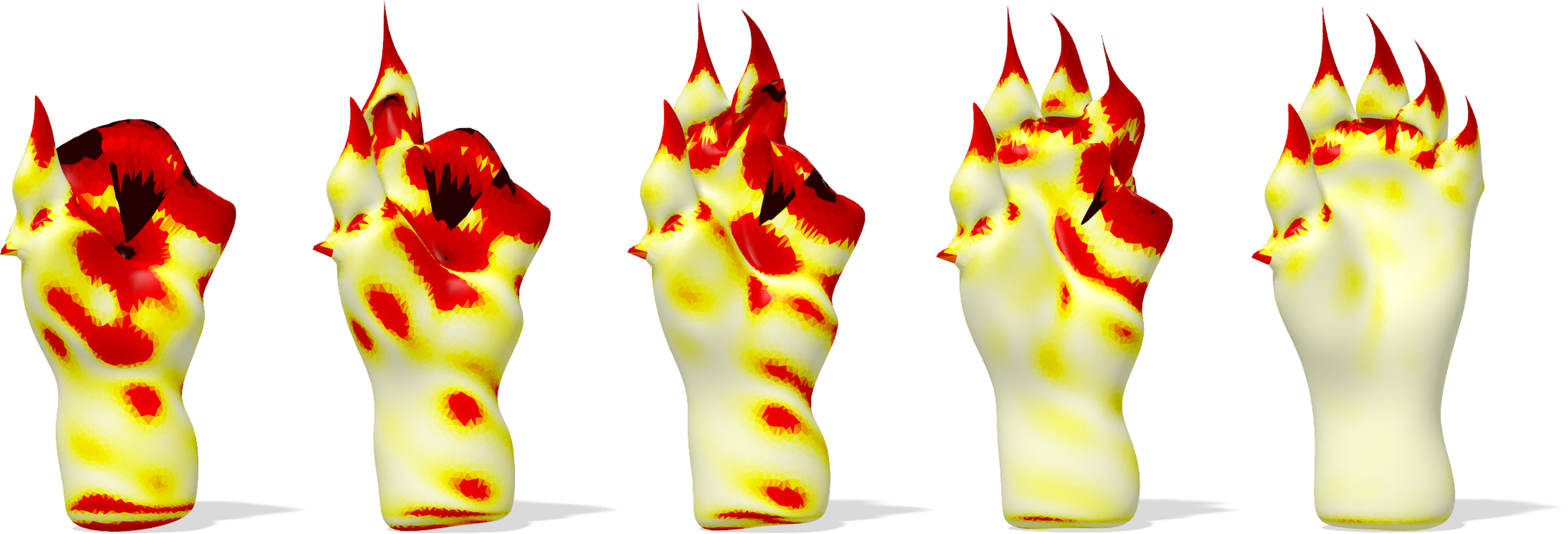}

  \put(86, -3.5){\footnotesize $100$}
  \put(67, -3.5){\footnotesize $90$}
  \put(47, -3.5){\footnotesize $80$}  
  \put(27, -3.5){\footnotesize $70$}
  \put(4, -3.5){\footnotesize $k=60$}
   \end{overpic}
   \vspace{-0.17cm}
\caption{\label{fig:hand_kimmel_pipeline}\rev{Incremental hand reconstruction using EO basis functions \cite{elliptic} and the binary regions of Figure~\ref{fig:hand_reconstruction}.}}
\end{figure}

\rev{
\noindent\textbf{Comparison to other pipelines. }
Differently from our method, both CMM \cite{Neumann} and elliptic operator (EO) \cite{elliptic} do not allow to build upon and enrich a given set of basis functions. In particular, CMM relies on the assumption that the set of localized basis functions covers the entire surface, while EO employs a soft potential with global support.
In Figure~\ref{fig:all_pipelines} we compare the reconstruction error of standard MH, EO using the potential defined in \cite{elliptic}, our method using the latter potential as a soft region $v_\mathrm{EO}$, our method using the binary regions of Figure~\ref{fig:hand_reconstruction}, and CMM using a covering set of compressed modes. 


For completeness, we also test the performance of EO when fed with a sequence of binary potentials (the finger regions in Figure~\ref{fig:hand_reconstruction}). We emphasize that EO is not designed to operate in conjunction with an existing basis, hence there is no natural way to implement an incremental update as the one shown in Figure~\ref{fig:hand_reconstruction}: the harmonics computed on each binary region would not have an underlying global structure to attach to. For this reason, we provide an extra region (the palm) where 50 EO basis functions are computed, and this basis is incrementally updated with 10 EO basis functions per finger. The result is shown in Figure~\ref{fig:hand_kimmel_pipeline}.
}

\noindent\textbf{Shape correspondence. }
Ovsjanikov et al.~\cite{ovsjanikov2012functional} proposed to represent correspondences between shapes by a linear operator (called {\em functional map}) $T:L^2(\M)\to L^2(\N)$ mapping functions on $\M$ to functions on $\N$ (one can see that point-to-point mappings constitute a special case in which delta-functions are mapped to delta-functions).
As a linear operator, the functional map $T$ admits a matrix representation $\mathbf{C} = (c_{ij})$ w.r.t. bases $\{\phi_i^\M\}$ and $\{\phi_j^\N\}$ on $L^2(\M)$ and $L^2(\N)$ respectively,
\begin{equation}\label{eq:fmap}
Tf = \sum_{ij\ge 1} \langle \phi_i^\M , f \rangle_{L^2(\M)} \underbrace{\langle T \phi_i^\M , \phi_j^\N \rangle_{L^2(\N)}}_{c_{ji}} \phi_j^\N\,,
\end{equation}
for an arbitrary $f\in L^2(\M)$. By choosing the Laplacian eigenfunctions on $\M$ and $\N$ as the bases $\{\phi_i^\M\}$ and $\{\phi_j^\N\}$, one can truncate the series \eqref{eq:fmap} to the first $k'$ terms -- hence obtaining a compact representation which can be interpreted as a band-limited approximation of the full map. Correspondence problems can then be phrased as searching for a $k'\times k'$ matrix $\mathbf{C}$ minimizing simple data fidelity criteria \cite{ovsjanikov2012functional,dorian} or exhibiting a particular structure depending on the correspondence setting \cite{pokrass13,kovnatsky15,rodola16-partial}.

Similar to the previous experiments, the standard Laplacian eigenbasis may not be the best choice in the presence of fine details: the low-pass nature of the spectral representation of the map, embodied in matrix $\mathbf{C}$, negatively affects the quality of the representation at a point-wise level. Indeed, recovering a point-to-point map from a functional map is considered a difficult problem in itself \cite{rodola-vmv15,filter}, and is at the heart of several applications dealing with maps.

\begin{figure}[t]
  \centering
  \begin{overpic}
  [trim=0cm 0cm 0cm 0cm,clip,width=1\linewidth]{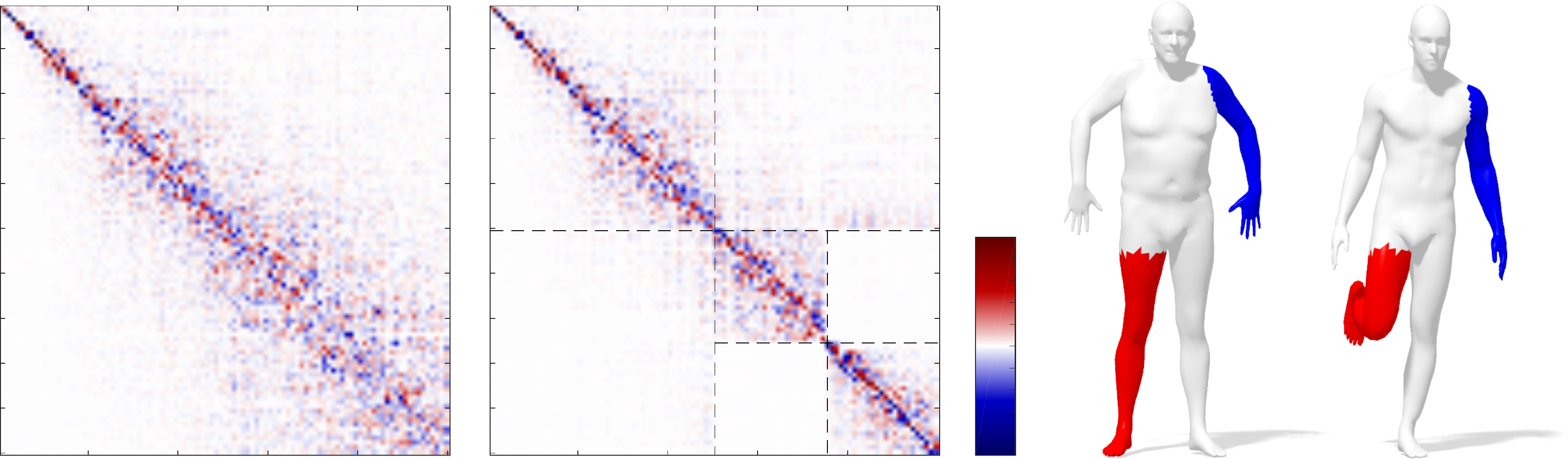}
  \put(64.5,0){\tiny $-1$}
  \put(65.5,6){\tiny $0$}
  \put(64.5,12.5){\tiny $+1$}
  \put(38,-4.5){\footnotesize $k'$}
  \put(48.5,-4.5){\footnotesize $k$}
  \put(55.5,-4.5){\footnotesize $k$}
  \put(10,-4.5){\footnotesize $k'+2k$}
  \put(0,-1){\line(1,0){28.5}}
  \put(31.5,-1){\line(1,0){13.5}}
  \put(46,-1){\line(1,0){6.5}}
  \put(53.5,-1){\line(1,0){6}}
  \end{overpic}
  \caption{\label{fig:cblock}Functional map matrices w.r.t. the standard Laplacian eigenbasis (left) and w.r.t a ``mixed'' basis composed of $k'$ Laplacian eigenfunctions and $k+k$ localized harmonics (middle). The maps encode the ground-truth correspondence between the two shapes shown on the right; the regions used for the computation of LMH are highlighted in red and blue. Note the block-diagonal structure of the second matrix, a manifestation of the capability of LMH to encode local information compactly.}
\end{figure}

LMH can be directly employed for representing functional correspondence {\em in conjunction} with the Laplacian eigenbasis:\rev{
\begin{align}\label{eq:fm2}
Tf = \sum_{\ell,m= 1}^{k+k'} \langle \omega_\ell^\M , f \rangle_{L^2(\M)} \langle T \omega_\ell^\M , \omega_m^\N \rangle_{L^2(\N)} \omega_m^\N\,,
\end{align}
%
where $\omega\in\Omega$ and $\Omega = \{\phi_i\}_{i=1}^{k'} \cup \{\psi_j\}_{j=1}^k$ is the union of the standard and localized manifold harmonics. Note that the formula above allows for `cross-talk' between the MH and LMH bases, and it} can be seen as a localized {\em refinement} to some initial correspondence represented in the (global) Laplacian eigenbasis. An example of the resulting correspondence matrix $\mathbf{C}$ is shown in Figure~\ref{fig:cblock}.

\begin{figure}[t]
  \centering
  \setlength{\tabcolsep}{0pt}
  \begin{tabular}{ccc}
  \hspace{-0.2cm}  
  
    \begin{minipage}{0.5\linewidth}

%
%
\definecolor{mycolor1}{rgb}{0.29412,0.54471,0.74941}%
\definecolor{mycolor2}{rgb}{0.90471,0.19176,0.19882}%
\begin{tikzpicture}

\begin{axis}[%
width=0.7\linewidth,
height=0.7\linewidth,
at={(0.758in,0.481in)},
scale only axis,
xmin=12,
xmax=51,
ymin=50, 
ymax=100, 
axis background/.style={fill=white},
legend style={at={(0.03,0.03)},anchor=south west,legend cell align=left,align=left,draw=white!15!black},
xmajorgrids,
ymajorgrids,
xlabel style={yshift=0.7em},
ylabel style={yshift=-1.3em},
xlabel = {\footnotesize{basis dimension}},
ylabel = {\footnotesize{mean geodesic error}},
title style={yshift= - 0.7em},
title = {\footnotesize{TOSCA dog: 20 pairs}},
xtick={12.00, 19.80, 27.60, 35.40,   43.20,   51.00},
xticklabels = {\footnotesize{$50$},\footnotesize{$60$},\footnotesize{$70$},\footnotesize{$80$},\footnotesize{$90$},\footnotesize{$100$}},
ytick={50, 60, 70, 80, 90, 100},
yticklabels = {\footnotesize{$0.5\xi$},\footnotesize{$0.6\xi$},\footnotesize{$0.7\xi$},\footnotesize{$0.8\xi$},\footnotesize{$0.9\xi$},\footnotesize{ $\xi$}},
]
\addplot [color=mycolor1,solid,line width=2.0pt]
  table[row sep=crcr]{%
12	100\\
13	99.5552098323961\\
14	98.7466455964445\\
15	99.1412448311252\\
16	98.2333787744689\\
17	96.4527705432628\\
18	95.3477633976159\\
19	94.1478499682933\\
20	94.2520097206367\\
21	94.6888132886559\\
22	94.2062400419683\\
23	93.8172580398196\\
24	92.9662099308092\\
25	92.1771488948943\\
26	91.8334866648913\\
27	91.7841444587789\\
28	91.0388048225351\\
29	90.395590034129\\
30	90.0909521279226\\
31	89.7070818680356\\
32	90.1561493463163\\
33	89.7115842681741\\
34	89.4422849068594\\
35	89.0288745975648\\
36	88.8296170981826\\
37	89.3385215228399\\
38	89.4281775250964\\
39	89.0754688871621\\
40	88.88520069211\\
41	89.4337307169195\\
42	89.4711745576223\\
43	89.3228051030573\\
44	88.838618292048\\
45	88.7254600115007\\
46	88.3950693606412\\
47	87.6340188581468\\
48	86.766357546017\\
49	85.8360387618111\\
50	85.3657237057672\\
51	85.2179762711611\\
};
\addlegendentry{\footnotesize{MH}};

\addplot [color=mycolor2,solid,line width=2.0pt]
  table[row sep=crcr]{%
12	100\\ 
13	97.2562362225664\\
14	96.829042725924\\
15	95.5072996563345\\
16	91.4689988987471\\
17	90.6040827207799\\
18	89.1361691773195\\
19	86.1510841785643\\
20	85.2208679483248\\
21	86.3406028502052\\
22	84.2438612037887\\
23	82.9589152643618\\
24	80.0638098176839\\
25	78.9359313731705\\
26	79.051995704561\\
27	76.5492813580513\\
28	74.9716256426325\\
29	74.5583555115715\\
30	74.2457117152366\\
31	73.3959961330231\\
32	73.2484185876942\\
33	73.0415289586138\\
34	71.9561332852165\\
35	70.368375774132\\
36	70.1889364166734\\
37	69.7787901552588\\
38	68.5928601436297\\
39	67.9251030705367\\
40	65.2446763858518\\
41	64.6418776914932\\
42	62.749973670271\\
43	62.4342121082388\\
44	62.2216246006317\\
45	61.7555401950828\\
46	60.5845687625212\\
47	60.1622962351747\\
48	58.8492877746237\\
49	57.4405747823762\\
50	57.1580812836671\\
51	57.0477955249922\\
};
\addlegendentry{\footnotesize{LMH}};

\end{axis}
\end{tikzpicture}%
%
    \end{minipage}
    & \hspace{-0.1cm}  
    
      \begin{minipage}{0.5\linewidth}
%
%
\definecolor{mycolor1}{rgb}{0.29412,0.54471,0.74941}%
\definecolor{mycolor2}{rgb}{0.90471,0.19176,0.19882}%
\begin{tikzpicture}

\begin{axis}[%
width=0.7\linewidth,
height=0.7\linewidth,
at={(0.758in,0.481in)},
scale only axis,
xmin=3,
xmax=51,
ymin=50, 
ymax=100,
axis background/.style={fill=white},
legend style={at={(0.03,0.03)},anchor=south west,legend cell align=left,align=left,draw=white!15!black},
xmajorgrids,
ymajorgrids,
xlabel style={yshift=0.7em},
ylabel style={yshift=-1.3em},
xlabel = {\footnotesize{basis dimension}},
ylabel = {\footnotesize{mean geodesic error}},
title style={yshift= - 0.7em},
title = {\footnotesize{TOSCA cat: 20 pairs}},
xtick={3.00,   12.60,   22.20,   31.80,   41.40,   51.00},
xticklabels = {\footnotesize{$50$},\footnotesize{$60$},\footnotesize{$70$},\footnotesize{$80$},\footnotesize{$90$},\footnotesize{$100$}},
ytick={50, 60, 70, 80, 90, 100},
yticklabels = {\footnotesize{$0.5\xi$},\footnotesize{$0.6\xi$},\footnotesize{$0.7\xi$},\footnotesize{$0.8\xi$},\footnotesize{$0.9\xi$},\footnotesize{ $\xi$}},
]
\addplot [color=mycolor1,solid,line width=2.0pt]
  table[row sep=crcr]{%
3	100\\
4	99.6455412682708\\
5	99.5053870414806\\
6	99.4470358930433\\
7	99.601359406324\\
8	99.175490725602\\
9	99.1259361731946\\
10	98.8728426109671\\
11	98.2491660843024\\
12	97.8987842023276\\
13	97.6616458034519\\
14	96.6872994192059\\
15	95.4286154427254\\
16	94.7703998063951\\
17	95.013050073341\\
18	95.1157469647341\\
19	95.0963980159004\\
20	93.6581846339818\\
21	92.4931159976287\\
22	91.8493520270072\\
23	90.8955200375971\\
24	89.8204102777369\\
25	89.067705477483\\
26	88.5924914970644\\
27	87.5883174206641\\
28	87.0369526820911\\
29	86.5101527795692\\
30	86.1052071067715\\
31	85.071994523631\\
32	84.0194631506808\\
33	82.4636762124604\\
34	81.8246071103623\\
35	81.0540712116219\\
36	80.949055258435\\
37	80.7456495188301\\
38	79.2147243158122\\
39	78.3755344130858\\
40	77.4780813680381\\
41	76.8710532590006\\
42	75.2188834517349\\
43	74.4633508411669\\
44	73.2596145499979\\
45	72.5559414399213\\
46	71.413610281751\\
47	71.1973903765429\\
48	69.492734040825\\
49	68.996003127188\\
50	68.4260518477565\\
51	67.2479907556502\\
};

\addplot [color=mycolor2,solid,line width=2.0pt]
  table[row sep=crcr]{%
3	100\\
4	97.2748836914725\\
5	94.9095322544327\\
6	93.0442204689612\\
7	90.0936300046621\\
8	87.6568441963786\\
9	86.9816097666786\\
10	83.5597651428765\\
11	81.1950004054154\\
12	78.8633364078404\\
13	75.2382079005576\\
14	73.6721171661523\\
15	72.3194903132278\\
16	70.0167391378489\\
17	69.3668023953774\\
18	68.5953749959151\\
19	67.9628577264519\\
20	67.3188327442078\\
21	66.724091707673\\
22	66.1007573030164\\
23	64.6457417184961\\
24	64.095076977759\\
25	63.4085145647112\\
26	63.0373634327652\\
27	62.745773640966\\
28	62.3712912416071\\
29	62.1779590009627\\
30	61.5470098920146\\
31	61.1709117160269\\
32	60.6895768944874\\
33	59.9214110897395\\
34	59.4904950349827\\
35	59.391766784404\\
36	59.0543580958878\\
37	58.9721373296505\\
38	58.1933065104715\\
39	57.5274786786907\\
40	57.0583361636589\\
41	56.6352384605731\\
42	56.1784175638981\\
43	55.7230044320946\\
44	55.4362280974373\\
45	54.5232402978061\\
46	54.2164988813817\\
47	53.6819338862103\\
48	53.4265041457673\\
49	53.4258031129527\\
50	53.2401449298321\\
51	52.9938251944136\\
};

\end{axis}
\end{tikzpicture}%
%
    \end{minipage}
  \end{tabular}
\caption{\label{fig:corr_err}Mean geodesic error vs. number of basis functions used in the functional map representation. Note that LMH lead to an increase in accuracy, resulting in turn in a more compact representation of the correspondence. Here $\xi$ is defined as the mean geodesic error of MH at $k'=50$.}
\end{figure}
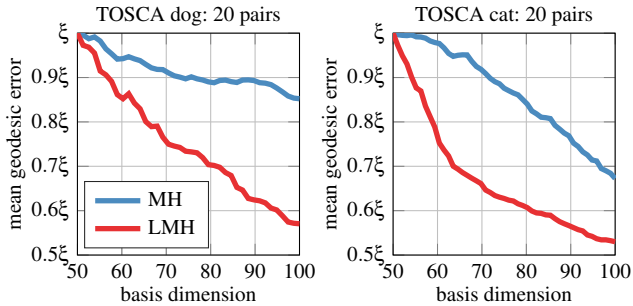

\begin{figure}[t]
  \centering
  \setlength{\tabcolsep}{0pt}
  \begin{tabular}{lr}
  \hspace{-0.3cm}
  
    \begin{minipage}{0.49\columnwidth}
%
%
\definecolor{mycolor1}{rgb}{0.29412,0.54471,0.74941}%
\definecolor{mycolor2}{rgb}{0.37176,0.71765,0.36118}%
\definecolor{mycolor3}{rgb}{0.90471,0.19176,0.19882}%
\begin{tikzpicture}

\begin{axis}[%
width=0.7\linewidth,
height=0.7\linewidth,
at={(0.758in,0.481in)},
scale only axis,
xmin=1,
xmax=100,
ymode=log,
ymin=0,
ymax=0.4,
yminorticks=true,
xmajorgrids,
ymajorgrids,
xtick={1, 25, 50, 75, 100},
every x tick label/.append style={font=\color{black}, font=\footnotesize},
every y tick label/.append style={font=\color{black}, font=\footnotesize},
xticklabels = {1,25,50,75,100},
ytick={0.01, 0.02, 0.05, 0.1, 0.2, 0.4},
yticklabels = {0.01, 0.02, 0.05, 0.1, 0.2, 0.4},
xlabel style={yshift=0.7em},
ylabel style={yshift=-1.3em},
xlabel = {\footnotesize{basis dimension}},
ylabel = {\footnotesize{mean geodesic error}},
axis background/.style={fill=white},
legend style={legend cell align=left,align=left,draw=white!15!black}
]
\addplot [color=mycolor1,solid,line width=3.0pt]
  table[row sep=crcr]{%
1	0.397124275484773\\
2	0.208158969910865\\
3	0.167229893978309\\
4	0.0723768964403664\\
5	0.079105714535983\\
6	0.0523610540126959\\
7	0.0374701466001024\\
8	0.0338504302641759\\
9	0.0409133855171308\\
10	0.0474054177253668\\
11	0.0425518175371106\\
12	0.0320893077834093\\
13	0.0334966289032448\\
14	0.0314799165523864\\
15	0.0286562139543462\\
16	0.0288963817678428\\
17	0.0275295241784549\\
18	0.0238426428458218\\
19	0.0240322338467212\\
20	0.0234733537504353\\
21	0.0221557430143773\\
22	0.0220590460459702\\
23	0.0212580287560152\\
24	0.0208480747029964\\
25	0.0208446957104612\\
26	0.0212369299347563\\
27	0.0208645857217341\\
28	0.0209840400419285\\
29	0.0200397321817476\\
30	0.0192002626737508\\
31	0.0198899099266094\\
32	0.0189913716474799\\
33	0.0220460761721612\\
34	0.022047947482808\\
35	0.0225549372044397\\
36	0.0224573898746903\\
37	0.0219625046729646\\
38	0.0213019872802764\\
39	0.0209365899139315\\
40	0.0206748325411201\\
41	0.0200709080183727\\
42	0.0192741048041036\\
43	0.0193369932122691\\
44	0.0189865484983748\\
45	0.0186353416013497\\
46	0.0184697654472967\\
47	0.0185600009822723\\
48	0.0184434621196081\\
49	0.0177478978433276\\
50	0.0177686945503482\\
51	0.0173579454785764\\
52	0.0176026418549163\\
53	0.017505034854306\\
54	0.0179392016026257\\
55	0.0175239738247529\\
56	0.0172971626622688\\
57	0.0172358717710081\\
58	0.0171455360902306\\
59	0.0167438352440046\\
60	0.01656522186785\\
61	0.0167034007994132\\
62	0.0170706089697268\\
63	0.0178132205320812\\
64	0.0171265981710923\\
65	0.016955760002876\\
66	0.0168269753693261\\
67	0.0167927840529975\\
68	0.0166032550529045\\
69	0.0170076895408444\\
70	0.0165454994650602\\
71	0.016530769941355\\
72	0.0159823428342091\\
73	0.0156904128995287\\
74	0.0154606646968789\\
75	0.0150758900130081\\
76	0.0149056220307914\\
77	0.0150109788435749\\
78	0.0153272213445306\\
79	0.0152154922860234\\
80	0.0148829914368114\\
81	0.0147064917447387\\
82	0.0130831317422019\\
83	0.0136069018852612\\
84	0.0129806769870869\\
85	0.0134063937883966\\
86	0.0137834481862383\\
87	0.0134965242498521\\
88	0.0132737323291691\\
89	0.0134418230456503\\
90	0.0136587000064951\\
91	0.0135813087612407\\
92	0.0135547186505856\\
93	0.0135266830207641\\
94	0.0132945707066043\\
95	0.0133296986580816\\
96	0.0130111822025667\\
97	0.0127145530141978\\
98	0.01259638328977\\
99	0.0123971687923411\\
100	0.0123748078502184\\
};
\addlegendentry{{\footnotesize MH}};

\addplot [color=mycolor3,solid,line width=2.0pt]
  table[row sep=crcr]{%
1	0.397124275484773\\
2	0.208158969910865\\
3	0.167229893978309\\
4	0.0723768964403664\\
5	0.079105714535983\\
6	0.0523610540126959\\
7	0.0374701466001024\\
8	0.0338504302641759\\
9	0.0409133855171308\\
10	0.0474054177253668\\
11	0.0425518175371106\\
12	0.0320893077834093\\
13	0.0334966289032448\\
14	0.0314799165523864\\
15	0.0286562139543462\\
16	0.0288963817678428\\
17	0.0275295241784549\\
18	0.0238426428458218\\
19	0.0240322338467212\\
20	0.0234733537504353\\
21	0.0221557430143773\\
22	0.0220590460459702\\
23	0.0212580287560152\\
24	0.0208480747029964\\
25	0.0208446957104612\\
26	0.0212369299347563\\
27	0.0208645857217341\\
28	0.0209840400419285\\
29	0.0200397321817476\\
30	0.0192002626737508\\
31	0.0198899099266094\\
32	0.0189913716474799\\
33	0.0220460761721612\\
34	0.022047947482808\\
35	0.0225549372044397\\
36	0.0224573898746903\\
37	0.0219625046729646\\
38	0.0213019872802764\\
39	0.0209365899139315\\
40	0.0206748325411201\\
41	0.0200709080183727\\
42	0.0192741048041036\\
43	0.0193369932122691\\
44	0.0189865484983748\\
45	0.0186353416013497\\
46	0.0184697654472967\\
47	0.0185600009822723\\
48	0.0184434621196081\\
49	0.0177478978433276\\
50	0.0177686945503482\\
51	0.0166558753829915\\
52	0.0144002166878451\\
53	0.0140888387639063\\
54	0.0141024425859073\\
55	0.0135830752992899\\
56	0.0130548692648624\\
57	0.013172202564984\\
58	0.0126051076279892\\
59	0.0123543108324087\\
60	0.0118185583824219\\
61	0.0112582306878041\\
62	0.0113495368589664\\
63	0.0117188164155011\\
64	0.0117885503385576\\
65	0.0113521419739417\\
66	0.0113472942339222\\
67	0.0111713459070823\\
68	0.0109993844257772\\
69	0.0108235812947049\\
70	0.0110068363318833\\
71	0.0109776218626762\\
72	0.0107954516501978\\
73	0.0107689266952558\\
74	0.0109998002097679\\
75	0.010786956512894\\
76	0.0106843491178399\\
77	0.0106297313333432\\
78	0.0104364884521684\\
79	0.0101372147275409\\
80	0.00996364122325792\\
81	0.00990499928067982\\
82	0.0096162259608035\\
83	0.00981164988720857\\
84	0.0100141404033888\\
85	0.00997995173093972\\
86	0.00988664952079044\\
87	0.0100156583405362\\
88	0.00990819874241302\\
89	0.00986768914232279\\
90	0.00974393263991792\\
91	0.00969282244754942\\
92	0.00961226675293286\\
93	0.00954520653645839\\
94	0.0095953370628052\\
95	0.00954587495638217\\
96	0.00955616506329074\\
97	0.0094433724696092\\
98	0.0094381879534057\\
99	0.00946450980653936\\
100	0.00944480729781041\\
};
\addlegendentry{{\footnotesize LMH}};

\end{axis}
\end{tikzpicture}%
      \vspace{-1.cm}
    \end{minipage}
        
 & \hspace{-0.1cm}
\vspace{0.7cm}
    
    \begin{minipage}{0.50\columnwidth}
      \begin{overpic}
      [trim=0cm 0cm 0cm 0cm,clip,width=1.07\linewidth]{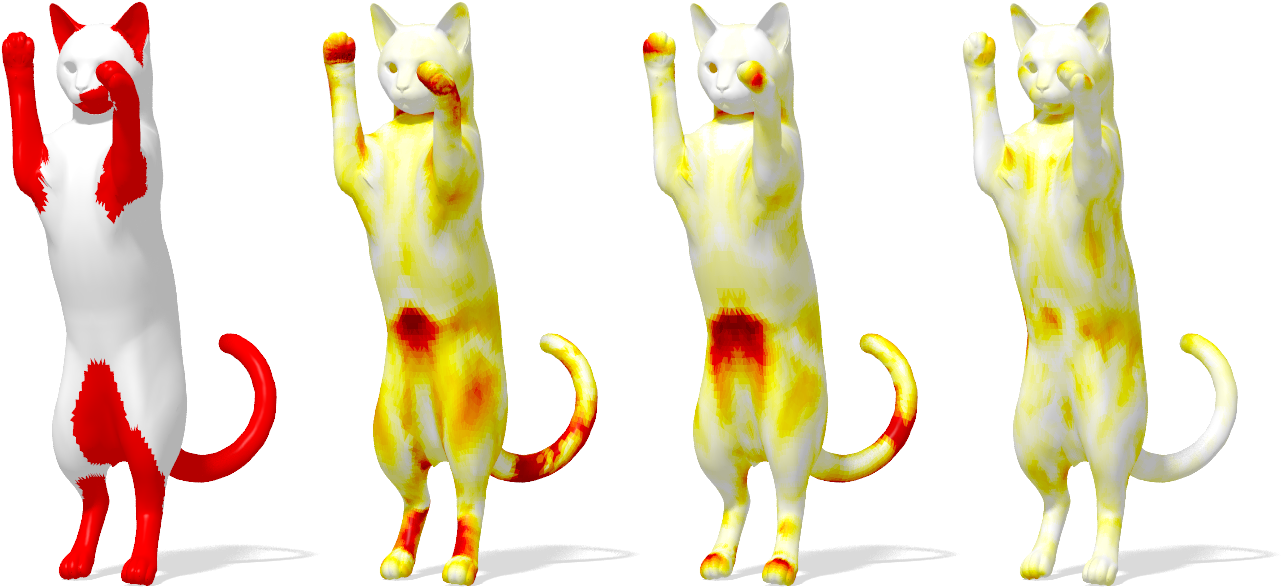}
      	
	  	\put(2.5,-7){\footnotesize region}
	  	\put(29,-7){\footnotesize MH}
		\put(26,-13){\footnotesize $k=50$}
		\put(54,-7){\footnotesize MH}
		\put(50,-13){\footnotesize $k=100$}
	  	\put(78,-7){\footnotesize LMH}
		\put(75,-13){\footnotesize $k=100$}

      \end{overpic}
    \end{minipage}
  \end{tabular}
\caption{\label{fig:cat_details}Accuracy improvement in the functional map representation, obtained by introducing LMH after $k'=50$ Laplacian eigenfunctions. On the right we show the regions used for LMH and the geodesic error (encoded as hot colors, growing from white to dark red) obtained for different configurations. Note the higher accuracy attained by LMH for the same amount of basis functions.}
\end{figure}

In Figure~\ref{fig:corr_err} we show a quantitative comparison between the two representations \eqref{eq:fmap} and \eqref{eq:fm2}. For this experiment we use near-isometric shapes from the TOSCA dataset \cite{TOSCA}. For each pair of shapes, we use their ground-truth point-to-point correspondence to construct functional maps of increasing size in the Laplacian eigenbasis and in the LMH basis. For the latter, we use Eq.~\eqref{eq:fm2} with $k'=50$ and $k$ increasing from 1 to 50. The harmonics $\{\psi_\ell^\M,\psi_m^\N\}_{\ell,m=1}^k$ are localized to the regions having large reconstruction error, computed as in the previous experiments. Note that even though these regions can be arbitrarily disconnected and irregular (see, e.g., Figure~\ref{fig:cat_details}), our framework can be applied without modifications.
A point-to-point map is recovered from each functional map using the nearest neighbor approach \cite{ovsjanikov2012functional}. We measure the correspondence quality via its {\em geodesic error}. Assume a point-to-point match $(x,y)\in\M\times\N$ is recovered, whereas the ground-truth correspondence is $(x,y^*)$; we compute the quantity
\begin{equation}
\epsilon(x) = \frac{d_\N (y,y^*)}{\sqrt{\mathrm{Area}(\N)}}\,,
\end{equation}
where $d_\N$ is the geodesic distance on $\N$.


For the second set of experiments we consider a challenging setting of shape correspondence known as {\em deformable object-in-clutter} \cite{cosmoclutter}. In this scenario, the task is to match a given model to a scene where the model appears in a different pose, and in the simultaneous presence of clutter (extra objects) and missing parts. The problem was recently tackled in \cite{cosmoclutter} using the functional map representation; to our knowledge, this method represents the current state of the art for this class of problems.

\begin{figure}[b]
  \centering
  \begin{overpic}
  [trim=0cm 0cm 0cm 0cm,clip,width=1\linewidth]{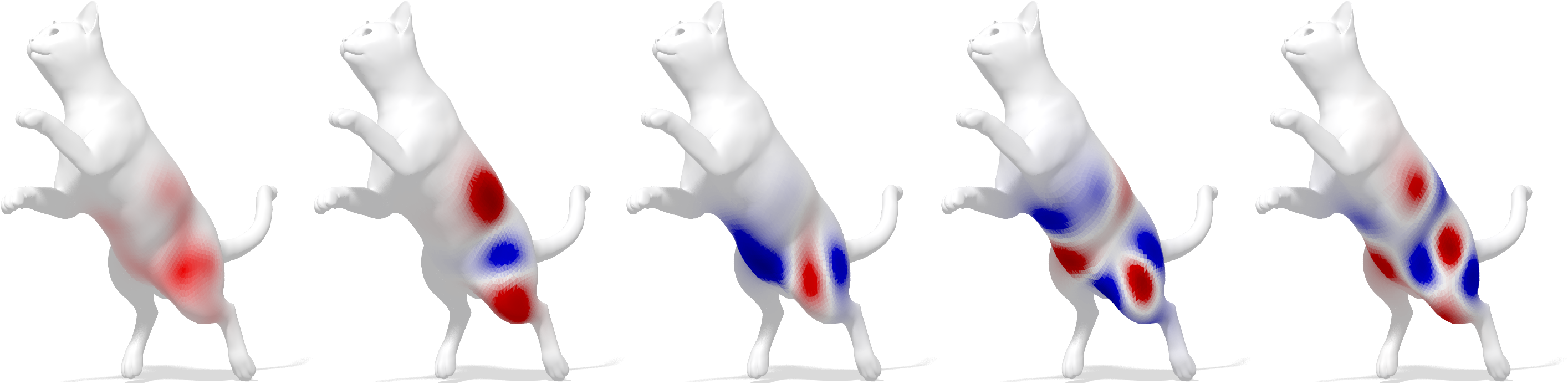}
  \put(1,-4){\footnotesize $u:\M\to[0,1]$}
  \put(30,-4){\footnotesize $\psi_4$}
  \put(50,-4){\footnotesize $\psi_5$}
  \put(70,-4){\footnotesize $\psi_9$}
  \put(90,-4){\footnotesize $\psi_{10}$}
  \end{overpic}
  \caption{\label{fig:soft}Localized manifold harmonics on a {\em soft} region encoded by function $u$. No thresholding is required in order to obtain a valid set of functions minimizing the energy \eqref{eq:ener}.}
\end{figure}

As data for these tests we use the entire dataset adopted for the comparisons in \cite{cosmoclutter}. The dataset consists of 3 TOSCA models ({\em cat, centaur, dog}) and 150 synthetic scenes in which the models appear. Sparse point-to-point matches (around 10) between models and scenes, obtained using the approach of \cite{rodola12}, are also provided. Given $m$ input matches, we construct a mixture of $m$ Gaussians with equal variance (set to 1\% of the shape diameter) to define a {\em soft} region $u$ on both model and scene (see Figure~\ref{fig:soft}). We then construct a functional map $\mathbf{C}$ upon the input sparse correspondence, and represent it w.r.t. $k=15$ localized manifold harmonics computed on the soft regions (note that here we do {\em not} use any global eigenfunction, i.e., we set $k'=0$ in \eqref{eq:fm2}). Finally, we recover from $\mathbf{C}$ a dense point-to-point map localized on $u$ using the intrinsic ICP approach of \cite{ovsjanikov2012functional}.

\begin{figure}[tb]
  \centering
  \input{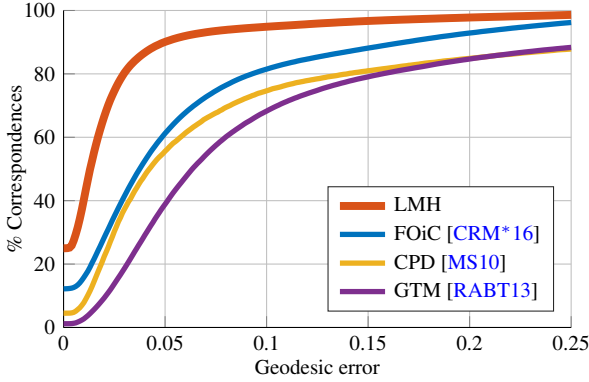}
\caption{\label{fig:3dv}Comparisons with the state of the art in deformable object-in-clutter. All methods use the same input data.}
\end{figure}

\begin{figure}[t]
  \centering
  \setlength{\tabcolsep}{0pt}
  \begin{tabular}{cc}
  \imagetop{
  \input{./same_k.tikz}
  }
  &
  \hspace{0.2cm}
  \imagetop{
  \begin{tabular}{c}
  \\
  \begin{overpic}
 [trim=0cm 0cm 0cm 0cm,width=0.2\linewidth]{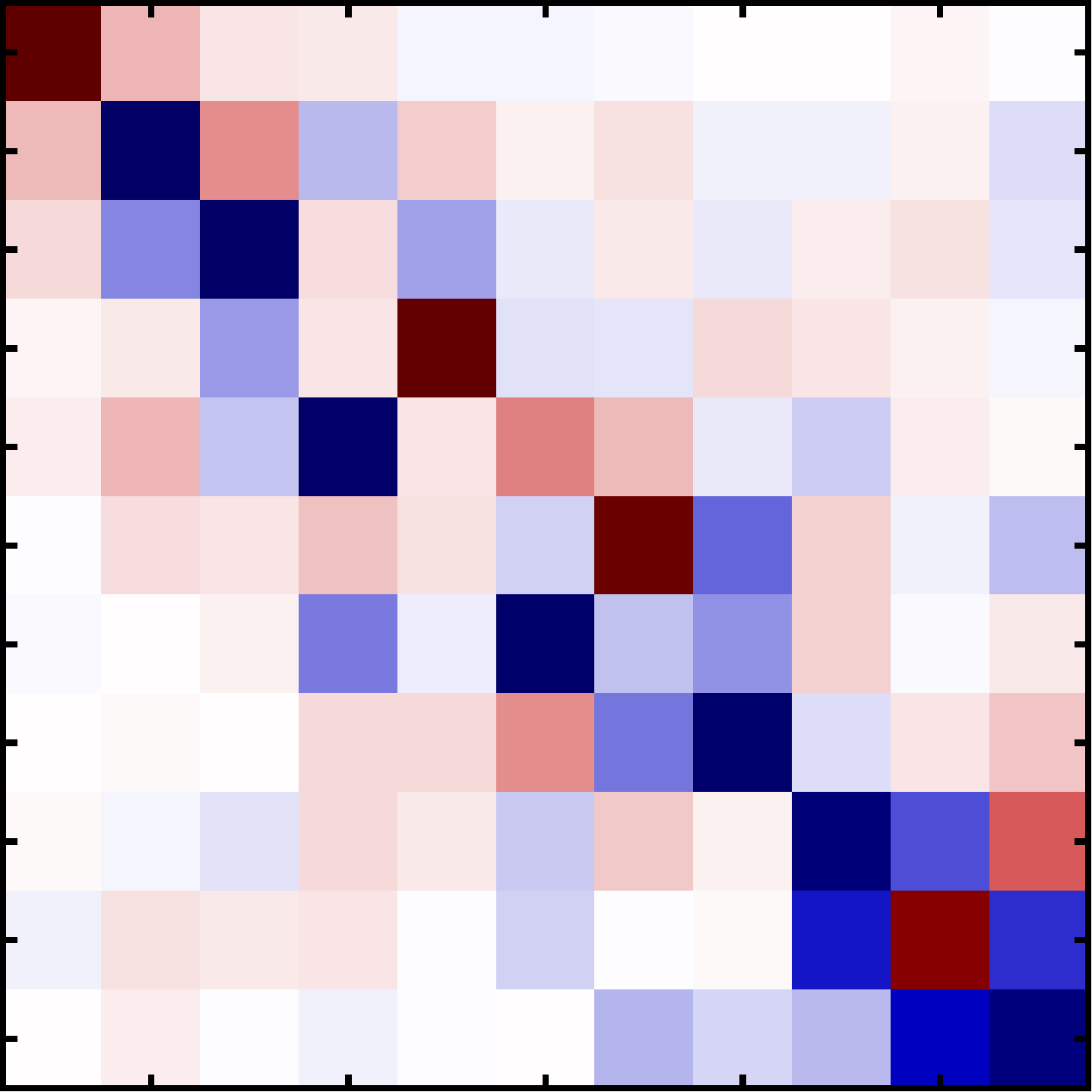}
  \put(15,106){\footnotesize $\langle \psi_i,T\psi_j \rangle_{L^2(\M)}$}
  \put(-10,90){\footnotesize 1}
  \put(-20,0){\footnotesize 10}
  \end{overpic}\\\\
  \begin{overpic}
 [trim=0cm 0cm 0cm 0cm,width=0.2\linewidth]{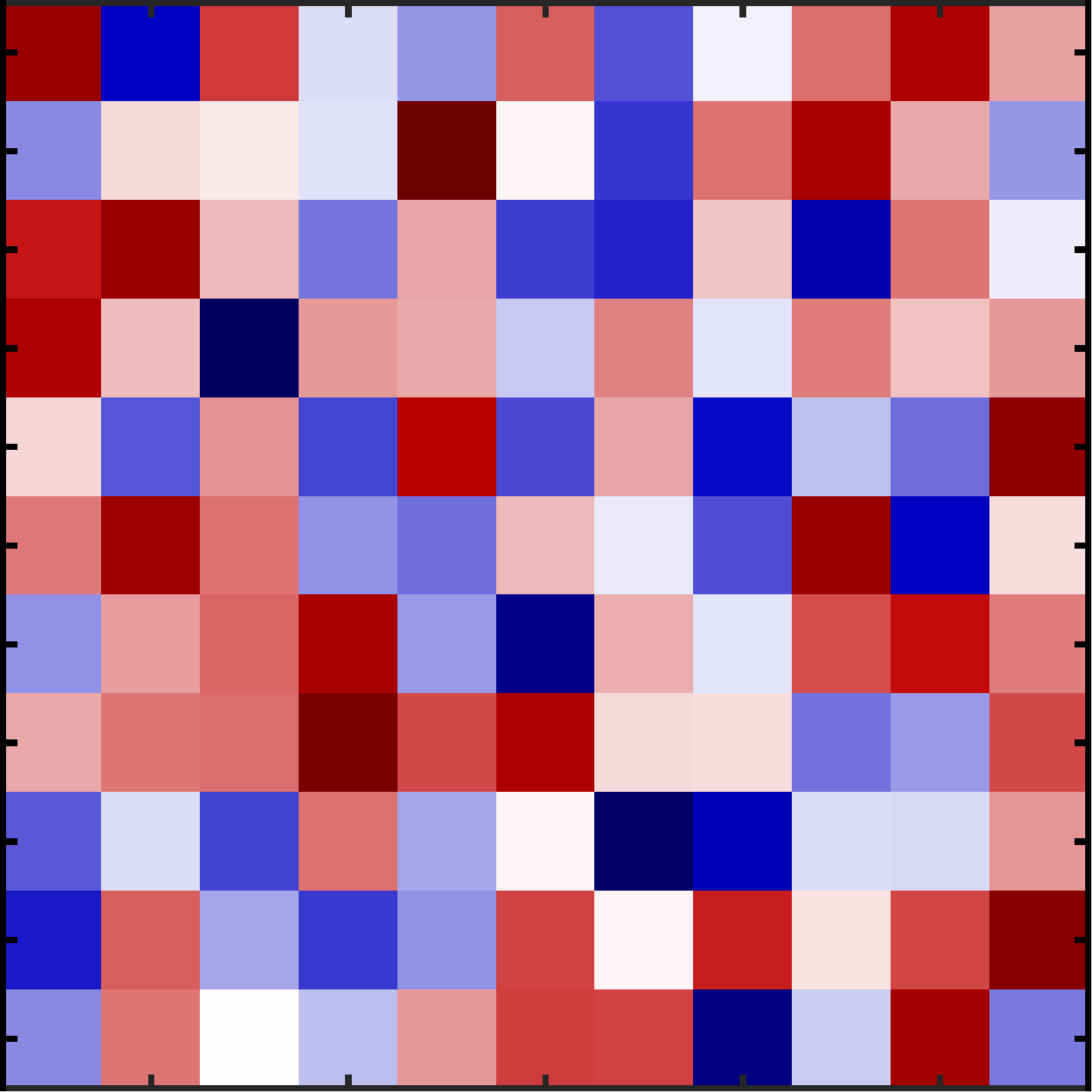}
 \put(15,106){\footnotesize $\langle \phi_i,T\phi_j \rangle_{L^2(\M)}$}
 \put(-10,90){\footnotesize 1}
 \put(-20,0){\footnotesize 10}
 \put(0,-15){\footnotesize 1}
 \put(85,-15){\footnotesize 10}
  \end{overpic}
  \end{tabular}
  }
  \end{tabular}
\caption{\label{fig:maxk}{\em Left}: Correspondence accuracy in the object-in-clutter setting. Each curve corresponds to a functional map expressed in a different basis, using the same input data. We show the performance when using $k=10$ (solid curves) and $k=50$ (dashed curves) basis functions. In this example, MH could not reach the quality of LMH for any choice of $k$. {\em Right}: Functional map matrices in the LMH basis $\{\psi\}$ (top) and in the MH basis $\{\phi\}$ (bottom).}
\end{figure}

\begin{figure*}[t]
  \centering
  \begin{overpic}
  [trim=0cm 0cm 0cm 0cm,clip,width=0.21\linewidth]{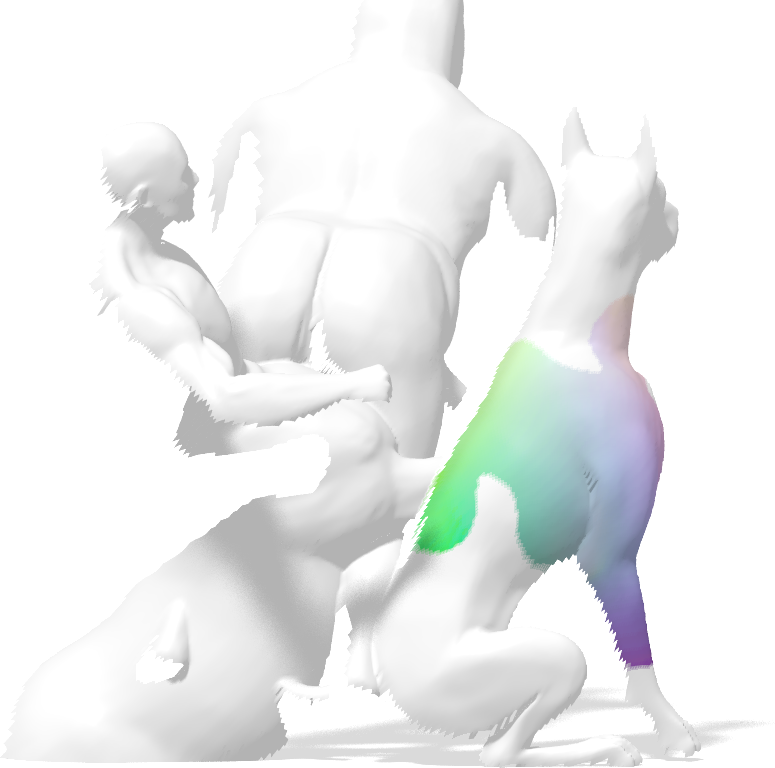}
  \put(40,-10){\footnotesize LMH}
  \end{overpic}
  \hspace{0.3cm}
  \begin{overpic}
  [trim=0cm 0cm 0cm 0cm,clip,width=0.21\linewidth]{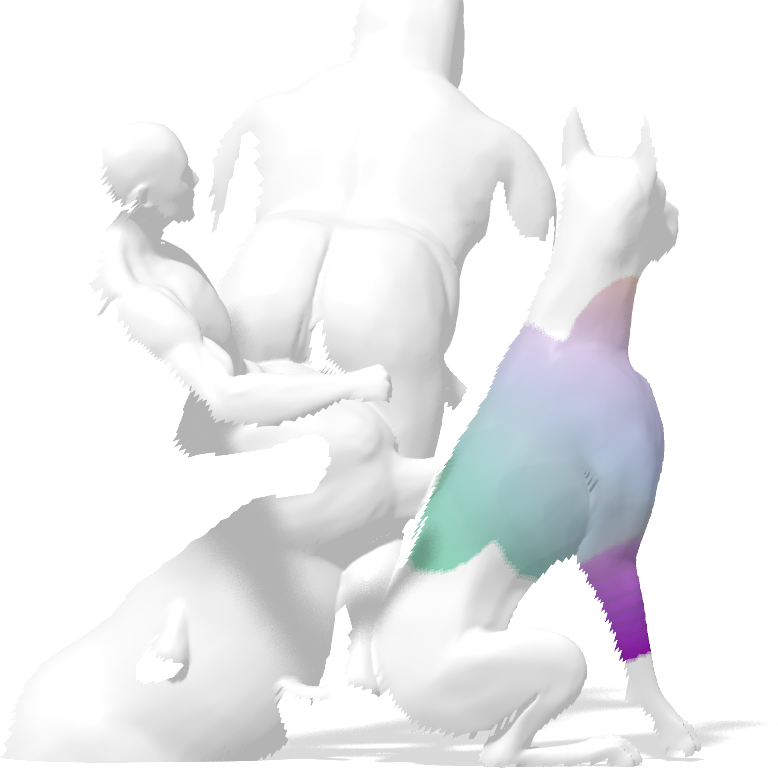}
  \put(40,-10){\footnotesize FOiC}
  \end{overpic}
  \begin{overpic}
  [trim=0cm 0cm 0cm 0cm,clip,width=0.26\linewidth]{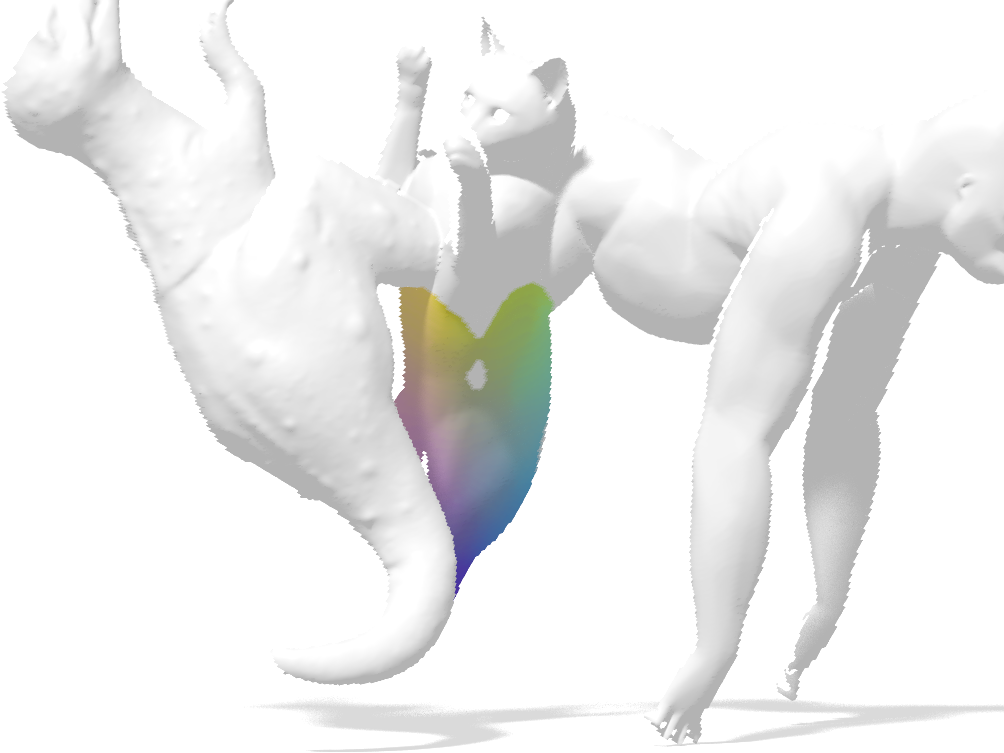}
  \end{overpic}
  \begin{overpic}
  [trim=0cm 0cm 0cm 0cm,clip,width=0.26\linewidth]{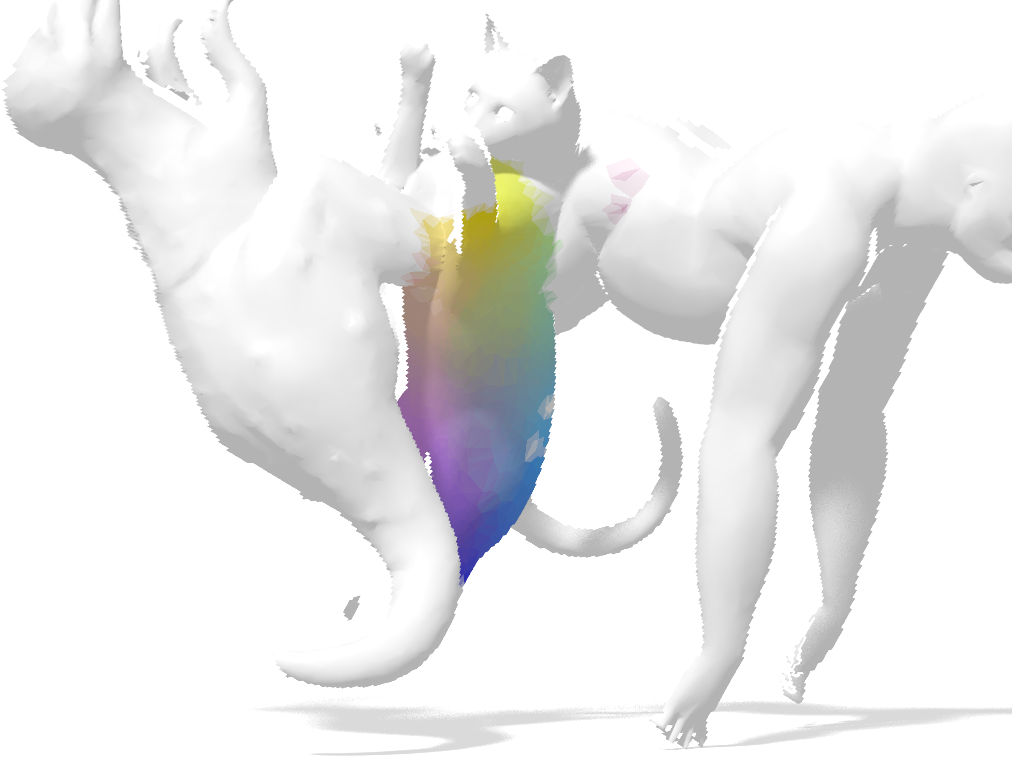}
  \end{overpic}\\
  \begin{overpic}
  [trim=0cm 0cm 0cm 0cm,clip,width=0.12\linewidth]{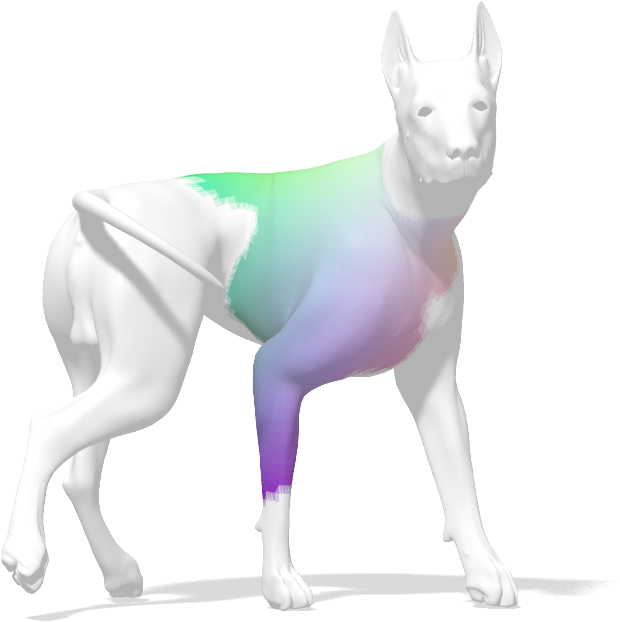}
  \end{overpic}
  \begin{overpic}
  [trim=0cm 0cm 0cm 0cm,clip,width=0.12\linewidth]{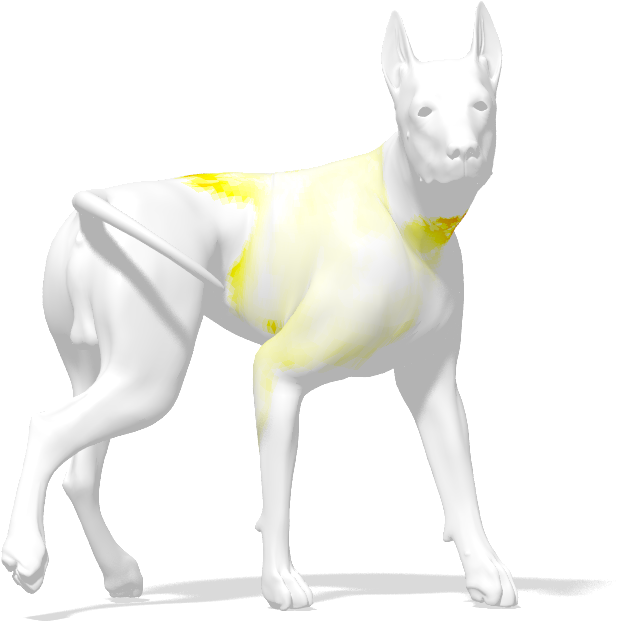}
  \end{overpic}
  \begin{overpic}
  [trim=0cm 0cm 0cm 0cm,clip,width=0.12\linewidth]{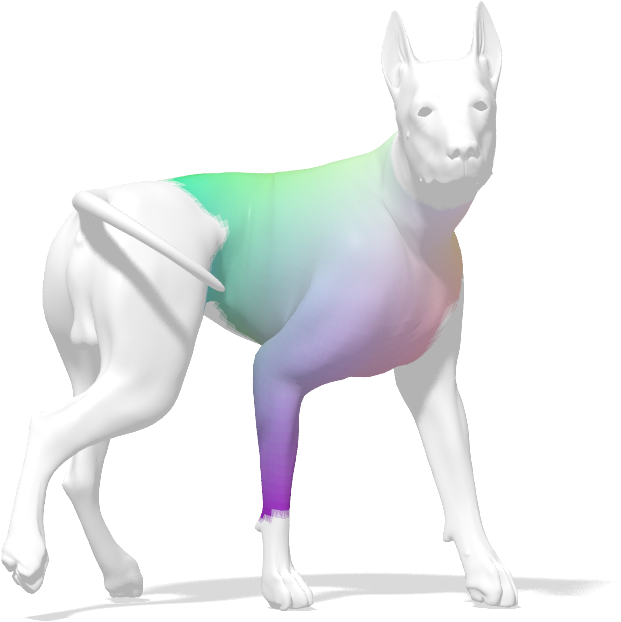}
  \end{overpic}
  \begin{overpic}
  [trim=0cm 0cm 0cm 0cm,clip,width=0.12\linewidth]{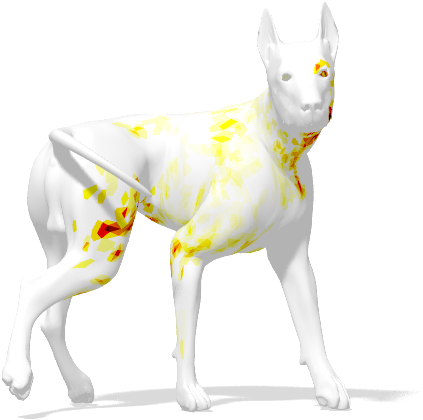}
  \end{overpic}
  \begin{overpic}
  [trim=0cm 0cm 0cm 0cm,clip,width=0.12\linewidth]{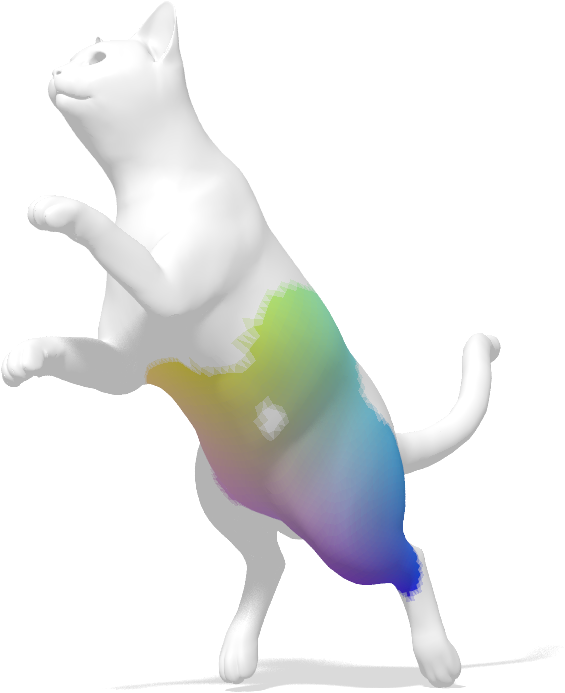}
  \put(60,90){\footnotesize LMH}
  \end{overpic}
  \begin{overpic}
  [trim=0cm 0cm 0cm 0cm,clip,width=0.12\linewidth]{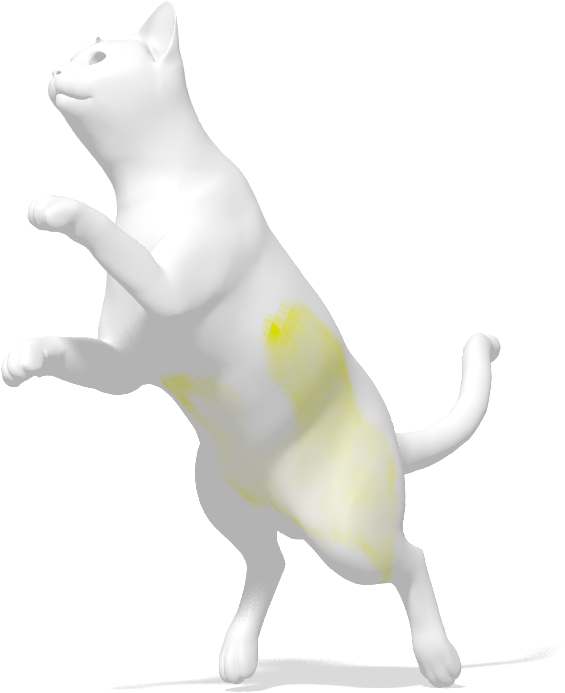}
  \end{overpic}
  \hspace{0.13cm}
  \begin{overpic}
  [trim=0cm 0cm 0cm 0cm,clip,width=0.12\linewidth]{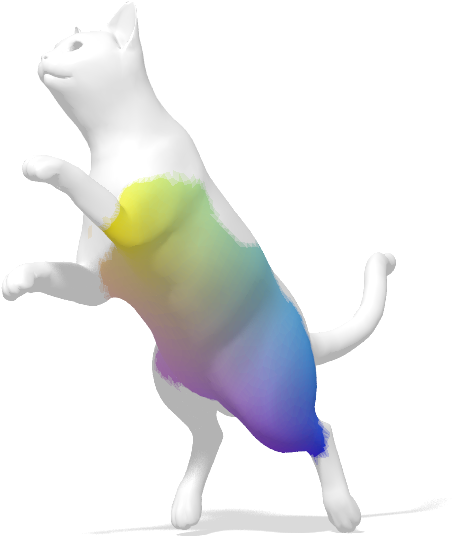}
  \put(60,95){\footnotesize FOiC}
  \end{overpic}
  \begin{overpic}
  [trim=0cm 0cm 0cm 0cm,clip,width=0.12\linewidth]{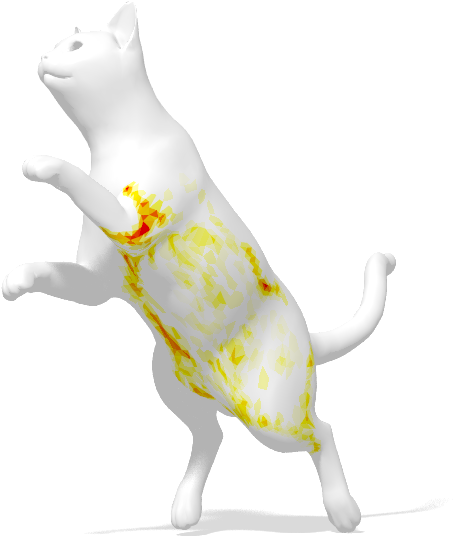}
  \end{overpic}
  \caption{\label{fig:clutter}Qualitative comparisons between our LMH-based approach for deformable shape correspondence in clutter and the state of the art \cite{cosmoclutter}. For each experiment we show the dense correspondence (corresponding points have same color) and the geodesic error (hot colors growing from white to dark red).}
\end{figure*}

The results of this experiment are reported in Figure~\ref{fig:3dv} quantitatively and in Figure~\ref{fig:clutter} qualitatively. Despite the simple approach, our method gains a significant improvement in accuracy, of up to 25\% upon the state of the art on this benchmark, highlighting the inherent robustness of LMH to missing parts and topological artifacts.
Finally, in Figure~\ref{fig:maxk} we compare (on a single pair of shapes) our pipeline with the counterparts obtained by replacing LMH with MH and PMH.

\section{Discussion and conclusions}
\label{sec:concl}
%

We introduced a new framework for spectral shape analysis and processing, allowing to perform operations which are localized to a given (possibly soft or disconnected) region of interest on the surface. Our framework is flexible, in that it can naturally enrich or fully replace the standard manifold harmonics in several tasks in graphics. We demonstrated its applicability in applications of geometry processing and shape correspondence, demonstrating a significant advantage if compared with the standard `global' constructions based on the eigendecomposition of the Laplace operator.

\paragraph*{Limitations. }
Perhaps the biggest limitation of our approach lies in the availability of regions (or soft counterparts thereof) upon which to carry out the localized spectral analysis. Such information may not be available in certain unsupervised applications, where it is often difficult to define a meaningful segmentation -- indeed, an inherently task-specific notion suggesting the use of data-driven approaches.
%
%
Further, similarly to the classical setting, it is not obvious how to choose the number of harmonics to employ for a given task. 

\paragraph*{Future directions. }
Despite the theoretical and empirical results provided within this paper, we feel that our study is still just `scratching the surface' of a much broader area of research, with potentially extensive applications in geometry processing and graphics. Even though quite elusive at this stage, we foresee connections with results in {\em localization theory} (see, e.g., \cite{mayboroda}), leading to a promising direction of research which we believe deserves a deeper exploration.

%

%

{\small
\vspace{1ex}\noindent\textbf{Acknowledgments. } The authors wish to thank Dorian Nogneng, Zorah L\"ahner, Haggai Maron and Nadav Dym for the technical support, and Klaus Glashoff, Or Litany, Christopher Brandt and Maks Ovsjanikov for useful discussions. Research was partially completed while ER was visiting the Institute for Mathematical Sciences, National University of Singapore in 2017. ER and MB are supported by ERC StG grant no. 307048 (COMET).
}

\paragraph*{Appendix A. }
We show how to compute a global solution to problem \eqref{eq:problem}. We start by observing that the hard constraints \eqref{eq:orthophi} require the desired basis functions $\bm{\Psi}$ to lie in the null space of the linear map $\mathbf{P}_{k'} := \bm{\Phi}\bm{\Phi}^\top\A$ (i.e., the projector onto $\mathrm{Im}(\bm{\Phi})$), or equivalently to lie in the range of $\mathbf{I}-\mathbf{P}_{k'}$ (i.e. the projector onto the orthogonal subspace). 
This is easily achieved by letting $\X=(\mathbf{I}-\mathbf{P}_{k'})\mathbf{Y}$, and solving the generalized eigenvalue problem:
\begin{align}\label{eq:eigha}
\tilde{\Q} \mathbf{Y} = \tilde{\A} \mathbf{Y} \bm{\Lambda}, 
\end{align}
where $\tilde{\Q} = (\mathbf{I}-\mathbf{P}_{k'})^\top (\mathbf{W}+\mu_R\A\mathrm{diag}(\mathbf{v})) (\mathbf{I}-\mathbf{P}_{k'})$ and $\tilde{\A} = (\mathbf{I}-\mathbf{P}_{k'})^\top \A (\mathbf{I}-\mathbf{P}_{k'}) = \A (\mathbf{I}-\mathbf{P}_{k'})$.
A similar trick was recently used in \cite{maron} for computing maximum magnitude eigenvalues of a large matrix. Note that solving problem \eqref{eq:eigha} involves the explicit construction of a dense $n \times n$ matrix $\mathbf{P}_{k'}$, becoming prohibitive for large meshes.

\paragraph*{Appendix B. }
We show how to {\em efficiently} compute a global solution to the generalized eigenvalue problem:
\begin{equation}\label{eq:appge}
\Q\X = \A\X\bm{\Lambda}\,,
\end{equation}
with $\Q = \mathbf{W}+\mu_R\A\mathrm{diag}(\mathbf{v})+\mu_\bot (\A\bm{\Phi})(\A\bm{\Phi})^\top$. 

Since the operator $\Q$ is real and symmetric w.r.t. the positive semi-definite mass matrix $\A$, we employ the (globally optimal) implicitly restarted Arnoldi method (IRAM) \cite{iram} (as implemented in the ARPACK suite \cite{arpack}) for computing its first $k$ eigenpairs. The application of IRAM involves iteratively solving linear systems of the form:
\begin{equation}
\Q\mathbf{x}^{(t)} = \A\mathbf{b}^{(t)}
\end{equation}
for some given $\mathbf{b}^{(t)}\in\mathbb{R}^n$. Expressing $\Q$ in terms of the matrices $\Z=\mathbf{W}+\mu_R\A\mathrm{diag}(\mathbf{v})$ and $\B=\A\bm{\Phi}\in\mathbb{R}^{n\times k'}$, we come to:
\begin{equation}\label{eq:lin}
(\Z+\mu_\bot \B\B^\top)\mathbf{x}^{(t)} = \A\mathbf{b}^{(t)}\,.
\end{equation}
Note that matrix $\B\B^\top$ can be interpreted as a rank-$k'$ update to $\Z$ (with $k'\ll n$), allowing us to apply the Sherman-Morrison-Woodbury identity \cite{Woodbury1950}:
\begin{equation*}
(\Z+\mu_\bot \B\B^\top)^{-1} = \Z^{-1} - \mu_\bot \Z^{-1}\B \underbrace{(\Id + \mu_\bot\B^\top\Z^{-1}\B)}_{\mathbf{Y}}~\hspace{-0.1cm}^{-1} \B^\top \Z^{-1}\,.
\end{equation*}
It is important to notice that the rhs does not involve the computation of $\B\B^\top$, and only involves efficient operations with a sparse $n\times n$ matrix $\Z$ and a dense $k'\times k'$ matrix $\mathbf{Y}$.
The application of this formula for the solution of problem \eqref{eq:lin}, and in turn \eqref{eq:appge} via IRAM, is illustrated for clarity in Algorithm~\ref{alg:wood}.
A similar procedure was followed in \cite{bronstein2016consistent} for the computation of CMM.

\begin{algorithm}
Solve sparse linear system $\Z\bm{\xi} = \A\mathbf{b}^{(t)}$ for $\bm{\xi}\in\mathbb{R}^n$;\\
Solve sparse linear system $\Z\bm{\Gamma} = \mu_\bot\B$ for $\bm{\Gamma}\in\mathbb{R}^{n\times k'}$;\\
Solve dense linear system $(\Id_{k'} + \B^\top\bm{\Gamma})\bm{\eta} = \B^\top\bm{\xi}$ for $\bm{\eta}\in\mathbb{R}^{k'}$;\\
Compute final solution $\mathbf{x}^{(t)} = \bm{\xi} - \bm{\Gamma}\bm{\eta}$.
\caption{\label{alg:wood}Efficient solution of problem \eqref{eq:lin}.}
\end{algorithm}

\bibliographystyle{eg-alpha-doi}

\bibliography{egbib}

\newpage
\section*{Supplementary Material}
These pages contain proofs for Theorems 1 and 2.

\paragraph*{Proof of Theorem 1. }
Let $ \mathbf{W} $,  $ \mu_{\bot}\mathbf{A}\mathbf{P}_{k'} $ and $ \mu_{R}\mathbf{A}\mathrm{diag}(\mathbf{v}) $ be real symmetric positive semidefinite matrices of dimension $ n \times n$, and define $\Q_{v,k'} = \mathbf{W} + \mu_{\bot}\mathbf{A}\mathbf{P}_{k'}+ \mu_{R}\mathbf{A}\mathrm{diag}(\mathbf{v}) $.
Let $ 0 = \lambda_{1}(\mathbf{W}) \leq \ldots \leq \lambda _{n} (\mathbf{W})$ be the eigenvalues for the generalized eigenvalue problem of $\mathbf{W}$ and $ \lambda_{1}( \mathbf{W} + \mu_{\bot}\mathbf{A}\mathbf{P}_{k'}) \leq \ldots \leq \lambda _{n}( \mathbf{W} + \mu_{\bot}\mathbf{A}\mathbf{P}_{k'})$ and $ \lambda_{1}(\Q_{v,k'}) \leq \ldots \leq \lambda _{n}(\Q_{v,k'})$ be the generalized eigenvalues of $  \mathbf{W} + \mu_{\bot}\mathbf{A}\mathbf{P}_{k'} $ and $\Q_{v,k'}$ respectively.
We aim to prove that 
\begin{equation}
\lambda _{k'}(\mathbf{W}) \le \lambda _1(\Q_{v,k'}) \,,
\end{equation}
for some $\mu_{\bot} , \mu_{R} \in \mathbb{R}$ and for every $k' \in \left\lbrace 0, \ldots , n-1 \right\rbrace$.

We start by observing that
\begin{equation}\label{r1}
\lambda _{k'}(\mathbf{W}) \leq \lambda _{k'+1}(\mathbf{W}) = \lambda _{1}( \mathbf{W} + \mu_{\bot}\mathbf{A}\mathbf{P}_{k'})\,,
\end{equation}
where the first inequality is given by the non-decreasing ordering of the eigenvalues, and the equality on the right follows from the fact that for some choice of $\mu_{\bot}> \lambda _{k'+1}(\mathbf{W})$, $\bm{\phi}_{k'+1}$ is the minimizer of $\mathbf{x}^{\top}( \mathbf{W} + \mu_{\bot}\mathbf{A}\mathbf{P}_{k'})\mathbf{x}$ under the orthogonality conditions $\left\langle \mathbf{x}, \mathbf{x} \right\rangle _{L^{2}(\mathcal{X})} = 1$ and $\left\langle \bm{\phi} _{l}, \mathbf{x} \right\rangle _{L^{2}(\mathcal{X})} = 0$, $\forall l \in \left\lbrace 1, \ldots , k' \right\rbrace $, i.e., $( \mu_{\bot}\mathbf{A}\mathbf{P}_{k'})\mathbf{x} = \mathbf{0}$.
%

Invoking a special case of Corollary 4.3.4b in \cite{horn2012matrix} and using the fact that $\mu_{R}\mathbf{A}\mathrm{diag}(\mathbf{v})$ only has non-negative eigenvalues (being a diagonal matrix with non-negative entries), we obtain the following inequality:
\begin{equation}\label{r2}
\lambda _{1}( \mathbf{W} + \mu_{\bot}\mathbf{A}\mathbf{P}_{k'}) \leq \lambda _{1}(\mathbf{W} + \mu_{\bot}\mathbf{A}\mathbf{P}_{k'}) + \mu_{R}\mathbf{A}\mathrm{diag}(\mathbf{v})) = \lambda _{1}(\Q_{v,k'}) \,.
\end{equation}
Furthermore, this inequality is an equality if and only if $\exists \mathbf{x} \in \mathbb{R}^{n}$ s.t. $\mathbf{x} \neq 0$ and the following three conditions are satisfied:
\begin{enumerate}
\item $( \mathbf{W} + \mu_{\bot}\mathbf{A}\mathbf{P}_{k'})\mathbf{x} = \lambda _{1}( \mathbf{W} + \mu_{\bot}\mathbf{A}\mathbf{P}_{k'})\mathbf{x} $;
\item $(\Q_{v,k'})\mathbf{x} = \lambda _{1}(\Q_{v,k'})\mathbf{x} $;
\item $(\mu_{R}\mathbf{A}\mathrm{diag}(\mathbf{v}))\mathbf{x} = 0 $.
\end{enumerate}
%
Putting together \eqref{r1} and \eqref{r2} we can conclude that:
\begin{equation}
\lambda _{k'}(\mathbf{W}) \le \lambda _{k'+1}(\mathbf{W}) \le \lambda _1(\mathbf{W} + \mu_{\bot}\mathbf{A}\mathbf{P}_{k'}) \le \lambda _1(\Q _{v,k'})\,.
\end{equation}
Note that the existence of a gap is given either by the violation of any of the three conditions above, or in the presence of simple spectra, i.e., whenever $\lambda _{k'}(\mathbf{W}) \neq \lambda _{k'+1}(\mathbf{W})$.

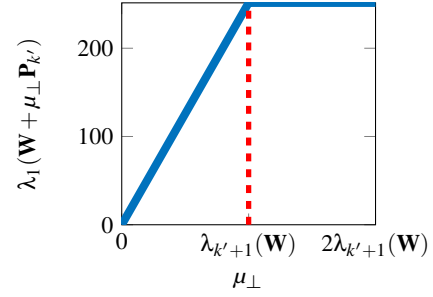
\begin{figure}[bt]
  \centering
%
%
\definecolor{mycolor1}{rgb}{0.00000,0.44700,0.74100}%
\begin{tikzpicture}

\begin{axis}[%
width=0.4\linewidth,
height=0.35\linewidth,
scale only axis,
xmin=0,
xmax=502.863493100948,
ymin=0,
ymax=251.431746550475,
axis background/.style={fill=white},
xlabel = {$\mu_{\perp}$},
ylabel = {$\lambda_{1}(\mathbf{W} + \mu_{\perp}\mathbf{P}_{k'})$},
xtick={0,251.431746550474,502.863493100948},
xticklabels = {0,$\lambda_{k'+1}(\mathbf{W})$,$2\lambda_{k'+1}(\mathbf{W})$},
]
\addplot [color=mycolor1,solid,line width=3.1pt,forget plot]
  table[row sep=crcr]{%
0	1.92652643784972e-13\\
50.2863493100948	50.2863493100949\\
100.57269862019	100.57269862019\\
150.859047930284	150.859047930284\\
201.145397240379	201.14539724038\\
251.431746550474	251.431746550473\\
301.718095860569	251.431746550475\\
352.004445170664	251.431746550474\\
402.290794480759	251.431746550474\\
452.577143790853	251.431746550474\\
502.863493100948	251.431746550474\\
};
\addplot [color=red,dashed,line width=2.0pt]
  table[row sep=crcr]{%
251.431746550474	0\\
251.431746550474	25.1431746550475\\
251.431746550474	50.2863493100949\\
251.431746550474	75.4295239651424\\
251.431746550474	100.57269862019\\
251.431746550474	125.715873275237\\
251.431746550474	150.859047930285\\
251.431746550474	176.002222585332\\
251.431746550474	201.14539724038\\
251.431746550474	226.288571895427\\
251.431746550474	251.431746550475\\
};
\end{axis}
\end{tikzpicture}%
  \caption{\label{fig:etas}Plot of $\lambda_{1}(\mathbf{W} + \mu_{\bot}\mathbf{A}\mathbf{P}_{k'})$ at increasing $\mu_{\bot}$. Note how for every $\mu_{\bot} \leq \lambda_{k'+1}(\mathbf{W})$ the frequency ($y$-axis) increases, converging at $\mu_{\bot} >\lambda_{k'+1}(\mathbf{W})$. At convergence, the orthogonality constraint (encoded in the penalty term $\mathcal{E}_\bot(\psi)$ in the LMH formulation) is satisfied.}
\end{figure}

\paragraph*{Choice of $\mu_{\bot}$.}
%
We aim to prove that for every $\mu_{\bot} > \gamma$ for some $\gamma \in \mathbb{R}^{+}$ we have:
\begin{equation}
\lambda_{1}(\mathbf{W} + \mu_{\bot}\mathbf{A}\mathbf{P}_{k'}) \geq \lambda_{k'+1}(\mathbf{W} ) \,.
\end{equation}
We can rewrite the two terms of this inequality as:
\begin{align}
\lambda_{1}(\mathbf{W} + \mu_{\bot}\mathbf{A}\mathbf{P}_{k'}) &= \underset{\left\langle \mathbf{x},\mathbf{x} \right\rangle _{L^{2}(\mathcal{X})} = 1}{\min} \mathbf{x}^{\top}(\mathbf{W} + \mu_{\bot}\mathbf{A}\mathbf{P}_{k'})\mathbf{x} \label{t1}\\
\lambda_{k'+1}(\mathbf{W} ) &= \underset{\underset{\left\langle \bm{\phi} _{i}, \mathbf{x} \right\rangle _{L^{2}(\mathcal{X})} = 0, \ \forall i = 1, \ldots, k' }{\left\langle \mathbf{x},\mathbf{x} \right\rangle _{L^{2}(\mathcal{X})} = 1}}{\min} \mathbf{x}^{\top}\mathbf{W}\mathbf{x} \,.
\end{align}
The objective in \eqref{t1} can be rewritten as:
\begin{equation}\label{sums}
\mathbf{x}^{\top}(\mathbf{W} + \mu_{\bot}\mathbf{A}\mathbf{P}_{k'})\mathbf{x} =  \mathbf{x}^{\top}\mathbf{W}\mathbf{x} + \mathbf{x}^{\top}(\mu_{\bot}\mathbf{A}\mathbf{P}_{k'})\mathbf{x} \,.
\end{equation}
We now express our vectors as the Fourier series $\mathbf{x} = \sum_{i=1}^{n}\alpha _{i} \bm{\phi}_{i}$, where $\alpha _{i} = \left\langle \bm{\phi}_i,\mathbf{x} \right\rangle _{L^{2}(\mathcal{X})}$. Noting that $\left\langle \mathbf{x},\mathbf{x} \right\rangle _{L^{2}(\mathcal{X})} = 1 $ implies $\sum_{i=1}^{n}\alpha_{i}^2 = 1$, we can write:
\begin{align}
\mathbf{x}^{\top}\mathbf{W}\mathbf{x} &= (\sum_{i=1}^{n}\alpha _{i} \bm{\phi}_{i})^{\top}\mathbf{W}(\sum_{i=1}^{n}\alpha _{i} \bm{\phi}_{i}) \\
&= (\sum_{i=1}^{n}\alpha _{i} \bm{\phi}_{i})^{\top}(\sum_{i=1}^{n}\lambda_{i}(\mathbf{W})\alpha _{i} \mathbf{A}\bm{\phi}_{i}) \\
&= \sum_{i=1}^{n}\lambda_{i}(\mathbf{W})\alpha _{i}^2 \label{rr1}\,.
\end{align}
Similarly, we can rewrite the second summand in \eqref{sums} as:
\begin{align}
\mathbf{x}^{\top}(\mu_{\bot}\mathbf{A}\mathbf{P}_{k'})\mathbf{x} &= (\sum_{i=1}^{n}\alpha _{i} \bm{\phi}_{i})^{\top}(\mu_{\bot}\mathbf{A}\mathbf{P}_{k'})(\sum_{i=1}^{n}\alpha _{i} \bm{\phi}_{i})\\
& = \mu_{\bot}(\sum_{i=1}^{n}\alpha _{i} \bm{\phi}_{i})^{\top}(\mathbf{A}\bm{\Phi}\bm{\Phi}^{\top}\mathbf{A})(\sum_{i=1}^{n}\alpha _{i} \bm{\phi}_{i}) \\
&= \mu_{\bot}\Big((\sum_{i=1}^{n}\alpha _{i} \bm{\phi}_{i})^{\top}\mathbf{A}\bm{\Phi}\Big) \Big(\bm{\Phi}^{\top}\mathbf{A}(\sum_{i=1}^{n}\alpha _{i} \bm{\phi}_{i})\Big)\\
& = \mu_{\bot}\left[ \alpha_{1} , \ldots , \alpha_{k'} \right] \left[\alpha_{1} , \ldots , \alpha_{k'} \right] ^{\top}\\
& = \mu_{\bot}\sum_{i=1}^{k'}\alpha _{i}^2 \label{rr2}\,.
\end{align}
From \eqref{rr1} and \eqref{rr2} we can conclude:
\begin{align}
\mathbf{x}^{\top}(\mathbf{W} + \mu_{\bot}\mathbf{A}\mathbf{P}_{k'})\mathbf{x} &= \mathbf{x}^{\top}\mathbf{W}\mathbf{x} + \mathbf{x}^{\top}(\mu_{\bot}\mathbf{A}\mathbf{P}_{k'})\mathbf{x} \\
&= \sum_{i=1}^{n}\lambda_{i}(\mathbf{W})\alpha _{i}^2 + \mu_{\bot}\sum_{i=1}^{k'}\alpha _{i}^2 \,.
\end{align}
At this point we split the proof in three different cases:
\begin{enumerate}
\item $\left\langle \bm{\phi} _{i}, \mathbf{x} \right\rangle _{L^{2}(\mathcal{X})} = 0, \ \forall i = 1, \ldots, k' $, that is equivalent to ask that $\mathbf{P}_{k'}\mathbf{x} = \mathbf{0}$. In this case we have:
\begin{align}
\lambda_{1}(\mathbf{W} &+ \mu_{\bot}\mathbf{A}\mathbf{P}_{k'}) = \underset{\left\langle \mathbf{x},\mathbf{x} \right\rangle _{L^{2}(\mathcal{X})} = 1}{\min} \mathbf{x}^{\top}(\mathbf{W} + \mu_{\bot}\mathbf{A}\mathbf{P}_{k'})\mathbf{x} \\
&=\underset{\underset{\left\langle \bm{\phi} _{i}, \mathbf{x} \right\rangle _{L^{2}(\mathcal{X})} = 0, \ \forall i = 1, \ldots, k' }{\left\langle \mathbf{x},\mathbf{x} \right\rangle _{L^{2}(\mathcal{X})} = 1}}{\min} (\mathbf{x}^{\top}(\mathbf{W} + \mu_{\bot}\mathbf{A}\mathbf{P}_{k'})\mathbf{x}) \\
&=
\underset{\underset{\left\langle \bm{\phi} _{i}, \mathbf{x} \right\rangle _{L^{2}(\mathcal{X})} = 0, \ \forall i = 1, \ldots, k' }{\left\langle \mathbf{x},\mathbf{x} \right\rangle _{L^{2}(\mathcal{X})} = 1}}{\min} \mathbf{x}^{\top}\mathbf{W}\mathbf{x} = \lambda_{k'+1}(\mathbf{W} )\,.
\end{align}
\item $\mathbf{x} \in span(\bm{\phi}_{1}, \ldots , \bm{\phi}_{k'})$, implying that $\alpha_{i} = 0 \ \forall i > k'$ and hence $\mathbf{x} = \sum_{i=1}^{k'}\alpha _{i} \bm{\phi}_{i}$. We get:
\begin{equation}
\mathbf{x}^{\top}(\mathbf{W} + \mu_{\bot}\mathbf{A}\mathbf{P}_{k'})\mathbf{x} = \sum_{i=1}^{k'}\lambda_{i}(\mathbf{W})\alpha _{i}^2 + \mu_{\bot}\sum_{i=1}^{k'}\alpha _{i}^2 \,.
\end{equation}
Since we take the minimum over the $\mathbf{x}$ s.t. $\left\langle \mathbf{x},\mathbf{x} \right\rangle _{L^{2}(\mathcal{X})} = 1$ we have $\sum_{i=1}^{k'}\alpha _{i}^2 = 1 $ and:
\begin{equation}
\mathbf{x}^{\top}(\mathbf{W} + \mu_{\bot}\mathbf{A}\mathbf{P}_{k'})\mathbf{x} = \sum_{i=1}^{k'}\lambda_{i}(\mathbf{W})\alpha _{i}^2 + \mu_{\bot} \geq \mu_{\bot}  \,,
\end{equation}
where the equality is realized for $\mathbf{x} = \bm{\phi}_{1}$ since $\lambda_{1}(\mathbf{W}) = 0$, and all other cases yield $\mu_{\bot}$ plus some non-negative quantity. We get to:
\begin{equation}
\lambda_{1}(\mathbf{W} + \mu_{\bot}\mathbf{A}\mathbf{P}_{k'}) = \underset{\left\langle \mathbf{x},\mathbf{x} \right\rangle _{L^{2}(\mathcal{X})} = 1}{\min} \mathbf{x}^{\top}(\mathbf{W} + \mu_{\bot}\mathbf{A}\mathbf{P}_{k'})\mathbf{x} = \mu_{\bot} \,.
\end{equation}
\item For the last case we have $\left\langle \bm{\phi} _{i}, \mathbf{x} \right\rangle _{L^{2}(\mathcal{X})} \neq 0 $ for at least one $i = 1, \ldots , k'$   and for at least one $i > k'$ at the same time. 
\begin{align}
\mathbf{x}^{\top}(\mathbf{W} &+ \mu_{\bot}\mathbf{A}\mathbf{P}_{k'})\mathbf{x} = \sum_{i=1}^{n}\lambda_{i}(\mathbf{W})\alpha _{i}^2 + \mu_{\bot}\sum_{i=1}^{k'}\alpha _{i}^2 \\
&= \sum_{i=1}^{k'}\lambda_{i}(\mathbf{W})\alpha _{i}^2 + \sum_{i=k'+1}^{n}\lambda_{i}(\mathbf{W})\alpha _{i}^2 + \mu_{\bot}\sum_{i=1}^{k'}\alpha _{i}^2 \\&= \sum_{i=1}^{k'}(\lambda_{i}(\mathbf{W}) + \mu_{\bot})\alpha _{i}^2 + \sum_{i=k'+1}^{n}\lambda_{i}(\mathbf{W})\alpha _{i}^2 \,.
\end{align}
Since $\lambda_{i}(\mathbf{W}) \geq \lambda_{k'+1}(\mathbf{W})$, $\forall i \geq k'+1$ we can write:
\begin{align}
\mathbf{x}^{\top}(\mathbf{W} &+ \mu_{\bot}\mathbf{A}\mathbf{P}_{k'})\mathbf{x} \\
&= \sum_{i=1}^{k'}(\lambda_{i}(\mathbf{W}) + \mu_{\bot})\alpha _{i}^2 + \sum_{i=k'+1}^{n}\lambda_{i}(\mathbf{W})\alpha _{i}^2 \\
&\geq
 \sum_{i=1}^{k'}(\lambda_{i}(\mathbf{W}) + \mu_{\bot})\alpha _{i}^2 + \lambda_{k'+1}(\mathbf{W})\sum_{i=k'+1}^{n}\alpha _{i}^2\\&\geq
\sum_{i=1}^{k'}\mu_{\bot}\alpha _{i}^2 + \lambda_{k'+1}(\mathbf{W})\sum_{i=k'+1}^{n}\alpha _{i}^2
\,.
\end{align}
If we take $\mu_{\bot} > \lambda_{k'+1}(\mathbf{W}) $ in order to satisfy the condition imposed by case 2, we get:
\begin{align}
\mathbf{x}^{\top}(\mathbf{W} &+ \mu_{\bot}\mathbf{A}\mathbf{P}_{k'})\mathbf{x}  \geq
\sum_{i=1}^{k'}\mu_{\bot}\alpha _{i}^2 + \lambda_{k'+1}(\mathbf{W})\sum_{i=k'+1}^{n}\alpha _{i}^2 \\
&> \lambda_{k'+1}(\mathbf{W})\sum_{i=1}^{k'}\alpha _{i}^2 + \lambda_{k'+1}(\mathbf{W})\sum_{i=k'+1}^{n}\alpha _{i}^2 \\&= \lambda_{k'+1}(\mathbf{W})\sum_{i=1}^{n}\alpha _{i}^2\\& = \lambda_{k'+1}(\mathbf{W})
\,.
\end{align}
We can therefore conclude that
\begin{align}
\lambda_{1}(\mathbf{W} + \mu_{\bot}\mathbf{A}\mathbf{P}_{k'}) &= \underset{\left\langle \mathbf{x},\mathbf{x} \right\rangle _{L^{2}(\mathcal{X})} = 1}{\min} \mathbf{x}^{\top}(\mathbf{W} + \mu_{\bot}\mathbf{A}\mathbf{P}_{k'})\mathbf{x}\\
& > \lambda_{k'+1}(\mathbf{W}) \ \mathit{if} \ \mu_{\bot} > \lambda_{k'+1}(\mathbf{W})\,.
\end{align}
\end{enumerate}
In Figure \ref{fig:etas} we show an empirical evaluation across several choices of $\mu _{\bot}$.
%
%
\paragraph*{Proof of Theorem 2. }
We want to show that $\forall k \in \left\lbrace 1, 2, \ldots ,n \right\rbrace $ we have the following upper bound:
$$ \lambda _{i}(\Q _{v,k'}) \leq \lambda_{i+k'}(\mathbf{W}^{R}) \ .$$
Similarly to Theorem 1, the proof follows directly from Corollary 4.3.4b in \cite{horn2012matrix}, which specialized to our case reads:
\begin{equation} 
\lambda _{i}(\mathbf{W}^{R}+\mu_{\bot}\mathbf{A}\mathbf{P}_{k'}) \leq \lambda_{i+\pi}(\mathbf{W}^{R})\,,
\end{equation}
where $\pi$ is the number of positive eigenvalues of $\mu_{\bot}\mathbf{A}\mathbf{P}_{k'}$. Since $\Q _{v,k'} = \mathbf{W}^{R}+\mu_{\bot}\mathbf{A}\mathbf{P}_{k'}$ and using the fact that $\mu_{\bot}\mathbf{A}\mathbf{P}_{k'}$ is a positive semidefinite matrix with rank $k'$, we have $\pi=k'$, leading to:
\begin{equation}
\lambda _{i}(\Q _{v,k'}) = \lambda _{i}(\mathbf{W}^{R}+\mu_{\bot}\mathbf{A}\mathbf{P}_{k'}) \leq \lambda_{i+\pi}(\mathbf{W}^{R}) = \lambda_{i+k'}(\mathbf{W}^{R})\,.
\end{equation}

\end{document}